
\catcode`\@=11


\message{Loading jyTeX fonts...}



\font\vptrm=cmr5 \font\vptmit=cmmi5 \font\vptsy=cmsy5 \font\vptbf=cmbx5

\skewchar\vptmit='177 \skewchar\vptsy='60 \fontdimen16
\vptsy=\the\fontdimen17 \vptsy

\def\vpt{\ifmmode\err@badsizechange\else
     \@mathfontinit
     \textfont0=\vptrm  \scriptfont0=\vptrm  \scriptscriptfont0=\vptrm
     \textfont1=\vptmit \scriptfont1=\vptmit \scriptscriptfont1=\vptmit
     \textfont2=\vptsy  \scriptfont2=\vptsy  \scriptscriptfont2=\vptsy
     \textfont3=\xptex  \scriptfont3=\xptex  \scriptscriptfont3=\xptex
     \textfont\bffam=\vptbf
     \scriptfont\bffam=\vptbf
     \scriptscriptfont\bffam=\vptbf
     \@fontstyleinit
     \def\rm{\vptrm\fam=\z@}%
     \def\bf{\vptbf\fam=\bffam}%
     \def\oldstyle{\vptmit\fam=\@ne}%
     \rm\fi}


\font\viptrm=cmr6 \font\viptmit=cmmi6 \font\viptsy=cmsy6
\font\viptbf=cmbx6

\skewchar\viptmit='177 \skewchar\viptsy='60 \fontdimen16
\viptsy=\the\fontdimen17 \viptsy

\def\vipt{\ifmmode\err@badsizechange\else
     \@mathfontinit
     \textfont0=\viptrm  \scriptfont0=\vptrm  \scriptscriptfont0=\vptrm
     \textfont1=\viptmit \scriptfont1=\vptmit \scriptscriptfont1=\vptmit
     \textfont2=\viptsy  \scriptfont2=\vptsy  \scriptscriptfont2=\vptsy
     \textfont3=\xptex   \scriptfont3=\xptex  \scriptscriptfont3=\xptex
     \textfont\bffam=\viptbf
     \scriptfont\bffam=\vptbf
     \scriptscriptfont\bffam=\vptbf
     \@fontstyleinit
     \def\rm{\viptrm\fam=\z@}%
     \def\bf{\viptbf\fam=\bffam}%
     \def\oldstyle{\viptmit\fam=\@ne}%
     \rm\fi}

\font\viiptrm=cmr7 \font\viiptmit=cmmi7 \font\viiptsy=cmsy7
\font\viiptit=cmti7 \font\viiptbf=cmbx7

\skewchar\viiptmit='177 \skewchar\viiptsy='60 \fontdimen16
\viiptsy=\the\fontdimen17 \viiptsy

\def\viipt{\ifmmode\err@badsizechange\else
     \@mathfontinit
     \textfont0=\viiptrm  \scriptfont0=\vptrm  \scriptscriptfont0=\vptrm
     \textfont1=\viiptmit \scriptfont1=\vptmit \scriptscriptfont1=\vptmit
     \textfont2=\viiptsy  \scriptfont2=\vptsy  \scriptscriptfont2=\vptsy
     \textfont3=\xptex    \scriptfont3=\xptex  \scriptscriptfont3=\xptex
     \textfont\itfam=\viiptit
     \scriptfont\itfam=\viiptit
     \scriptscriptfont\itfam=\viiptit
     \textfont\bffam=\viiptbf
     \scriptfont\bffam=\vptbf
     \scriptscriptfont\bffam=\vptbf
     \@fontstyleinit
     \def\rm{\viiptrm\fam=\z@}%
     \def\it{\viiptit\fam=\itfam}%
     \def\bf{\viiptbf\fam=\bffam}%
     \def\oldstyle{\viiptmit\fam=\@ne}%
     \rm\fi}


\font\viiiptrm=cmr8 \font\viiiptmit=cmmi8 \font\viiiptsy=cmsy8
\font\viiiptit=cmti8
\font\viiiptbf=cmbx8

\skewchar\viiiptmit='177 \skewchar\viiiptsy='60 \fontdimen16
\viiiptsy=\the\fontdimen17 \viiiptsy

\def\viiipt{\ifmmode\err@badsizechange\else
     \@mathfontinit
     \textfont0=\viiiptrm  \scriptfont0=\viptrm  \scriptscriptfont0=\vptrm
     \textfont1=\viiiptmit \scriptfont1=\viptmit \scriptscriptfont1=\vptmit
     \textfont2=\viiiptsy  \scriptfont2=\viptsy  \scriptscriptfont2=\vptsy
     \textfont3=\xptex     \scriptfont3=\xptex   \scriptscriptfont3=\xptex
     \textfont\itfam=\viiiptit
     \scriptfont\itfam=\viiptit
     \scriptscriptfont\itfam=\viiptit
     \textfont\bffam=\viiiptbf
     \scriptfont\bffam=\viptbf
     \scriptscriptfont\bffam=\vptbf
     \@fontstyleinit
     \def\rm{\viiiptrm\fam=\z@}%
     \def\it{\viiiptit\fam=\itfam}%
     \def\bf{\viiiptbf\fam=\bffam}%
     \def\oldstyle{\viiiptmit\fam=\@ne}%
     \rm\fi}


\def\getixpt{%
     \font\ixptrm=cmr9
     \font\ixptmit=cmmi9
     \font\ixptsy=cmsy9
     \font\ixptit=cmti9
     \font\ixptbf=cmbx9
     \skewchar\ixptmit='177 \skewchar\ixptsy='60
     \fontdimen16 \ixptsy=\the\fontdimen17 \ixptsy}

\def\ixpt{\ifmmode\err@badsizechange\else
     \@mathfontinit
     \textfont0=\ixptrm  \scriptfont0=\viiptrm  \scriptscriptfont0=\vptrm
     \textfont1=\ixptmit \scriptfont1=\viiptmit \scriptscriptfont1=\vptmit
     \textfont2=\ixptsy  \scriptfont2=\viiptsy  \scriptscriptfont2=\vptsy
     \textfont3=\xptex   \scriptfont3=\xptex    \scriptscriptfont3=\xptex
     \textfont\itfam=\ixptit
     \scriptfont\itfam=\viiptit
     \scriptscriptfont\itfam=\viiptit
     \textfont\bffam=\ixptbf
     \scriptfont\bffam=\viiptbf
     \scriptscriptfont\bffam=\vptbf
     \@fontstyleinit
     \def\rm{\ixptrm\fam=\z@}%
     \def\it{\ixptit\fam=\itfam}%
     \def\bf{\ixptbf\fam=\bffam}%
     \def\oldstyle{\ixptmit\fam=\@ne}%
     \rm\fi}


\font\xptrm=cmr10 \font\xptmit=cmmi10 \font\xptsy=cmsy10
\font\xptex=cmex10 \font\xptit=cmti10 \font\xptsl=cmsl10
\font\xptbf=cmbx10 \font\xpttt=cmtt10 \font\xptss=cmss10
\font\xptsc=cmcsc10 \font\xptbfs=cmb10 \font\xptbmit=cmmib10

\skewchar\xptmit='177 \skewchar\xptbmit='177 \skewchar\xptsy='60
\fontdimen16 \xptsy=\the\fontdimen17 \xptsy

\def\xpt{\ifmmode\err@badsizechange\else
     \@mathfontinit
     \textfont0=\xptrm  \scriptfont0=\viiptrm  \scriptscriptfont0=\vptrm
     \textfont1=\xptmit \scriptfont1=\viiptmit \scriptscriptfont1=\vptmit
     \textfont2=\xptsy  \scriptfont2=\viiptsy  \scriptscriptfont2=\vptsy
     \textfont3=\xptex  \scriptfont3=\xptex    \scriptscriptfont3=\xptex
     \textfont\itfam=\xptit
     \scriptfont\itfam=\viiptit
     \scriptscriptfont\itfam=\viiptit
     \textfont\bffam=\xptbf
     \scriptfont\bffam=\viiptbf
     \scriptscriptfont\bffam=\vptbf
     \textfont\bfsfam=\xptbfs
     \scriptfont\bfsfam=\viiptbf
     \scriptscriptfont\bfsfam=\vptbf
     \textfont\bmitfam=\xptbmit
     \scriptfont\bmitfam=\viiptmit
     \scriptscriptfont\bmitfam=\vptmit
     \@fontstyleinit
     \def\rm{\xptrm\fam=\z@}%
     \def\it{\xptit\fam=\itfam}%
     \def\sl{\xptsl}%
     \def\bf{\xptbf\fam=\bffam}%
     \def\tt{\xpttt}%
     \def\ss{\xptss}%
     \def\sc{\xptsc}%
     \def\bfs{\xptbfs\fam=\bfsfam}%
     \def\bmit{\fam=\bmitfam}%
     \def\oldstyle{\xptmit\fam=\@ne}%
     \rm\fi}


\def\getxipt{%
     \font\xiptrm=cmr10  scaled\magstephalf
     \font\xiptmit=cmmi10 scaled\magstephalf
     \font\xiptsy=cmsy10 scaled\magstephalf
     \font\xiptex=cmex10 scaled\magstephalf
     \font\xiptit=cmti10 scaled\magstephalf
     \font\xiptsl=cmsl10 scaled\magstephalf
     \font\xiptbf=cmbx10 scaled\magstephalf
     \font\xipttt=cmtt10 scaled\magstephalf
     \font\xiptss=cmss10 scaled\magstephalf
     \skewchar\xiptmit='177 \skewchar\xiptsy='60
     \fontdimen16 \xiptsy=\the\fontdimen17 \xiptsy}

\def\xipt{\ifmmode\err@badsizechange\else
     \@mathfontinit
     \textfont0=\xiptrm  \scriptfont0=\viiiptrm  \scriptscriptfont0=\viptrm
     \textfont1=\xiptmit \scriptfont1=\viiiptmit \scriptscriptfont1=\viptmit
     \textfont2=\xiptsy  \scriptfont2=\viiiptsy  \scriptscriptfont2=\viptsy
     \textfont3=\xiptex  \scriptfont3=\xptex     \scriptscriptfont3=\xptex
     \textfont\itfam=\xiptit
     \scriptfont\itfam=\viiiptit
     \scriptscriptfont\itfam=\viiptit
     \textfont\bffam=\xiptbf
     \scriptfont\bffam=\viiiptbf
     \scriptscriptfont\bffam=\viptbf
     \@fontstyleinit
     \def\rm{\xiptrm\fam=\z@}%
     \def\it{\xiptit\fam=\itfam}%
     \def\sl{\xiptsl}%
     \def\bf{\xiptbf\fam=\bffam}%
     \def\tt{\xipttt}%
     \def\ss{\xiptss}%
     \def\oldstyle{\xiptmit\fam=\@ne}%
     \rm\fi}


\font\xiiptrm=cmr12 \font\xiiptmit=cmmi12 \font\xiiptsy=cmsy10
scaled\magstep1 \font\xiiptex=cmex10  scaled\magstep1 \font\xiiptit=cmti12
\font\xiiptsl=cmsl12 \font\xiiptbf=cmbx12
\font\xiiptss=cmss12 \font\xiiptsc=cmcsc10 scaled\magstep1
\font\xiiptbfs=cmb10  scaled\magstep1 \font\xiiptbmit=cmmib10
scaled\magstep1

\skewchar\xiiptmit='177 \skewchar\xiiptbmit='177 \skewchar\xiiptsy='60
\fontdimen16 \xiiptsy=\the\fontdimen17 \xiiptsy

\def\xiipt{\ifmmode\err@badsizechange\else
     \@mathfontinit
     \textfont0=\xiiptrm  \scriptfont0=\viiiptrm  \scriptscriptfont0=\viptrm
     \textfont1=\xiiptmit \scriptfont1=\viiiptmit \scriptscriptfont1=\viptmit
     \textfont2=\xiiptsy  \scriptfont2=\viiiptsy  \scriptscriptfont2=\viptsy
     \textfont3=\xiiptex  \scriptfont3=\xptex     \scriptscriptfont3=\xptex
     \textfont\itfam=\xiiptit
     \scriptfont\itfam=\viiiptit
     \scriptscriptfont\itfam=\viiptit
     \textfont\bffam=\xiiptbf
     \scriptfont\bffam=\viiiptbf
     \scriptscriptfont\bffam=\viptbf
     \textfont\bfsfam=\xiiptbfs
     \scriptfont\bfsfam=\viiiptbf
     \scriptscriptfont\bfsfam=\viptbf
     \textfont\bmitfam=\xiiptbmit
     \scriptfont\bmitfam=\viiiptmit
     \scriptscriptfont\bmitfam=\viptmit
     \@fontstyleinit
     \def\rm{\xiiptrm\fam=\z@}%
     \def\it{\xiiptit\fam=\itfam}%
     \def\sl{\xiiptsl}%
     \def\bf{\xiiptbf\fam=\bffam}%
     \def\tt{\xiipttt}%
     \def\ss{\xiiptss}%
     \def\sc{\xiiptsc}%
     \def\bfs{\xiiptbfs\fam=\bfsfam}%
     \def\bmit{\fam=\bmitfam}%
     \def\oldstyle{\xiiptmit\fam=\@ne}%
     \rm\fi}


\def\getxiiipt{%
     \font\xiiiptrm=cmr12  scaled\magstephalf
     \font\xiiiptmit=cmmi12 scaled\magstephalf
     \font\xiiiptsy=cmsy9  scaled\magstep2
     \font\xiiiptit=cmti12 scaled\magstephalf
     \font\xiiiptsl=cmsl12 scaled\magstephalf
     \font\xiiiptbf=cmbx12 scaled\magstephalf
     \font\xiiipttt=cmtt12 scaled\magstephalf
     \font\xiiiptss=cmss12 scaled\magstephalf
     \skewchar\xiiiptmit='177 \skewchar\xiiiptsy='60
     \fontdimen16 \xiiiptsy=\the\fontdimen17 \xiiiptsy}

\def\xiiipt{\ifmmode\err@badsizechange\else
     \@mathfontinit
     \textfont0=\xiiiptrm  \scriptfont0=\xptrm  \scriptscriptfont0=\viiptrm
     \textfont1=\xiiiptmit \scriptfont1=\xptmit \scriptscriptfont1=\viiptmit
     \textfont2=\xiiiptsy  \scriptfont2=\xptsy  \scriptscriptfont2=\viiptsy
     \textfont3=\xivptex   \scriptfont3=\xptex  \scriptscriptfont3=\xptex
     \textfont\itfam=\xiiiptit
     \scriptfont\itfam=\xptit
     \scriptscriptfont\itfam=\viiptit
     \textfont\bffam=\xiiiptbf
     \scriptfont\bffam=\xptbf
     \scriptscriptfont\bffam=\viiptbf
     \@fontstyleinit
     \def\rm{\xiiiptrm\fam=\z@}%
     \def\it{\xiiiptit\fam=\itfam}%
     \def\sl{\xiiiptsl}%
     \def\bf{\xiiiptbf\fam=\bffam}%
     \def\tt{\xiiipttt}%
     \def\ss{\xiiiptss}%
     \def\oldstyle{\xiiiptmit\fam=\@ne}%
     \rm\fi}


\font\xivptrm=cmr12   scaled\magstep1 \font\xivptmit=cmmi12
scaled\magstep1 \font\xivptsy=cmsy10  scaled\magstep2
\font\xivptex=cmex10  scaled\magstep2 \font\xivptit=cmti12 scaled\magstep1
\font\xivptsl=cmsl12  scaled\magstep1 \font\xivptbf=cmbx12
scaled\magstep1
\font\xivptss=cmss12  scaled\magstep1 \font\xivptsc=cmcsc10
scaled\magstep2 \font\xivptbfs=cmb10  scaled\magstep2
\font\xivptbmit=cmmib10 scaled\magstep2

\skewchar\xivptmit='177 \skewchar\xivptbmit='177 \skewchar\xivptsy='60
\fontdimen16 \xivptsy=\the\fontdimen17 \xivptsy

\def\xivpt{\ifmmode\err@badsizechange\else
     \@mathfontinit
     \textfont0=\xivptrm  \scriptfont0=\xptrm  \scriptscriptfont0=\viiptrm
     \textfont1=\xivptmit \scriptfont1=\xptmit \scriptscriptfont1=\viiptmit
     \textfont2=\xivptsy  \scriptfont2=\xptsy  \scriptscriptfont2=\viiptsy
     \textfont3=\xivptex  \scriptfont3=\xptex  \scriptscriptfont3=\xptex
     \textfont\itfam=\xivptit
     \scriptfont\itfam=\xptit
     \scriptscriptfont\itfam=\viiptit
     \textfont\bffam=\xivptbf
     \scriptfont\bffam=\xptbf
     \scriptscriptfont\bffam=\viiptbf
     \textfont\bfsfam=\xivptbfs
     \scriptfont\bfsfam=\xptbfs
     \scriptscriptfont\bfsfam=\viiptbf
     \textfont\bmitfam=\xivptbmit
     \scriptfont\bmitfam=\xptbmit
     \scriptscriptfont\bmitfam=\viiptmit
     \@fontstyleinit
     \def\rm{\xivptrm\fam=\z@}%
     \def\it{\xivptit\fam=\itfam}%
     \def\sl{\xivptsl}%
     \def\bf{\xivptbf\fam=\bffam}%
     \def\tt{\xivpttt}%
     \def\ss{\xivptss}%
     \def\sc{\xivptsc}%
     \def\bfs{\xivptbfs\fam=\bfsfam}%
     \def\bmit{\fam=\bmitfam}%
     \def\oldstyle{\xivptmit\fam=\@ne}%
     \rm\fi}


\font\xviiptrm=cmr17 \font\xviiptmit=cmmi12 scaled\magstep2
\font\xviiptsy=cmsy10 scaled\magstep3 \font\xviiptex=cmex10
scaled\magstep3 \font\xviiptit=cmti12 scaled\magstep2
\font\xviiptbf=cmbx12 scaled\magstep2 \font\xviiptbfs=cmb10
scaled\magstep3

\skewchar\xviiptmit='177 \skewchar\xviiptsy='60 \fontdimen16
\xviiptsy=\the\fontdimen17 \xviiptsy

\def\xviipt{\ifmmode\err@badsizechange\else
     \@mathfontinit
     \textfont0=\xviiptrm  \scriptfont0=\xiiptrm  \scriptscriptfont0=\viiiptrm
     \textfont1=\xviiptmit \scriptfont1=\xiiptmit \scriptscriptfont1=\viiiptmit
     \textfont2=\xviiptsy  \scriptfont2=\xiiptsy  \scriptscriptfont2=\viiiptsy
     \textfont3=\xviiptex  \scriptfont3=\xiiptex  \scriptscriptfont3=\xptex
     \textfont\itfam=\xviiptit
     \scriptfont\itfam=\xiiptit
     \scriptscriptfont\itfam=\viiiptit
     \textfont\bffam=\xviiptbf
     \scriptfont\bffam=\xiiptbf
     \scriptscriptfont\bffam=\viiiptbf
     \textfont\bfsfam=\xviiptbfs
     \scriptfont\bfsfam=\xiiptbfs
     \scriptscriptfont\bfsfam=\viiiptbf
     \@fontstyleinit
     \def\rm{\xviiptrm\fam=\z@}%
     \def\it{\xviiptit\fam=\itfam}%
     \def\bf{\xviiptbf\fam=\bffam}%
     \def\bfs{\xviiptbfs\fam=\bfsfam}%
     \def\oldstyle{\xviiptmit\fam=\@ne}%
     \rm\fi}


\font\xxiptrm=cmr17  scaled\magstep1


\def\xxipt{\ifmmode\err@badsizechange\else
     \@mathfontinit
     \@fontstyleinit
     \def\rm{\xxiptrm\fam=\z@}%
     \rm\fi}


\font\xxvptrm=cmr17  scaled\magstep2


\def\xxvpt{\ifmmode\err@badsizechange\else
     \@mathfontinit
     \@fontstyleinit
     \def\rm{\xxvptrm\fam=\z@}%
     \rm\fi}




\message{Loading jyTeX macros...}

\message{modifications to plain.tex,}


\def\newcount{\alloc@0\count\countdef\insc@unt}
\def\newdimen{\alloc@1\dimen\dimendef\insc@unt}
\def\newskip{\alloc@2\skip\skipdef\insc@unt}
\def\newmuskip{\alloc@3\muskip\muskipdef\@cclvi}
\def\newbox{\alloc@4\box\chardef\insc@unt}
\def\newtoks{\alloc@5\toks\toksdef\@cclvi}
\def\newhelp#1#2{\newtoks#1\global#1\expandafter{\csname#2\endcsname}}
\def\newread{\alloc@6\read\chardef\sixt@@n}
\def\newwrite{\alloc@7\write\chardef\sixt@@n}
\def\newfam{\alloc@8\fam\chardef\sixt@@n}
\def\newinsert#1{\global\advance\insc@unt by\m@ne
     \ch@ck0\insc@unt\count
     \ch@ck1\insc@unt\dimen
     \ch@ck2\insc@unt\skip
     \ch@ck4\insc@unt\box
     \allocationnumber=\insc@unt
     \global\chardef#1=\allocationnumber
     \wlog{\string#1=\string\insert\the\allocationnumber}}
\def\newif#1{\count@\escapechar \escapechar\m@ne
     \expandafter\expandafter\expandafter
          \xdef\@if#1{true}{\let\noexpand#1=\noexpand\iftrue}%
     \expandafter\expandafter\expandafter
          \xdef\@if#1{false}{\let\noexpand#1=\noexpand\iffalse}%
     \global\@if#1{false}\escapechar=\count@}


\newlinechar=`\^^J
\overfullrule=0pt




\let\itfam=\undefined

\let\bffam=\undefined

\count18=3


\chardef\sharps="19


\mathchardef\alpha="710B \mathchardef\beta="710C
\mathchardef\gamma="710D \mathchardef\delta="710E
\mathchardef\epsilon="710F \mathchardef\zeta="7110
\mathchardef\eta="7111 \mathchardef\theta="7112 \mathchardef\iota="7113
\mathchardef\kappa="7114 \mathchardef\lambda="7115
\mathchardef\mu="7116 \mathchardef\nu="7117 \mathchardef\xi="7118
\mathchardef\pi="7119 \mathchardef\rho="711A \mathchardef\sigma="711B
\mathchardef\tau="711C \mathchardef\upsilon="711D
\mathchardef\phi="711E \mathchardef\chi="711F \mathchardef\psi="7120
\mathchardef\omega="7121 \mathchardef\varepsilon="7122
\mathchardef\vartheta="7123 \mathchardef\varpi="7124
\mathchardef\varrho="7125 \mathchardef\varsigma="7126
\mathchardef\varphi="7127 \mathchardef\imath="717B
\mathchardef\jmath="717C \mathchardef\ell="7160 \mathchardef\wp="717D
\mathchardef\partial="7140 \mathchardef\flat="715B
\mathchardef\natural="715C \mathchardef\sharp="715D



\def\angle{{\vbox{\ialign{$\m@th\scriptstyle##$\crcr
     \not\mathrel{\mkern14mu}\crcr
     \noalign{\nointerlineskip}
     \mkern2.5mu\leaders\hrule height.34\rp@\hfill\mkern2.5mu\crcr}}}}
\def\vdots{\vbox{\baselineskip4\rp@ \lineskiplimit\z@
     \kern6\rp@\hbox{.}\hbox{.}\hbox{.}}}
\def\ddots{\mathinner{\mkern1mu\raise7\rp@\vbox{\kern7\rp@\hbox{.}}\mkern2mu
     \raise4\rp@\hbox{.}\mkern2mu\raise\rp@\hbox{.}\mkern1mu}}
\def\overrightarrow#1{\vbox{\ialign{##\crcr
     \rightarrowfill\crcr
     \noalign{\kern-\rp@\nointerlineskip}
     $\hfil\displaystyle{#1}\hfil$\crcr}}}
\def\overleftarrow#1{\vbox{\ialign{##\crcr
     \leftarrowfill\crcr
     \noalign{\kern-\rp@\nointerlineskip}
     $\hfil\displaystyle{#1}\hfil$\crcr}}}
\def\overbrace#1{\mathop{\vbox{\ialign{##\crcr
     \noalign{\kern3\rp@}
     \downbracefill\crcr
     \noalign{\kern3\rp@\nointerlineskip}
     $\hfil\displaystyle{#1}\hfil$\crcr}}}\limits}
\def\underbrace#1{\mathop{\vtop{\ialign{##\crcr
     $\hfil\displaystyle{#1}\hfil$\crcr
     \noalign{\kern3\rp@\nointerlineskip}
     \upbracefill\crcr
     \noalign{\kern3\rp@}}}}\limits}
\def\big#1{{\hbox{$\left#1\vbox to8.5\rp@ {}\right.\n@space$}}}
\def\Big#1{{\hbox{$\left#1\vbox to11.5\rp@ {}\right.\n@space$}}}
\def\bigg#1{{\hbox{$\left#1\vbox to14.5\rp@ {}\right.\n@space$}}}
\def\Bigg#1{{\hbox{$\left#1\vbox to17.5\rp@ {}\right.\n@space$}}}
\def\@vereq#1#2{\lower.5\rp@\vbox{\baselineskip\z@skip\lineskip-.5\rp@
     \ialign{$\m@th#1\hfil##\hfil$\crcr#2\crcr=\crcr}}}
\def\rlh@#1{\vcenter{\hbox{\ooalign{\raise2\rp@
     \hbox{$#1\rightharpoonup$}\crcr
     $#1\leftharpoondown$}}}}
\def\bordermatrix#1{\begingroup\m@th
     \setbox\z@\vbox{%
          \def\cr{\crcr\noalign{\kern2\rp@\global\let\cr\endline}}%
          \ialign{$##$\hfil\kern2\rp@\kern\p@renwd
               &\thinspace\hfil$##$\hfil&&\quad\hfil$##$\hfil\crcr
               \omit\strut\hfil\crcr
               \noalign{\kern-\baselineskip}%
               #1\crcr\omit\strut\cr}}%
     \setbox\tw@\vbox{\unvcopy\z@\global\setbox\@ne\lastbox}%
     \setbox\tw@\hbox{\unhbox\@ne\unskip\global\setbox\@ne\lastbox}%
     \setbox\tw@\hbox{$\kern\wd\@ne\kern-\p@renwd\left(\kern-\wd\@ne
          \global\setbox\@ne\vbox{\box\@ne\kern2\rp@}%
          \vcenter{\kern-\ht\@ne\unvbox\z@\kern-\baselineskip}%
          \,\right)$}%
     \null\;\vbox{\kern\ht\@ne\box\tw@}\endgroup}
\def\endinsert{\egroup
     \if@mid\dimen@\ht\z@
          \advance\dimen@\dp\z@
          \advance\dimen@12\rp@
          \advance\dimen@\pagetotal
          \ifdim\dimen@>\pagegoal\@midfalse\p@gefalse\fi
     \fi
     \if@mid\bigskip\box\z@
          \bigbreak
     \else\insert\topins{\penalty100 \splittopskip\z@skip
               \splitmaxdepth\maxdimen\floatingpenalty\z@
               \ifp@ge\dimen@\dp\z@
                    \vbox to\vsize{\unvbox\z@\kern-\dimen@}%
               \else\box\z@\nobreak\bigskip
               \fi}%
     \fi
     \endgroup}


\def\cases#1{\left\{\,\vcenter{\m@th
     \ialign{$##\hfil$&\quad##\hfil\crcr#1\crcr}}\right.}
\def\matrix#1{\null\,\vcenter{\m@th
     \ialign{\hfil$##$\hfil&&\quad\hfil$##$\hfil\crcr
          \mathstrut\crcr
          \noalign{\kern-\baselineskip}
          #1\crcr
          \mathstrut\crcr
          \noalign{\kern-\baselineskip}}}\,}


\newif\ifraggedbottom

\def\raggedbottom{\ifraggedbottom\else
     \advance\topskip by\z@ plus60pt \raggedbottomtrue\fi}%
\def\normalbottom{\ifraggedbottom
     \advance\topskip by\z@ plus-60pt \raggedbottomfalse\fi}

\message{hacks,}


\toksdef\toks@i=1 \toksdef\toks@ii=2


\def\TeX{T\kern-.1667em \lower.5ex \hbox{E}\kern-.125em X\null}
\def\jyTeX{{\leavevmode
     \raise.587ex \hbox{\it\j}\kern-.1em \lower.048ex \hbox{\it y}\kern-.12em
     \TeX}}

\let\then=\iftrue
\def\ifnoarg#1\then{\def\hack@{#1}\ifx\hack@\empty}
\def\ifundefined#1\then{%
     \expandafter\ifx\csname\expandafter\blank\string#1\endcsname\relax}
\def\useif#1\then{\csname#1\endcsname}
\def\usename#1{\csname#1\endcsname}
\def\useafter#1#2{\expandafter#1\csname#2\endcsname}

\long\def\loop#1\repeat{\def\@iterate{#1\expandafter\@iterate\fi}\@iterate
     \let\@iterate=\relax}

\let\TeXend=\end
\def\begin#1{\begingroup\def\@@blockname{#1}\usename{begin#1}}
\def\end#1{\usename{end#1}\def\hack@{#1}%
     \ifx\@@blockname\hack@
          \endgroup
     \else\err@badgroup\hack@\@@blockname
     \fi}
\def\@@blockname{}

\def\defaultoption[#1]#2{%
     \def\hack@{\ifx\hack@ii[\toks@={#2}\else\toks@={#2[#1]}\fi\the\toks@}%
     \futurelet\hack@ii\hack@}

\def\markup#1{\let\@@marksf=\empty
     \ifhmode\edef\@@marksf{\spacefactor=\the\spacefactor\relax}\/\fi
     ${}^{\hbox{\subscriptfonts#1}}$\@@marksf}


\newtoks\shortyear
\newtoks\militaryhour
\newtoks\standardhour
\newtoks\minute
\newtoks\amorpm

\def\settime{\count@=\time\divide\count@ by60
     \militaryhour=\expandafter{\number\count@}%
     {\multiply\count@ by-60 \advance\count@ by\time
          \xdef\hack@{\ifnum\count@<10 0\fi\number\count@}}%
     \minute=\expandafter{\hack@}%
     \ifnum\count@<12
          \amorpm={am}
     \else\amorpm={pm}
          \ifnum\count@>12 \advance\count@ by-12 \fi
     \fi
     \standardhour=\expandafter{\number\count@}%
     \def\hack@19##1##2{\shortyear={##1##2}}%
          \expandafter\hack@\the\year}

\def\monthword#1{%
     \ifcase#1
          $\bullet$\err@badcountervalue{monthword}%
          \or January\or February\or March\or April\or May\or June%
          \or July\or August\or September\or October\or November\or December%
     \else$\bullet$\err@badcountervalue{monthword}%
     \fi}

\def\monthabbr#1{%
     \ifcase#1
          $\bullet$\err@badcountervalue{monthabbr}%
          \or Jan\or Feb\or Mar\or Apr\or May\or Jun%
          \or Jul\or Aug\or Sep\or Oct\or Nov\or Dec%
     \else$\bullet$\err@badcountervalue{monthabbr}%
     \fi}

\def\militarytime{\the\militaryhour:\the\minute}
\def\standardtime{\the\standardhour:\the\minute}


\def\@setnumstyle#1#2{\expandafter\global\expandafter\expandafter
     \expandafter\let\expandafter\expandafter
     \csname @\expandafter\blank\string#1style\endcsname
     \csname#2\endcsname}
\def\numstyle#1{\usename{@\expandafter\blank\string#1style}#1}
\def\ifblank#1\then{\useafter\ifx{@\expandafter\blank\string#1}\blank}

\def\blank#1{}

\def\Roman#1{\expandafter\uppercase\expandafter{\romannumeral#1}}
\def\alphabetic#1{%
     \ifcase#1
          $\bullet$\err@badcountervalue{alphabetic}%
          \or a\or b\or c\or d\or e\or f\or g\or h\or i\or j\or k\or l\or m%
          \or n\or o\or p\or q\or r\or s\or t\or u\or v\or w\or x\or y\or z%
     \else$\bullet$\err@badcountervalue{alphabetic}%
     \fi}
\def\Alphabetic#1{\expandafter\uppercase\expandafter{\alphabetic{#1}}}
\def\symbols#1{%
     \ifcase#1
          $\bullet$\err@badcountervalue{symbols}%
          \or*\or\dag\or\ddag\or\S\or$\|$%
          \or**\or\dag\dag\or\ddag\ddag\or\S\S\or$\|\|$%
     \else$\bullet$\err@badcountervalue{symbols}%
     \fi}


\catcode`\^^?=13 \def^^?{\relax}

\def\trimleading#1\to#2{\edef#2{#1}%
     \expandafter\@trimleading\expandafter#2#2^^?^^?}
\def\@trimleading#1#2#3^^?{\ifx#2^^?\def#1{}\else\def#1{#2#3}\fi}

\def\trimtrailing#1\to#2{\edef#2{#1}%
     \expandafter\@trimtrailing\expandafter#2#2^^? ^^?\relax}
\def\@trimtrailing#1#2 ^^?#3{\ifx#3\relax\toks@={}%
     \else\def#1{#2}\toks@={\trimtrailing#1\to#1}\fi
     \the\toks@}

\def\trim#1\to#2{\trimleading#1\to#2\trimtrailing#2\to#2}

\catcode`\^^?=15


\long\def\additemL#1\to#2{\toks@={\^^\{#1}}\toks@ii=\expandafter{#2}%
     \xdef#2{\the\toks@\the\toks@ii}}

\long\def\additemR#1\to#2{\toks@={\^^\{#1}}\toks@ii=\expandafter{#2}%
     \xdef#2{\the\toks@ii\the\toks@}}

\def\getitemL#1\to#2{\expandafter\@getitemL#1\hack@#1#2}
\def\@getitemL\^^\#1#2\hack@#3#4{\def#4{#1}\def#3{#2}}

\message{font macros,}


\newdimen\rp@
\newcount\@@sizeindex \@@sizeindex=0
\newcount\@@factori
\newcount\@@factorii
\newcount\@@factoriii
\newcount\@@factoriv

\countdef\maxfam=18
\newfam\itfam
\newfam\bffam
\newfam\bfsfam
\newfam\bmitfam

\def\@mathfontinit{\count@=4
     \loop\textfont\count@=\nullfont
          \scriptfont\count@=\nullfont
          \scriptscriptfont\count@=\nullfont
          \ifnum\count@<\maxfam\advance\count@ by\@ne
     \repeat}

\def\@fontstyleinit{%
     \def\it{\err@fontnotavailable\it}%
     \def\bf{\err@fontnotavailable\bf}%
     \def\bfs{\err@bfstobf}%
     \def\bmit{\err@fontnotavailable\bmit}%
     \def\sc{\err@fontnotavailable\sc}%
     \def\sl{\err@sltoit}%
     \def\ss{\err@fontnotavailable\ss}%
     \def\tt{\err@fontnotavailable\tt}}

\def\@parameterinit#1{\rm\rp@=.1em \@getscaling{#1}%
     \let\^^\=\@doscaling\scalingskipslist
     \setbox\strutbox=\hbox{\vrule
          height.708\baselineskip depth.292\baselineskip width\z@}}

\def\@getfactor#1#2#3#4{\@@factori=#1 \@@factorii=#2
     \@@factoriii=#3 \@@factoriv=#4}

\def\@getscaling#1{\count@=#1 \advance\count@ by-\@@sizeindex\@@sizeindex=#1
     \ifnum\count@<0
          \let\@mulordiv=\divide
          \let\@divormul=\multiply
          \multiply\count@ by\m@ne
     \else\let\@mulordiv=\multiply
          \let\@divormul=\divide
     \fi
     \edef\@@scratcha{\ifcase\count@                {1}{1}{1}{1}\or
          {1}{7}{23}{3}\or     {2}{5}{3}{1}\or      {9}{89}{13}{1}\or
          {6}{25}{6}{1}\or     {8}{71}{14}{1}\or    {6}{25}{36}{5}\or
          {1}{7}{53}{4}\or     {12}{125}{108}{5}\or {3}{14}{53}{5}\or
          {6}{41}{17}{1}\or    {13}{31}{13}{2}\or   {9}{107}{71}{2}\or
          {11}{139}{124}{3}\or {1}{6}{43}{2}\or     {10}{107}{42}{1}\or
          {1}{5}{43}{2}\or     {5}{69}{65}{1}\or    {11}{97}{91}{2}\fi}%
     \expandafter\@getfactor\@@scratcha}

\def\@doscaling#1{\@mulordiv#1by\@@factori\@divormul#1by\@@factorii
     \@mulordiv#1by\@@factoriii\@divormul#1by\@@factoriv}


\newskip\headskip
\newskip\footskip

\def\typesize=#1pt{\count@=#1 \advance\count@ by-10
     \ifcase\count@
          \@setsizex\or\err@badtypesize\or
          \@setsizexii\or\err@badtypesize\or
          \@setsizexiv
     \else\err@badtypesize
     \fi}

\def\@setsizex{\getixpt
     \def\subsubscriptfonts{\vpt}%
          \def\subsubscriptsize{\vpt\@parameterinit{-8}}%
     \def\subscriptfonts{\viipt}\def\subscriptsize{\viipt\@parameterinit{-4}}%
     \def\footnotefonts{\viiipt}\def\footnotesize{\viiipt\@parameterinit{-2}}%
     \def\smallfonts{\ixpt}\def\smallsize{\ixpt\@parameterinit{-1}}%
     \def\normalfonts{\xpt}\def\normalsize{\xpt\@parameterinit{0}}%
     \def\bigfonts{\xiipt}\def\bigsize{\xiipt\@parameterinit{2}}%
     \def\Bigfonts{\xivpt}\def\Bigsize{\xivpt\@parameterinit{4}}%
     \def\biggfonts{\xviipt}\def\biggsize{\xviipt\@parameterinit{6}}%
     \def\Biggfonts{\xxipt}\def\Biggsize{\xxipt\@parameterinit{8}}%
     \def\tinyfonts{\vpt}\def\tinysize{\vpt\@parameterinit{-8}}%
     \def\HUGEFONTS{\xxvpt}\def\HUGESIZE{\xxvpt\@parameterinit{10}}%
     \normalsize\fixedskipslist}

\def\@setsizexii{\getxipt
     \def\subsubscriptfonts{\vipt}%
          \def\subsubscriptsize{\vipt\@parameterinit{-6}}%
     \def\subscriptfonts{\viiipt}%
          \def\subscriptsize{\viiipt\@parameterinit{-2}}%
     \def\footnotefonts{\xpt}\def\footnotesize{\xpt\@parameterinit{0}}%
     \def\smallfonts{\xipt}\def\smallsize{\xipt\@parameterinit{1}}%
     \def\normalfonts{\xiipt}\def\normalsize{\xiipt\@parameterinit{2}}%
     \def\bigfonts{\xivpt}\def\bigsize{\xivpt\@parameterinit{4}}%
     \def\Bigfonts{\xviipt}\def\Bigsize{\xviipt\@parameterinit{6}}%
     \def\biggfonts{\xxipt}\def\biggsize{\xxipt\@parameterinit{8}}%
     \def\Biggfonts{\xxvpt}\def\Biggsize{\xxvpt\@parameterinit{10}}%
     \def\tinyfonts{\vpt}\def\tinysize{\vpt\@parameterinit{-8}}%
     \def\HUGEFONTS{\xxvpt}\def\HUGESIZE{\xxvpt\@parameterinit{10}}%
     \normalsize\fixedskipslist}

\def\@setsizexiv{\getxiiipt
     \def\subsubscriptfonts{\viipt}%
          \def\subsubscriptsize{\viipt\@parameterinit{-4}}%
     \def\subscriptfonts{\xpt}\def\subscriptsize{\xpt\@parameterinit{0}}%
     \def\footnotefonts{\xiipt}\def\footnotesize{\xiipt\@parameterinit{2}}%
     \def\smallfonts{\xiiipt}\def\smallsize{\xiiipt\@parameterinit{3}}%
     \def\normalfonts{\xivpt}\def\normalsize{\xivpt\@parameterinit{4}}%
     \def\bigfonts{\xviipt}\def\bigsize{\xviipt\@parameterinit{6}}%
     \def\Bigfonts{\xxipt}\def\Bigsize{\xxipt\@parameterinit{8}}%
     \def\biggfonts{\xxvpt}\def\biggsize{\xxvpt\@parameterinit{10}}%
     \def\Biggfonts{\err@sizetoolarge\Biggfonts\HUGEFONTS}%
          \def\Biggsize{\err@sizetoolarge\Biggsize\HUGESIZE}%
     \def\tinyfonts{\vpt}\def\tinysize{\vpt\@parameterinit{-8}}%
     \def\HUGEFONTS{\xxvpt}\def\HUGESIZE{\xxvpt\@parameterinit{10}}%
     \normalsize\fixedskipslist}

\def\subsubscriptfonts{\vpt} \def\subsubscriptsize{\vpt\@parameterinit{-8}}
\def\subscriptfonts{\viipt}  \def\subscriptsize{\viipt\@parameterinit{-4}}
\def\footnotefonts{\viiipt}  \def\footnotesize{\viiipt\@parameterinit{-2}}
\def\smallfonts{\err@sizenotavailable\smallfonts}
                             \def\smallsize{\ixpt\@parameterinit{-1}}
\def\normalfonts{\xpt}       \def\normalsize{\xpt\@parameterinit{0}}
\def\bigfonts{\xiipt}        \def\bigsize{\xiipt\@parameterinit{2}}
\def\Bigfonts{\xivpt}        \def\Bigsize{\xivpt\@parameterinit{4}}
\def\biggfonts{\xviipt}      \def\biggsize{\xviipt\@parameterinit{6}}
\def\Biggfonts{\xxipt}       \def\Biggsize{\xxipt\@parameterinit{8}}
\def\tinyfonts{\vpt}         \def\tinysize{\vpt\@parameterinit{-8}}
\def\HUGEFONTS{\xxvpt}       \def\HUGESIZE{\xxvpt\@parameterinit{10}}

\message{document layout,}


\newtoks\everyoutput \everyoutput={}
\newdimen\depthofpage
\newcount\pagenum \pagenum=0

\newdimen\oddtopmargin  \newdimen\eventopmargin
\newdimen\oddleftmargin \newdimen\evenleftmargin
\newtoks\oddhead        \newtoks\evenhead
\newtoks\oddfoot        \newtoks\evenfoot

\def\topmargin{\afterassignment\@seteventop\oddtopmargin}
\def\leftmargin{\afterassignment\@setevenleft\oddleftmargin}
\def\head{\afterassignment\@setevenhead\oddhead}
\def\foot{\afterassignment\@setevenfoot\oddfoot}

\def\@seteventop{\eventopmargin=\oddtopmargin}
\def\@setevenleft{\evenleftmargin=\oddleftmargin}
\def\@setevenhead{\evenhead=\oddhead}
\def\@setevenfoot{\evenfoot=\oddfoot}

\def\pagenumstyle#1{\@setnumstyle\pagenum{#1}}

\newif\ifdraft
\def\draft{\drafttrue\leftmargin=.5in \overfullrule=5pt }

\def\outputstyle#1{\global\expandafter\let\expandafter
          \@outputstyle\csname#1output\endcsname
     \usename{#1setup}}

\output={\@outputstyle}

\def\normaloutput{\the\everyoutput
     \global\advance\pagenum by\@ne
     \ifodd\pagenum
          \voffset=\oddtopmargin \hoffset=\oddleftmargin
     \else\voffset=\eventopmargin \hoffset=\evenleftmargin
     \fi
     \advance\voffset by-1in  \advance\hoffset by-1in
     \count0=\pagenum
     \expandafter\shipout\pagebox
     \ifnum\outputpenalty>-\@MM\else\dosupereject\fi}

\newdimen\fullhsize
\newbox\leftpage
\newcount\leftpagenum
\newcount\outputpagenum \outputpagenum=0
\let\leftorright=L

\def\twoupoutput{\the\everyoutput
     \global\advance\pagenum by\@ne
     \if L\leftorright
          \global\setbox\leftpage=\leftline{\pagebox}%
          \global\leftpagenum=\pagenum
          \global\let\leftorright=R%
     \else\global\advance\outputpagenum by\@ne
          \ifodd\outputpagenum
               \voffset=\oddtopmargin \hoffset=\oddleftmargin
          \else\voffset=\eventopmargin \hoffset=\evenleftmargin
          \fi
          \advance\voffset by-1in  \advance\hoffset by-1in
          \count0=\leftpagenum \count1=\pagenum
          \shipout\vbox{\hbox to\fullhsize
               {\box\leftpage\hfil\leftline{\pagebox}}}%
          \global\let\leftorright=L%
     \fi
     \ifnum\outputpenalty>-\@MM
     \else\dosupereject
          \if R\leftorright
               \globaldefs=\@ne\head={\hfil}\foot={\hfil}\globaldefs=\z@
               \null\newpage
          \fi
     \fi}

\def\pagebox{\vbox{\makeheadline\pagebody\makefootline}}

\def\makeheadline{%
     \vbox to\z@{\baselinestretch=\@m
          \vskip\topskip\vskip-.708\baselineskip\vskip-\headskip
          \line{\vbox to\ht\strutbox{}%
               \ifodd\pagenum\the\oddhead\else\the\evenhead\fi}%
          \vss}%
     \nointerlineskip}

\def\pagebody{\vbox to\vsize{%
     \boxmaxdepth\maxdepth
     \ifvoid\topins\else\unvbox\topins\fi
     \depthofpage=\dp255
     \unvbox255
     \ifraggedbottom\kern-\depthofpage\vfil\fi
     \ifvoid\footins
     \else\vskip\skip\footins
          \footnoterule
          \unvbox\footins
          \vskip-\footnoteskip
     \fi}}

\def\makefootline{\baselineskip=\footskip
     \line{\ifodd\pagenum\the\oddfoot\else\the\evenfoot\fi}}


\newskip\abovechapterskip
\newskip\belowchapterskip
\newskip\abovesectionskip
\newskip\belowsectionskip
\newskip\abovesubsectionskip
\newskip\belowsubsectionskip

\def\chapterstyle#1{\global\expandafter\let\expandafter\@chapterstyle
     \csname#1text\endcsname}
\def\sectionstyle#1{\global\expandafter\let\expandafter\@sectionstyle
     \csname#1text\endcsname}
\def\subsectionstyle#1{\global\expandafter\let\expandafter\@subsectionstyle
     \csname#1text\endcsname}

\def\chapter#1{%
     \ifdim\lastskip=17sp \else\chapterbreak\vskip\abovechapterskip\fi
     \@chapterstyle{\ifblank\chapternumstyle\then
          \else\newchapternum=\next\chapternumformat\ \fi#1}%
     \nobreak\vskip\belowchapterskip\vskip17sp }

\def\section#1{%
     \ifdim\lastskip=17sp \else\sectionbreak\vskip\abovesectionskip\fi
     \@sectionstyle{\ifblank\sectionnumstyle\then
          \else\newsectionnum=\next\sectionnumformat\ \fi#1}%
     \nobreak\vskip\belowsectionskip\vskip17sp }

\def\subsection#1{%
     \ifdim\lastskip=17sp \else\subsectionbreak\vskip\abovesubsectionskip\fi
     \@subsectionstyle{\ifblank\subsectionnumstyle\then
          \else\newsubsectionnum=\next\subsectionnumformat\ \fi#1}%
     \nobreak\vskip\belowsubsectionskip\vskip17sp }


\let\TeXunderline=\underline
\let\TeXoverline=\overline
\def\underline#1{\relax\ifmmode\TeXunderline{#1}\else
     $\TeXunderline{\hbox{#1}}$\fi}
\def\overline#1{\relax\ifmmode\TeXoverline{#1}\else
     $\TeXoverline{\hbox{#1}}$\fi}

\def\baselinestretch{\afterassignment\@baselinestretch\count@}
\def\@baselinestretch{\baselineskip=\normalbaselineskip
     \divide\baselineskip by\@m\baselineskip=\count@\baselineskip
     \setbox\strutbox=\hbox{\vrule
          height.708\baselineskip depth.292\baselineskip width\z@}%
     \bigskipamount=\the\baselineskip
          plus.25\baselineskip minus.25\baselineskip
     \medskipamount=.5\baselineskip
          plus.125\baselineskip minus.125\baselineskip
     \smallskipamount=.25\baselineskip
          plus.0625\baselineskip minus.0625\baselineskip}

\def\\{\ifhmode\ifnum\lastpenalty=-\@M\else\hfil\penalty-\@M\fi\fi
     \ignorespaces}
\def\newpage{\vfil\break}

\def\lefttext#1{\par{\@text\leftskip=\z@\rightskip=\centering
     \noindent#1\par}}
\def\righttext#1{\par{\@text\leftskip=\centering\rightskip=\z@
     \noindent#1\par}}
\def\centertext#1{\par{\@text\leftskip=\centering\rightskip=\centering
     \noindent#1\par}}
\def\@text{\parindent=\z@ \parfillskip=\z@ \everypar={}%
     \spaceskip=.3333em \xspaceskip=.5em
     \def\\{\ifhmode\ifnum\lastpenalty=-\@M\else\penalty-\@M\fi\fi
          \ignorespaces}}

\def\beginleft{\par\@text\leftskip=\z@ \rightskip=\centering}
     
\def\beginright{\par\@text\leftskip=\centering\rightskip=\z@ }
     
\def\begincenter{\par\@text\leftskip=\centering\rightskip=\centering}

\def\beginnarrow{\defaultoption[\parindent]\@beginnarrow}
\def\@beginnarrow[#1]{\par\advance\leftskip by#1\advance\rightskip by#1}

\begingroup
\catcode`\[=1 \catcode`\{=11 \gdef\beginignore[\endgroup\bgroup
     \catcode`\e=0 \catcode`\\=12 \catcode`\{=11 \catcode`\f=12 \let\or=\relax
     \let\nd{ignor=\fi \let\}=\egroup
     \iffalse}
\endgroup

\long\def\marginnote#1{\leavevmode
     \edef\@marginsf{\spacefactor=\the\spacefactor\relax}%
     \ifdraft\strut\vadjust{%
          \hbox to\z@{\hskip\hsize\hskip.1in
               \vbox to\z@{\vskip-\dp\strutbox
                    \marginnoteformat
                    \vskip-\ht\strutbox
                    \noindent\strut#1\par
                    \vss}%
               \hss}}%
     \fi
     \@marginsf}


\newtoks\everybye \everybye={\par\vfil}
\outer\def\bye{\the\everybye
     \footnotecheck
     \prelabelcheck
     \streamcheck
     \supereject
     \TeXend}

\message{footnotes,}

\newcount\footnotenum \footnotenum=0
\newskip\footnoteskip
\let\@footnotelist=\empty

\def\footnotenumstyle#1{\@setnumstyle\footnotenum{#1}%
     \useafter\ifx{@footnotenumstyle}\symbols
          \global\let\@footup=\empty
     \else\global\let\@footup=\markup
     \fi}

\def\footnote{\footnotecheck\defaultoption[]\@footnote}
\def\@footnote[#1]{\@footnotemark[#1]\@footnotetext}

\def\footnotemark{\defaultoption[]\@footnotemark}
\def\@footnotemark[#1]{\let\@footsf=\empty
     \ifhmode\edef\@footsf{\spacefactor=\the\spacefactor\relax}\/\fi
     \ifnoarg#1\then
          \global\advance\footnotenum by\@ne
          \@footup{\footnotenumformat}%
          \edef\@@foota{\footnotenum=\the\footnotenum\relax}%
          \expandafter\additemR\expandafter\@footup\expandafter
               {\@@foota\footnotenumformat}\to\@footnotelist
          \global\let\@footnotelist=\@footnotelist
     \else\markup{#1}%
          \additemR\markup{#1}\to\@footnotelist
          \global\let\@footnotelist=\@footnotelist
     \fi
     \@footsf}

\def\footnotetext{%
     \ifx\@footnotelist\empty\err@extrafootnotetext\else\@footnotetext\fi}
\def\@footnotetext{%
     \getitemL\@footnotelist\to\@@foota
     \global\let\@footnotelist=\@footnotelist
     \insert\footins\bgroup
     \footnoteformat
     \splittopskip=\ht\strutbox\splitmaxdepth=\dp\strutbox
     \interlinepenalty=\interfootnotelinepenalty\floatingpenalty=\@MM
     \noindent\llap{\@@foota}\strut
     \bgroup\aftergroup\@footnoteend
     \let\@@scratcha=}
\def\@footnoteend{\strut\par\vskip\footnoteskip\egroup}

\def\footnoterule{\normalfonts
     \kern-.3em \hrule width2in height.04em \kern .26em }

\def\footnotecheck{%
     \ifx\@footnotelist\empty
     \else\err@extrafootnotemark
          \global\let\@footnotelist=\empty
     \fi}

\message{labels,}

\let\@@labeldef=\xdef
\newif\if@labelfile
\newwrite\@labelfile
\let\@prelabellist=\empty

\def\label#1#2{\trim#1\to\@@labarg\edef\@@labtext{#2}%
     \edef\@@labname{lab@\@@labarg}%
     \useafter\ifundefined\@@labname\then\else\@yeslab\fi
     \useafter\@@labeldef\@@labname{#2}%
     \ifstreaming
          \expandafter\toks@\expandafter\expandafter\expandafter
               {\csname\@@labname\endcsname}%
          \immediate\write\streamout{\noexpand\label{\@@labarg}{\the\toks@}}%
     \fi}
\def\@yeslab{%
     \useafter\ifundefined{if\@@labname}\then
          \err@labelredef\@@labarg
     \else\useif{if\@@labname}\then
               \err@labelredef\@@labarg
          \else\global\usename{\@@labname true}%
               \useafter\ifundefined{pre\@@labname}\then
               \else\useafter\ifx{pre\@@labname}\@@labtext
                    \else\err@badlabelmatch\@@labarg
                    \fi
               \fi
               \if@labelfile
               \else\global\@labelfiletrue
                    \immediate\write\sixt@@n{--> Creating file \jobname.lab}%
                    \immediate\openout\@labelfile=\jobname.lab
               \fi
               \immediate\write\@labelfile
                    {\noexpand\prelabel{\@@labarg}{\@@labtext}}%
          \fi
     \fi}

\def\putlab#1{\trim#1\to\@@labarg\edef\@@labname{lab@\@@labarg}%
     \useafter\ifundefined\@@labname\then\@nolab\else\usename\@@labname\fi}
\def\@nolab{%
     \useafter\ifundefined{pre\@@labname}\then
          \undefinedlabelformat
          \err@needlabel\@@labarg
          \useafter\xdef\@@labname{\undefinedlabelformat}%
     \else\usename{pre\@@labname}%
          \useafter\xdef\@@labname{\usename{pre\@@labname}}%
     \fi
     \useafter\newif{if\@@labname}%
     \expandafter\additemR\@@labarg\to\@prelabellist}

\def\prelabel#1{\useafter\gdef{prelab@#1}}

\def\ifundefinedlabel#1\then{%
     \expandafter\ifx\csname lab@#1\endcsname\relax}
\def\useiflab#1\then{\csname iflab@#1\endcsname}

\def\prelabelcheck{{%
     \def\^^\##1{\useiflab{##1}\then\else\err@undefinedlabel{##1}\fi}%
     \@prelabellist}}

\message{equation numbering,}

\newcount\chapternum
\newcount\sectionnum
\newcount\subsectionnum
\newcount\equationnum
\newcount\subequationnum
\newcount\figurenum
\newcount\subfigurenum
\newcount\tablenum
\newcount\subtablenum

\newif\if@subeqncount
\newif\if@subfigcount
\newif\if@subtblcount

\def\newchapternum{\newsectionnum=\z@\@resetnum\chapternum}
\def\newsectionnum{\newsubsectionnum=\z@\@resetnum\sectionnum}
\def\newsubsectionnum{\newequationnum=\z@\newfigurenum=\z@\newtablenum=\z@
     \@resetnum\subsectionnum}
\def\newequationnum{\newsubequationnum=\z@\@resetnum\equationnum}
\def\newsubequationnum{\@resetnum\subequationnum}
\def\newfigurenum{\newsubfigurenum=\z@\@resetnum\figurenum}
\def\newsubfigurenum{\@resetnum\subfigurenum}
\def\newtablenum{\newsubtablenum=\z@\@resetnum\tablenum}
\def\newsubtablenum{\@resetnum\subtablenum}

\def\@resetnum#1{\global\advance#1by1 \edef\next{\the#1\relax}\global#1}

\newchapternum=0

\def\chapternumstyle#1{\@setnumstyle\chapternum{#1}}
\def\sectionnumstyle#1{\@setnumstyle\sectionnum{#1}}
\def\subsectionnumstyle#1{\@setnumstyle\subsectionnum{#1}}
\def\equationnumstyle#1{\@setnumstyle\equationnum{#1}}
\def\subequationnumstyle#1{\@setnumstyle\subequationnum{#1}%
     \ifblank\subequationnumstyle\then\global\@subeqncountfalse\fi
     \ignorespaces}
\def\figurenumstyle#1{\@setnumstyle\figurenum{#1}}
\def\subfigurenumstyle#1{\@setnumstyle\subfigurenum{#1}%
     \ifblank\subfigurenumstyle\then\global\@subfigcountfalse\fi
     \ignorespaces}
\def\tablenumstyle#1{\@setnumstyle\tablenum{#1}}
\def\subtablenumstyle#1{\@setnumstyle\subtablenum{#1}%
     \ifblank\subtablenumstyle\then\global\@subtblcountfalse\fi
     \ignorespaces}

\def\eqnlabel#1{%
     \if@subeqncount
          \newsubequationnum=\next
     \else\newequationnum=\next
          \ifblank\subequationnumstyle\then
          \else\global\@subeqncounttrue
               \newsubequationnum=\@ne
          \fi
     \fi
     \label{#1}{\puteqnformat}(\puteqn{#1})%
     \ifdraft\rlap{\hskip.1in{\tt#1}}\fi}

\let\puteqn=\putlab

\def\equation#1#2{\useafter\gdef{eqn@#1}{#2\eqno\eqnlabel{#1}}}
\def\Equation#1{\useafter\gdef{eqn@#1}}

\def\putequation#1{\useafter\ifundefined{eqn@#1}\then
     \err@undefinedeqn{#1}\else\usename{eqn@#1}\fi}

\def\eqnseriesstyle#1{\gdef\@eqnseriesstyle{#1}}
\def\begineqnseries{\subequationnumstyle{\@eqnseriesstyle}%
     \defaultoption[]\@begineqnseries}
\def\@begineqnseries[#1]{\edef\@@eqnname{#1}}
\def\endeqnseries{\subequationnumstyle{blank}%
     \expandafter\ifnoarg\@@eqnname\then
     \else\label\@@eqnname{\puteqnformat}%
     \fi
     \aftergroup\ignorespaces}

\def\figlabel#1{%
     \if@subfigcount
          \newsubfigurenum=\next
     \else\newfigurenum=\next
          \ifblank\subfigurenumstyle\then
          \else\global\@subfigcounttrue
               \newsubfigurenum=\@ne
          \fi
     \fi
     \label{#1}{\putfigformat}\putfig{#1}%
     {\def\marginnoteformat{\tt}\marginnote{#1}}}

\let\putfig=\putlab

\def\figseriesstyle#1{\gdef\@figseriesstyle{#1}}
\def\beginfigseries{\subfigurenumstyle{\@figseriesstyle}%
     \defaultoption[]\@beginfigseries}
\def\@beginfigseries[#1]{\edef\@@figname{#1}}
\def\endfigseries{\subfigurenumstyle{blank}%
     \expandafter\ifnoarg\@@figname\then
     \else\label\@@figname{\putfigformat}%
     \fi
     \aftergroup\ignorespaces}

\def\tbllabel#1{%
     \if@subtblcount
          \newsubtablenum=\next
     \else\newtablenum=\next
          \ifblank\subtablenumstyle\then
          \else\global\@subtblcounttrue
               \newsubtablenum=\@ne
          \fi
     \fi
     \label{#1}{\puttblformat}\puttbl{#1}%
     {\def\marginnoteformat{\tt}\marginnote{#1}}}

\let\puttbl=\putlab

\def\tblseriesstyle#1{\gdef\@tblseriesstyle{#1}}
\def\begintblseries{\subtablenumstyle{\@tblseriesstyle}%
     \defaultoption[]\@begintblseries}
\def\@begintblseries[#1]{\edef\@@tblname{#1}}
\def\endtblseries{\subtablenumstyle{blank}%
     \expandafter\ifnoarg\@@tblname\then
     \else\label\@@tblname{\puttblformat}%
     \fi
     \aftergroup\ignorespaces}

\message{reference numbering,}

\newcount\referencenum \referencenum=0
\newcount\@@prerefcount \@@prerefcount=0
\newcount\@@thisref
\newcount\@@lastref
\newcount\@@loopref
\newcount\@@refseq
\newdimen\refnumindent
\let\@undefreflist=\empty

\def\referencenumstyle#1{\@setnumstyle\referencenum{#1}}

\def\referencestyle#1{\usename{@ref#1}}

\def\@refsequential{%
     \gdef\@refpredef##1{\global\advance\referencenum by\@ne
          \let\^^\=0\label{##1}{\^^\{\the\referencenum}}%
          \useafter\gdef{ref@\the\referencenum}{{##1}{\undefinedlabelformat}}}%
     \gdef\@reference##1##2{%
          \ifundefinedlabel##1\then
          \else\def\^^\####1{\global\@@thisref=####1\relax}\putlab{##1}%
               \useafter\gdef{ref@\the\@@thisref}{{##1}{##2}}%
          \fi}%
     \gdef\endputreferences{%
          \loop\ifnum\@@loopref<\referencenum
                    \advance\@@loopref by\@ne
                    \expandafter\expandafter\expandafter\@printreference
                         \csname ref@\the\@@loopref\endcsname
          \repeat
          \par}}

\def\@refpreordered{%
     \gdef\@refpredef##1{\global\advance\referencenum by\@ne
          \additemR##1\to\@undefreflist}%
     \gdef\@reference##1##2{%
          \ifundefinedlabel##1\then
          \else\global\advance\@@loopref by\@ne
               {\let\^^\=0\label{##1}{\^^\{\the\@@loopref}}}%
               \@printreference{##1}{##2}%
          \fi}
     \gdef\endputreferences{%
          \def\^^\####1{\useiflab{####1}\then
               \else\reference{####1}{\undefinedlabelformat}\fi}%
          \@undefreflist
          \par}}

\def\beginprereferences{\par
     \def\reference##1##2{\global\advance\referencenum by1\@ne
          \let\^^\=0\label{##1}{\^^\{\the\referencenum}}%
          \useafter\gdef{ref@\the\referencenum}{{##1}{##2}}}}
\def\endprereferences{\global\@@prerefcount=\the\referencenum\par}

\def\beginputreferences{\par
     \refnumindent=\z@\@@loopref=\z@
     \loop\ifnum\@@loopref<\referencenum
               \advance\@@loopref by\@ne
               \setbox\z@=\hbox{\referencenum=\@@loopref
                    \referencenumformat\enskip}%
               \ifdim\wd\z@>\refnumindent\refnumindent=\wd\z@\fi
     \repeat
     \putreferenceformat
     \@@loopref=\z@
     \loop\ifnum\@@loopref<\@@prerefcount
               \advance\@@loopref by\@ne
               \expandafter\expandafter\expandafter\@printreference
                    \csname ref@\the\@@loopref\endcsname
     \repeat
     \let\reference=\@reference}

\def\@printreference#1#2{\ifx#2\undefinedlabelformat\err@undefinedref{#1}\fi
     \noindent\ifdraft\rlap{\hskip\hsize\hskip.1in \tt#1}\fi
     \llap{\referencenum=\@@loopref\referencenumformat\enskip}#2\par}

\def\reference#1#2{{\par\refnumindent=\z@\putreferenceformat\noindent#2\par}}

\def\putref#1{\trim#1\to\@@refarg
     \expandafter\ifnoarg\@@refarg\then
          \toks@={\relax}%
     \else\@@lastref=-\@m\def\@@refsep{}\def\@more{\@nextref}%
          \toks@={\@nextref#1,,}%
     \fi\the\toks@}
\def\@nextref#1,{\trim#1\to\@@refarg
     \expandafter\ifnoarg\@@refarg\then
          \let\@more=\relax
     \else\ifundefinedlabel\@@refarg\then
               \expandafter\@refpredef\expandafter{\@@refarg}%
          \fi
          \def\^^\##1{\global\@@thisref=##1\relax}%
          \global\@@thisref=\m@ne
          \setbox\z@=\hbox{\putlab\@@refarg}%
     \fi
     \advance\@@lastref by\@ne
     \ifnum\@@lastref=\@@thisref\advance\@@refseq by\@ne\else\@@refseq=\@ne\fi
     \ifnum\@@lastref<\z@
     \else\ifnum\@@refseq<\thr@@
               \@@refsep\def\@@refsep{,}%
               \ifnum\@@lastref>\z@
                    \advance\@@lastref by\m@ne
                    {\referencenum=\@@lastref\putrefformat}%
               \else\undefinedlabelformat
               \fi
          \else\def\@@refsep{--}%
          \fi
     \fi
     \@@lastref=\@@thisref
     \@more}

\message{streaming,}

\newif\ifstreaming

\def\streamto{\defaultoption[\jobname]\@streamto}
\def\@streamto[#1]{\global\streamingtrue
     \immediate\write\sixt@@n{--> Streaming to #1.str}%
     \newwrite\streamout\immediate\openout\streamout=#1.str }

\def\streamfrom{\defaultoption[\jobname]\@streamfrom}
\def\@streamfrom[#1]{\newread\streamin\openin\streamin=#1.str
     \ifeof\streamin
          \expandafter\err@nostream\expandafter{#1.str}%
     \else\immediate\write\sixt@@n{--> Streaming from #1.str}%
          \let\@@labeldef=\gdef
          \ifstreaming
               \edef\@elc{\endlinechar=\the\endlinechar}%
               \endlinechar=\m@ne
               \loop\read\streamin to\@@scratcha
                    \ifeof\streamin
                         \streamingfalse
                    \else\toks@=\expandafter{\@@scratcha}%
                         \immediate\write\streamout{\the\toks@}%
                    \fi
                    \ifstreaming
               \repeat
               \@elc
               \input #1.str
               \streamingtrue
          \else\input #1.str
          \fi
          \let\@@labeldef=\xdef
     \fi}

\def\streamcheck{\ifstreaming
     \immediate\write\streamout{\pagenum=\the\pagenum}%
     \immediate\write\streamout{\footnotenum=\the\footnotenum}%
     \immediate\write\streamout{\referencenum=\the\referencenum}%
     \immediate\write\streamout{\chapternum=\the\chapternum}%
     \immediate\write\streamout{\sectionnum=\the\sectionnum}%
     \immediate\write\streamout{\subsectionnum=\the\subsectionnum}%
     \immediate\write\streamout{\equationnum=\the\equationnum}%
     \immediate\write\streamout{\subequationnum=\the\subequationnum}%
     \immediate\write\streamout{\figurenum=\the\figurenum}%
     \immediate\write\streamout{\subfigurenum=\the\subfigurenum}%
     \immediate\write\streamout{\tablenum=\the\tablenum}%
     \immediate\write\streamout{\subtablenum=\the\subtablenum}%
     \immediate\closeout\streamout
     \fi}


\def\err@badtypesize{%
     \errhelp={The limited availability of certain fonts requires^^J%
          that the base type size be 10pt, 12pt, or 14pt.^^J}%
     \errmessage{--> Illegal base type size}}

\def\err@badsizechange{\immediate\write\sixt@@n
     {--> Size change not allowed in math mode, ignored}}

\def\err@sizetoolarge#1{\immediate\write\sixt@@n
     {--> \noexpand#1 too big, substituting HUGE}}

\def\err@sizenotavailable#1{\immediate\write\sixt@@n
     {--> Size not available, \noexpand#1 ignored}}

\def\err@fontnotavailable#1{\immediate\write\sixt@@n
     {--> Font not available, \noexpand#1 ignored}}

\def\err@sltoit{\immediate\write\sixt@@n
     {--> Style \noexpand\sl not available, substituting \noexpand\it}%
     \it}

\def\err@bfstobf{\immediate\write\sixt@@n
     {--> Style \noexpand\bfs not available, substituting \noexpand\bf}%
     \bf}

\def\err@badgroup#1#2{%
     \errhelp={The block you have just tried to close was not the one^^J%
          most recently opened.^^J}%
     \errmessage{--> \noexpand\end{#1} doesn't match \noexpand\begin{#2}}}

\def\err@badcountervalue#1{\immediate\write\sixt@@n
     {--> Counter (#1) out of bounds}}

\def\err@extrafootnotemark{\immediate\write\sixt@@n
     {--> \noexpand\footnotemark command
          has no corresponding \noexpand\footnotetext}}

\def\err@extrafootnotetext{%
     \errhelp{You have given a \noexpand\footnotetext command without first
          specifying^^Ja \noexpand\footnotemark.^^J}%
     \errmessage{--> \noexpand\footnotetext command has no corresponding
          \noexpand\footnotemark}}

\def\err@labelredef#1{\immediate\write\sixt@@n
     {--> Label "#1" redefined}}

\def\err@badlabelmatch#1{\immediate\write\sixt@@n
     {--> Definition of label "#1" doesn't match value in \jobname.lab}}

\def\err@needlabel#1{\immediate\write\sixt@@n
     {--> Label "#1" cited before its definition}}

\def\err@undefinedlabel#1{\immediate\write\sixt@@n
     {--> Label "#1" cited but never defined}}

\def\err@undefinedeqn#1{\immediate\write\sixt@@n
     {--> Equation "#1" not defined}}

\def\err@undefinedref#1{\immediate\write\sixt@@n
     {--> Reference "#1" not defined}}

\def\err@nostream#1{%
     \errhelp={You have tried to input a stream file that doesn't exist.^^J}%
     \errmessage{--> Stream file #1 not found}}

\message{jyTeX initialization}

\everyjob{\immediate\write16{--> jyTeX version \fmtversion}%
     \edef\@@jobname{\jobname}%
     \edef\jobname{\@@jobname}%
     \settime
     \openin0=\jobname.lab
     \ifeof0
     \else\closein0
          \immediate\write16{--> Getting labels from file \jobname.lab}%
          \input\jobname.lab
     \fi}


\def\fixedskipslist{%
     \^^\{\topskip}%
     \^^\{\splittopskip}%
     \^^\{\maxdepth}%
     \^^\{\skip\topins}%
     \^^\{\skip\footins}%
     \^^\{\headskip}%
     \^^\{\footskip}}

\def\scalingskipslist{%
     \^^\{\p@renwd}%
     \^^\{\delimitershortfall}%
     \^^\{\nulldelimiterspace}%
     \^^\{\scriptspace}%
     \^^\{\jot}%
     \^^\{\normalbaselineskip}%
     \^^\{\normallineskip}%
     \^^\{\normallineskiplimit}%
     \^^\{\baselineskip}%
     \^^\{\lineskip}%
     \^^\{\lineskiplimit}%
     \^^\{\bigskipamount}%
     \^^\{\medskipamount}%
     \^^\{\smallskipamount}%
     \^^\{\parskip}%
     \^^\{\parindent}%
     \^^\{\abovedisplayskip}%
     \^^\{\belowdisplayskip}%
     \^^\{\abovedisplayshortskip}%
     \^^\{\belowdisplayshortskip}%
     \^^\{\abovechapterskip}%
     \^^\{\belowchapterskip}%
     \^^\{\abovesectionskip}%
     \^^\{\belowsectionskip}%
     \^^\{\abovesubsectionskip}%
     \^^\{\belowsubsectionskip}}


\def\twoupsetup{
     \topmargin=.75in
     \leftmargin=.5in
     \vsize=6.9in
     \hsize=4.75in
     \fullhsize=10in
     \let\draft=\relax}

\outputstyle{normal}                             

\def\marginnoteformat{\subscriptsize             
     \hsize=1in \baselinestretch=1000 \everypar={}%
     \tolerance=5000 \hbadness=5000 \parskip=0pt \parindent=0pt
     \leftskip=0pt \rightskip=0pt \raggedright}

\head={\ifdraft\normalfonts\it\hfil DRAFT\hfil   
     \llap{\number\day\ \monthword\month\ \militarytime}\else\hfil\fi}
\foot={\hfil\normalfonts\numstyle\pagenum\hfil}  

\normalbaselineskip=12pt                         
\normallineskip=0pt                              
\normallineskiplimit=0pt                         
\normalbaselines                                 

\topskip=.85\baselineskip \splittopskip=\topskip \headskip=2\baselineskip
\footskip=\headskip

\pagenumstyle{arabic}                            

\parskip=0pt                                     
\parindent=20pt                                  

\baselinestretch=1000                            


\chapterstyle{left}                              
\chapternumstyle{blank}                          
\def\chapterbreak{\newpage}                      
\abovechapterskip=0pt                            
\belowchapterskip=1.5\baselineskip               
     plus.38\baselineskip minus.38\baselineskip
\def\chapternumformat{\numstyle\chapternum.}     

\sectionstyle{left}                              
\sectionnumstyle{blank}                          
\def\sectionbreak{\vskip0pt plus4\baselineskip\penalty-100
     \vskip0pt plus-4\baselineskip}              
\abovesectionskip=1.5\baselineskip               
     plus.38\baselineskip minus.38\baselineskip
\belowsectionskip=\the\baselineskip              
     plus.25\baselineskip minus.25\baselineskip
\def\sectionnumformat{
     \ifblank\chapternumstyle\then\else\numstyle\chapternum.\fi
     \numstyle\sectionnum.}

\subsectionstyle{left}                           
\subsectionnumstyle{blank}                       
\def\subsectionbreak{\vskip0pt plus4\baselineskip\penalty-100
     \vskip0pt plus-4\baselineskip}              
\abovesubsectionskip=\the\baselineskip           
     plus.25\baselineskip minus.25\baselineskip
\belowsubsectionskip=.75\baselineskip            
     plus.19\baselineskip minus.19\baselineskip
\def\subsectionnumformat{
     \ifblank\chapternumstyle\then\else\numstyle\chapternum.\fi
     \ifblank\sectionnumstyle\then\else\numstyle\sectionnum.\fi
     \numstyle\subsectionnum.}


\footnotenumstyle{symbols}                       
\footnoteskip=0pt                                
\def\footnotenumformat{\numstyle\footnotenum}    
\def\footnoteformat{\footnotesize                
     \everypar={}\parskip=0pt \parfillskip=0pt plus1fil
     \leftskip=1em \rightskip=0pt
     \spaceskip=0pt \xspaceskip=0pt
     \def\\{\ifhmode\ifnum\lastpenalty=-10000
          \else\hfil\penalty-10000 \fi\fi\ignorespaces}}


\def\undefinedlabelformat{$\bullet$}             


\equationnumstyle{arabic}                        
\subequationnumstyle{blank}                      
\figurenumstyle{arabic}                          
\subfigurenumstyle{blank}                        
\tablenumstyle{arabic}                           
\subtablenumstyle{blank}                         

\eqnseriesstyle{alphabetic}                      
\figseriesstyle{alphabetic}                      
\tblseriesstyle{alphabetic}                      

\def\puteqnformat{\hbox{
     \ifblank\chapternumstyle\then\else\numstyle\chapternum.\fi
     \ifblank\sectionnumstyle\then\else\numstyle\sectionnum.\fi
     \ifblank\subsectionnumstyle\then\else\numstyle\subsectionnum.\fi
     \numstyle\equationnum
     \numstyle\subequationnum}}
\def\putfigformat{\hbox{
     \ifblank\chapternumstyle\then\else\numstyle\chapternum.\fi
     \ifblank\sectionnumstyle\then\else\numstyle\sectionnum.\fi
     \ifblank\subsectionnumstyle\then\else\numstyle\subsectionnum.\fi
     \numstyle\figurenum
     \numstyle\subfigurenum}}
\def\puttblformat{\hbox{
     \ifblank\chapternumstyle\then\else\numstyle\chapternum.\fi
     \ifblank\sectionnumstyle\then\else\numstyle\sectionnum.\fi
     \ifblank\subsectionnumstyle\then\else\numstyle\subsectionnum.\fi
     \numstyle\tablenum
     \numstyle\subtablenum}}


\referencestyle{sequential}                      
\referencenumstyle{arabic}                       
\def\putrefformat{\numstyle\referencenum}        
\def\referencenumformat{\numstyle\referencenum.} 
\def\putreferenceformat{
     \everypar={\hangindent=1em \hangafter=1 }%
     \def\\{\hfil\break\null\hskip-1em \ignorespaces}%
     \leftskip=\refnumindent\parindent=0pt \interlinepenalty=1000 }


\normalsize


\def\fmtversion{2.6M (June 1992)}

\catcode`\@=12

\typesize=10pt \magnification=1200 \baselineskip17truept
\footnotenumstyle{arabic} \hsize=6truein\vsize=8.5truein

\sectionnumstyle{blank}
\chapternumstyle{blank}
\chapternum=1
\sectionnum=1
\pagenum=0

\def\begintitle{\pagenumstyle{blank}\parindent=0pt
\begin{narrow}[0.4in]}
\def\endtitle{\end{narrow}\newpage\pagenumstyle{arabic}}


\def\beginexercise{\vskip 20truept\parindent=0pt\begin{narrow}[10
truept]}
\def\endexercise{\vskip 10truept\end{narrow}}


\def\eql#1{\eqno\eqnlabel{#1}}
\def\ref{\reference}
\def\peq{\puteqn}
\def\pref{\putref}

\def\mgn{\marginnote}
\def\bex{\begin{exercise}}
\def\eex{\end{exercise}}


\font\open=msbm10 

 
\def\StretchRtArr#1{{\count255=0\loop\relbar\joinrel\advance\count255 by1
\ifnum\count255<#1\repeat\rightarrow}}
\def\StretchLtArr#1{\,{\leftarrow\!\!\count255=0\loop\relbar
\joinrel\advance\count255 by1\ifnum\count255<#1\repeat}}

\def\StretchLRtArr#1{\,{\leftarrow\!\!\count255=0\loop\relbar\joinrel\advance
\count255 by1\ifnum\count255<#1\repeat\rightarrow\,\,}}

\def\mbox#1{{\leavevmode\hbox{#1}}}

\def\hspace#1{{\phantom{\mbox#1}}}
\def\oZ{\mbox{\open\char90}}

\def\al{\alpha}
\def\Up{\Upsilon}
\def\be{\beta}
\def\ga{\gamma}
\def\de{\delta}

\def\la{\lambda}

\def\th{\theta}
\def\Th{\Theta}
\def\ze{\zeta}

\def\De{\Delta}

\def\det{{\rm det\,}}

\def\Det{{\rm Det}}

\def\sc{{\rm sc }}

\def\zf{$\zeta$--function}


\def\frac#1/#2{\leavevmode\kern.1em
\raise.5ex\hbox{\the\scriptfont0 #1}\kern-.1em/\kern-.15em
\lower.25ex\hbox{\the\scriptfont0 #2}}
\def\sfrac#1/#2{\leavevmode\kern.1em
\raise.5ex\hbox{\the\scriptscriptfont0 #1}\kern-.1em/\kern-.15em
\lower.25ex\hbox{\the\scriptscriptfont0 #2}}

\def\gtorder{\mathrel{\raise.3ex\hbox{$>$}\mkern-14mu
             \lower0.6ex\hbox{$\sim$}}}
\def\ltorder{\mathrel{\raise.3ex\hbox{$<$}\mkern-14mu
             \lower0.6ex\hbox{$\sim$}}}

\def\semidirprod{\rlap{\ss C}\raise1pt\hbox{$\mkern.75mu\times$}}
\def\for{\lower6pt\hbox{$\Big|$}}
\def\fish{\kern-.25em{\phantom{abcde}\over \phantom{abcde}}\kern-.25em}


\def\boxit#1{\vbox{\hrule\hbox{\vrule\kern3pt
        \vbox{\kern3pt#1\kern3pt}\kern3pt\vrule}\hrule}}
\def\dalemb#1#2{{\vbox{\hrule height .#2pt
        \hbox{\vrule width.#2pt height#1pt \kern#1pt \vrule
                width.#2pt} \hrule height.#2pt}}}

\def\ol{\overline}
\def\frac#1#2{{{#1}\over{#2}}}

\def\noin{\noindent}

\def\comb#1#2{{\left(#1\atop#2\right)}}

\def\cosec{{\rm cosec\,}}
\def\cot{{\rm cot\,}}

\def\etc{{\it etc. }}

\def\eg{{\it e.g.}}
\def\ie{{\it i.e. }}
\def\cf{{\it cf }}

\def\me#1#2#3{\langle{#1}\mid\!{#2}\!\mid{#3}\rangle}  


\def\Tr{{\rm Tr\,}}

\def\wt{\widetilde}

\def\3j#1#2#3#4#5#6{\left\lgroup\matrix{#1&#2&#3\cr#4&#5&#6\cr}
\right\rgroup}

\def\m?{\mgn{?}}

\def\beq{\begin{eqnarray}}
\def\eeq{\end{eqnarray}}


\def\aop#1#2#3{{\it Ann. Phys.} {\bf {#1}} ({#2}) #3}
\def\cjp#1#2#3{{\it Can. J. Phys.} {\bf {#1}} ({#2}) #3}
\def\cmp#1#2#3{{\it Comm. Math. Phys.} {\bf {#1}} ({#2}) #3}
\def\cqg#1#2#3{{\it Class. Quant. Grav.} {\bf {#1}} ({#2}) #3}

\def\ijmp#1#2#3{{\it Int. J. Mod. Phys.} {\bf {#1}} ({#2}) #3}

\def\jmp#1#2#3{{\it J. Math. Phys.} {\bf {#1}} ({#2}) #3}
\def\jpa#1#2#3{{\it J. Phys.} {\bf A{#1}} ({#2}) #3}
\def\jpc#1#2#3{{\it J. Phys.} {\bf C{#1}} ({#2}) #3}
\def\lnm#1#2#3{{\it Lect. Notes Math.} {\bf {#1}} ({#2}) #3}

\def\np#1#2#3{{\it Nucl. Phys.} {\bf B{#1}} ({#2}) #3}
\def\npa#1#2#3{{\it Nucl. Phys.} {\bf A{#1}} ({#2}) #3}
\def\pl#1#2#3{{\it Phys. Lett.} {\bf {#1}} ({#2}) #3}
\def\phm#1#2#3{{\it Phil.Mag.} {\bf {#1}} ({#2}) #3}
\def\prp#1#2#3{{\it Phys. Rep.} {\bf {#1}} ({#2}) #3}
\def\pr#1#2#3{{\it Phys. Rev.} {\bf {#1}} ({#2}) #3}
\def\prA#1#2#3{{\it Phys. Rev.} {\bf A{#1}} ({#2}) #3}

\def\prD#1#2#3{{\it Phys. Rev.} {\bf D{#1}} ({#2}) #3}
\def\prE#1#2#3{{\it Phys. Rev.} {\bf E{#1}} ({#2}) #3}
\def\prl#1#2#3{{\it Phys. Rev. Lett.} {\bf #1} ({#2}) #3}

\def\rmp#1#2#3{{\it Rev. Mod. Phys.} {\bf {#1}} ({#2}) #3}

\def\zfp#1#2#3{{\it Z. f. Phys.} {\bf {#1}} ({#2}) #3}

\def\cras#1#2#3{{\it Comptes Rend. Acad. Sci. (Paris)} {\bf{#1}} (#2) #3}
\def\prs#1#2#3{{\it Proc. Roy. Soc.} {\bf A{#1}} ({#2}) #3}
\def\pcps#1#2#3{{\it Proc. Camb. Phil. Soc.} {\bf{#1}} ({#2}) #3}
\def\mpcps#1#2#3{{\it Math. Proc. Camb. Phil. Soc.} {\bf{#1}} ({#2}) #3}

\def\amsh#1#2#3{{\it Abh. Math. Sem. Ham.} {\bf {#1}} ({#2}) #3}
\def\am#1#2#3{{\it Acta Mathematica} {\bf {#1}} ({#2}) #3}
\def\aim#1#2#3{{\it Adv. in Math.} {\bf {#1}} ({#2}) #3}
\def\ajm#1#2#3{{\it Am. J. Math.} {\bf {#1}} ({#2}) #3}
\def\amm#1#2#3{{\it Am. Math. Mon.} {\bf {#1}} ({#2}) #3}

\def\aom#1#2#3{{\it Ann. of Math.} {\bf {#1}} ({#2}) #3}
\def\cjm#1#2#3{{\it Can. J. Math.} {\bf {#1}} ({#2}) #3}
\def\bams#1#2#3{{\it Bull.Am.Math.Soc.} {\bf {#1}} ({#2}) #3}

\def\cmh#1#2#3{{\it Comm. Math. Helv.} {\bf {#1}} ({#2}) #3}

\def\dmj#1#2#3{{\it Duke Math. J.} {\bf {#1}} ({#2}) #3}
\def\invm#1#2#3{{\it Invent. Math.} {\bf {#1}} ({#2}) #3}

\def\jdg#1#2#3{{\it J. Diff. Geom.} {\bf {#1}} ({#2}) #3}

\def\joa#1#2#3{{\it J. of Algebra} {\bf {#1}} ({#2}) #3}
\def\jram#1#2#3{{\it J. f. Reine u. Angew. Math.} {\bf {#1}} ({#2}) #3}
\def\jims#1#2#3{{\it J. Indian. Math. Soc.} {\bf {#1}} ({#2}) #3}
\def\jlms#1#2#3{{\it J. Lond. Math. Soc.} {\bf {#1}} ({#2}) #3}
\def\jmpa#1#2#3{{\it J. Math. Pures. Appl.} {\bf {#1}} ({#2}) #3}
\def\ma#1#2#3{{\it Math. Ann.} {\bf {#1}} ({#2}) #3}

\def\mz#1#2#3{{\it Math. Zeit.} {\bf {#1}} ({#2}) #3}
\def\ojm#1#2#3{{\it Osaka J.Math.} {\bf {#1}} ({#2}) #3}
\def\pams#1#2#3{{\it Proc. Am. Math. Soc.} {\bf {#1}} ({#2}) #3}
\def\pems#1#2#3{{\it Proc. Edin. Math. Soc.} {\bf {#1}} ({#2}) #3}

\def\plb#1#2#3{{\it Phys. Letts.} {\bf {B#1}} ({#2}) #3}
\def\pla#1#2#3{{\it Phys. Letts.} {\bf {A#1}} ({#2}) #3}
\def\plms#1#2#3{{\it Proc. Lond. Math. Soc.} {\bf {#1}} ({#2}) #3}
\def\pgma#1#2#3{{\it Proc. Glasgow Math. Ass.} {\bf {#1}} ({#2}) #3}
\def\qjm#1#2#3{{\it Quart. J. Math.} {\bf {#1}} ({#2}) #3}
\def\qjpam#1#2#3{{\it Quart. J. Pure and Appl. Math.} {\bf {#1}} ({#2}) #3}

\def\rmjm#1#2#3{{\it Rocky Mountain J. Math.} {\bf {#1}} ({#2}) #3}

\def\tams#1#2#3{{\it Trans.Am.Math.Soc.} {\bf {#1}} ({#2}) #3}

\begin{title}
\vglue 0.5truein
\vskip15truept
\centertext {\Bigfonts \bf Discrete determinants and the} \vskip7truept
\vskip10truept\centertext{\Bigfonts \bf Gel'fand--Yaglom formula}
 \vskip 20truept
\centertext{J.S.Dowker\footnote{dowker@man.ac.uk}} \vskip 7truept
\centertext{\it Theory Group,} \centertext{\it School of Physics and
Astronomy,} \centertext{\it The University of Manchester,} \centertext{\it
Manchester, England} \vskip 7truept \centertext{}

\vskip 7truept

\vskip40truept
\begin{narrow}
I present a partly pedagogic discussion of the Gel'fand--Yaglom formula for the functional
determinant of a linear, one--dimensional, second order difference operator, in the
simplest settings. The formula is a textbook one in discrete Sturm--Liouville theory and
orthogonal polynomials. A two by two matrix approach is developed and applied to Robin
boundary conditions. Euler--Rayleigh sums of eigenvalues are computed. A delta
potential is introduced as a simple, non-trivial example and extended, in an appendix, to
the general case. The continuum limit is considered in a non--rigorous way and a rough
comparison with zeta regularised values is made. Vacuum energies are also considered in
the free case. Chebyshev polynomials act as free propagators and their properties are
developed using the two--matrix formulation, which appears to be novel and has some
advantages. A trace formula, rather than the more usual determinant one, is derived for
the Gel'fand--Yaglom function.

\end{narrow}
\vskip 5truept
\vskip 60truept
\vfil
\end{title}
\pagenum=0
\newpage

\section {\bf 1. Introduction}

Finite structures are very common in science either as approximations to some
continuous arrangement, perhaps for numerical purposes, or because of some inherent
discreteness or, again, for regularisation. They have also gained a certain currency in
elementary particle models.

In this communication I wish to make some rather elementary computations of one or
two quantum field theory quantities using finite difference notions.\footnote{ There are
numerous texts on finite difference equations. An unusual one is Bleich and Melan,
[\pref{BlandM}] and a modern one is Elaydi, [\pref{Elaydi}]. The classic work by
Atkinson, [\pref{Atkinson}], is a central reference.} I restrict myself to the simplest
one--dimensional systems \ie fields on the interval or circle.

Although these have been discussed, almost {\it ad nauseam}, I could not find this
particular development completely in the literature. The interesting work by Actor,
Bender and Reingruber, [\pref{ABR}], contains a detailed  treatment of the Casimir
effect on the lattice, and, while I cannot add too much to their extensive results, I will
recover some of the formulae for completeness.  I will also compute the discrete
determinants for the free field case and I will include a mass here too. Although the
explicit results are rather trivial, and just examples of general expressions, I believe they
have some didactic merit. As something more substantial, I also treat the case of a delta
potential.

Functional determinants appear in many areas and their computation is important
physically. An early method is the Gel'fand--Yaglom technique which is a means of
finding an operator determinant without knowing the eigenvalues explicitly. The
continuum case (originating with Gel'fand and Yaglom, [\pref{GeandY}], and Levit and
Smilansky, [\pref{LandS}]) has been analysed fairly extensively. The work by Kirsten
and McKane, [\pref{KandM}], contains a brief historical survey plus a contour integral
proof of the theorem and a discussion of the zero mode problem. In the quantum field
theory context, Dunne, [\pref{Dunne}], can be consulted for orientation and further
references.

There has been less work on discrete systems, although there is a considerable body of
work concerned with graphs, which, though relevant, I will not consider, {\it per se}.
Very general theorems have been derived by Forman, [\pref{Forman}], for the situation
when a potential is present. He proves and employs a discrete Gel'fand--Yaglom theorem.
In the following sections, I give a simple justification of the formulae  by standard
spectral means. The original treatment by Gel'fand and Yaglom involves a limit process
from a discretisation approach to functional integration, which, in content, is partly
equivalent to the remarks here.

I treat, at least initially,  the simplest set--up that allows me to illustrate the essentials.
This will be the uniform continuous string of length $L$ vibrating transversally. An
approximation by (equal) mass points takes us back to the precursor of Fourier analysis,
the subject of countless historical surveys and textbook explanations. For reference I
mention only the classic Rayleigh, [\pref{Rayleigh}], and Morse and Feshbach,
[\pref{MandF}]. The modes of this discrete system are, therefore, ancient but I will
develope them again. Some are given, relevantly, in the basic finite difference text by
Fort, [\pref{Fort}]. There will necessarily be a certain amount of repetition.

A summary of the discretization, of relevance to the present topic, is given by de
Verdi\'ere, [\pref{deverdiere}], \S9.2.

In the course of the calculation, I encounter Chebyshev polynomials and develope their
properties via a two-matrix technique which is convenient, and might be novel.

\section{\bf2. The discrete Gel'fand--Yaglom theorem.}

To make the situation precise, replace the interval $[0,L]$ by $\nu+2$ equally spaced
points, or vertices, two being end, or boundary, points. Label the points by $j$, $0\le j\le
\nu+1$ and consider some scalar function, $y(j)$, satisfying either Dirichlet (D) or
Neumann (N) conditions at the ends, (\eg\ [\pref{MandF}]),\footnote{ For convenience,
I will assume that all my functions, eigenfunctions \etc are real, except when considering
a twisted periodic field later.}
  $$\eqalign{
  y(0)&=y(\nu+1)=0\,,\quad D\cr
  y(0)&=y(1),\,\,\,y(\nu)=y(\nu+1)\,,\quad N\,.
  }
  \eql{bcs}
  $$

I discuss the Sturm--Liouville problem which, in its simplest formulation, involves the
eigenvalue recurrence, (\eg\ [\pref{Fort}]),
  $$
   y(j+1)+\big(\la-V(j)-2\big)\, y(j)+y(j-1)=0\,,
  \eql{lin3}
  $$
subject to boundary conditions, say (\peq{bcs}).

I refer to $V(j)$ as the potential because  (\peq{lin3}) can be rewritten as the more
familar looking Laplacian eigenvalue equation,\footnote{ $\nabla$ is the backwards
difference operator.}
  $$
  \bigg[-{1\over h^2}\nabla\De +\ol V(j)\bigg]y(j)=\ol\la\,y(j)\,.
  \eql{lap}
  $$

The lattice spacing, $h=L/(\nu+1)$, has been introduced by scaling to give a `physical'
Laplacian and one has the dimensionless quantities, $\la=h^2\ol\la$ and $V=h^2 \ol V$.

The procedure is textbook. Taking D conditions for definiteness, iteration from the $j=0$
end point, assuming any value of $y(1)$, except zero, yields  all the $y(j)$ as
polynomials in $\la$, in particular, the terminal value, $y(\nu+1,\la)$.  The eigenvalues
are thus the roots of this polynomial, $y(\nu+1,\la)=0$ (\eg\ Atkinson,
[\pref{Atkinson}]) and the determinant (\ie the product of all the $\la$) of the operator
is its constant term, $y(\nu+1,0)$, up to a factor, which is the essence of the
Gel'fand--Yaglom formula\footnote{ The nature of this constant is where the problem lies
in the continuum case.}. The factor involved is unity if the starting term is chosen to be
$y(1)=1$, as can be seen by looking at the $\la\to\infty$ limit (see later).

The product of all the physical $\ol\la$ is only a scaling factor different and one arrives at
the discrete Dirichlet result, \eg\ [\pref{Forman}],
  $$
  \Det_D={1\over h^{2\nu}}\,y(\nu+1,0)\,.
  \eql{fdetd}
  $$
This formula  is thus part and parcel of the standard eigenvalue problem. The resolvent of
(\peq{lin3}) is,
  $$
  R(\la)={d\over d\la}\log\,y(\nu+1,\la)\,,
  $$
with the usual machinery. For example, the sums of the inverse powers of the roots
follow, {\it \`a la} Euler and Rayleigh, [\pref{Rayleigh}] I, p.279 , as,
 $$
  -R(\la)=\sum_n{1\over\la_n}+
  \la\sum_n{1\over\la ^2_n}+\la^2\sum_n{1\over\la ^3_n}+\ldots\,\,.
  \eql{psums}
 $$

The rigorous proof that the discrete formula leads to the original continuous one of
Gel'fand and Yaglom and of Levit and Smilansky, [\pref{LandS}], is given by Forman. de
Verdi\'ere, [\pref{deverdiere}], also discusses the nature of this limit.

\section{\bf 3. Dirichlet constant potential}

Before continuing to other boundary conditions, I give the simplest application of
(\peq{fdetd}) which is when the potential is constant and equivalent to a mass term,
$\mu^2$. I then rewrite (\peq{lin3}),
  $$
  y(j+1)-2\cosh2\ga\,\, y(j)+ y(j-1)=0\,,
  \eql{lin2}
  $$
where I have set $\mu=h\ol\mu$ and $\mu^2-\la=4\sinh^2\ga$ and which I must
solve subject to the initial conditions $y(0)=0$, $ y(1)=1$. The roots of the auxiliary
equation are,
  $$\eqalign{
  m_{\pm}&=\cosh2\ga\pm\sinh2\ga=e^{\pm2\ga}\,,
  }
  $$
and the general solution is,
  $$
  y(j)=A m_+^j+Bm_-^j\,,
  $$
with
  $$
  A=-B={1\over2\sinh2\ga}\,.
  $$
This implies that the discrete Gel'fand--Yaglom function (more conventionally called the
fundamental solution) is,\mgn{Notation}
  $$
  y(j)={\sinh2\ga j\over\sinh2\ga}\,,
  \eql{gy}
  $$
evaluated at the terminal point, $j=\nu+1$, which is, perhaps, no surprise in view of the
textbook continuum analogue. The functions, $y(j)$, written $y(j,\la)$, are polynomials
in $4\sinh^2\ga$ (and hence in $\la$) which can be proved in many ways, one of which
is the direct iteration of (\peq{lin2}). Equation (\peq{lin3}) is a recursion formula for
these polynomials, which are Chebyshev polynomials, as is well known, the definition
being,
  $$
  U_{\nu}(\cosh2\ga)\equiv{\sinh 2\ga(\nu+1)\over\sinh2\ga}\,.
  \eql{cheb}
  $$

Pursuing the calculation, the determinant is obtained by setting $\la=0$,
  $$
  \Det_D(\ol\mu)={1\over h^{2\nu}}\,{\sinh2\ga_0 (\nu+1)\over\sinh2\ga_0}\,.
  \quad \mu=2\sinh\ga_0\,.
  \eql{detd2}
  $$
The constant of proportionality is settled by the infinite $\la$ limit when the
Gel'fand--Yaglom function has the explicit behaviour,
  $$
  {\sinh2\ga (\nu+1)\over\sinh2\ga}\to (2\cosh2\ga)^{\nu}\sim (-\la)^\nu\,.
  $$

The eigenvalues themselves are determined by,
  $$
  \sinh2\ga(\nu+1)=0
  $$
or
  $$
  \ga=\ga_n\equiv{n\pi i\over2(\nu+1)}\,,
  $$
and so
  $$
  \ol\la_n=\ol\mu^2+{4\over h^2}\sin^2 \!{\pi n\over 2(\nu+1)}\,,\quad n=1,\ldots,\nu\,,
  \eql{Deigen}
  $$
which is the textbook result, \eg\ [\pref{Fort}]. Equating the determinant (\peq{detd2})
to $\prod_n \la_n$ gives a standard product formula, \eg\ Bromwich,
[\pref{Bromwich}], p.211, which comes up later in \S13. Furthermore, the sums of
inverse powers of the roots, (\peq{psums}), yields finite summations for powers of
cosecants, \eg, typically,
  $$
  \sum_{n=1}^{p-1} \cosec^2 {\pi n\over 2p}={2\over3}(p^2-1)\,,
  \eql{cosecs}
  $$
which are very old and are simple examples of a wide class of trigonometric summations
obtainable in many ways.\footnote{ I attempted a few comments and gave some
references in [\pref{Dow7}]. See also Berndt and Yeap, [\pref{BandY}].} As $p\to\infty$
(the continuum limit) this sum becomes Euler's result, $\ze_R(2)=\pi^2/6$.

The eigenfunctions follow by noting that the fundamental solution, $y(j,\la)$,
(\peq{gy}), satisfies the equation (\peq{lin2}) with $\la=\la_n$ and obeys the Dirichlet
conditions. The eigenfunctions are therefore,
  $$
  y_n(j)=\sin {j\pi n\over \nu+1}\,,\quad n=1,\ldots,\nu\,,
  $$
of which there are $\nu$, this being the number of `dynamical' points. We therefore
reach the standard mode properties, \eg\ Fort, [\pref{Fort}], Spiegel, [\pref{Spiegel}].
This route is not a novel one.

\section{\bf 4. Neumann conditions}

As a warm--up for the Robin case, I consider Neumann boundary conditions (\peq{bcs})
which can be written, $\De y(0)=0$, $\De y(\nu)=0$. If $V(j)$ were uniform,
(\peq{lin3}) would be satisfied by $\De y(j)$ and the problem translated into a Dirichlet
one (see \S8) but, because of the $j$ dependence, this is not possible and it is necessary
to treat $(\peq{lin3})$ and its difference together.\footnote{ This is a common device in
the theory of ordinary differential equations. For difference equations see \eg\ Porter,
[\pref{Porter}], Goldberg, [\pref{Goldberg}], p.233 Ex.4., Elaydi, [\pref{Elaydi}].} This
is most neatly expressed using $2\times2$ matrices as in [\pref{Atkinson}],
[\pref{Forman}] and elsewhere, \eg\ [\pref{KandM}], although a little differently. It
might be considered a `phase space' representation.

Defining,
  $$
  \Upsilon(j)=\left(\matrix{y(j)\cr y(j+1)}\right)\,,
  \eql{matrix}
  $$
the recurrence under study is the first order one,
  $$
  \Up(j)- M(j)\Up(j-1)=0\,,
  \eql{nrec}
  $$
where
  $$\eqalign{
  M(j)&=\left(\matrix{0&1\cr-1 &V(j)+2-\la}\right)\cr
  &=\left(\matrix{0&1\cr-1 &2\cosh2\ga_j}\right)\,,
  }
  \eql{em}
  $$
whereby $\ga_j$ is defined. One of the equations is just an identity. Note that $\det
M(j)=1$.

The Sturm-Liouville Neumann boundary condition is $\De  y(0)=0$, $\De y(\nu)=0$
which means, choosing a normalisation,
  $$
\Up(0)=\left(\matrix{1\cr 1}\right)\,,\quad \Up(\nu)\propto\left(\matrix{1\cr 1}\right)\,.
\eql{cond}
  $$

The eigenvalue procedure is to iterate (\peq{nrec}) up to $\Up(\nu)$ starting from,
$\Up(0)$ \ie,
  $$
  \Up(\nu)=M(\nu)M(\nu-1)\ldots M(1)\Up(0)\,,
  \eql{miter}
  $$
and then impose the condition, (\peq{cond}), on $\Up(\nu)$. (Remember, the $M$s are
functions of $\la$.)

The roots of the polynomial,
  $$
  P(\nu,\la)=\left(\matrix{1,&-1}\right)\Up(\nu)=\De y(\nu,\la)\,,
  $$
are then the eigenvalues and the determinant is $P(\nu,0)$, which is the required
theorem, [\pref{Forman}],
  $$
  \Det_N={1\over h^{2\nu}}\,\De y(\nu,0)\,.
  $$
\section{\bf 5. Robin boundary conditions}

Having treated pure Neumann, it is not much more difficult to sort out Robin conditions,
called General Local in [\pref{Forman}]. The recurrence is still (\peq{nrec}) but with the
boundary conditions,
  $$
  \Up(0)=\left(\matrix{1\cr 1+\al}\right)\equiv\Up_{in}\,,
  \quad \Up(\nu)\propto\left(\matrix{1+\be\cr 1}\right)\equiv\Up_{out}\,,
  \eql{robbcs}
  $$
where $\al$ and $\be$ are the Robin parameters defined by,
  $$
  \De y(0)=\al\, y(0)\,,\quad \De y(\nu)=-\be\, y(\nu+1)\,.
  $$
The first condition in (\peq{robbcs}) is chosen, and the second is imposed.

Defining an `adjoint',
 $$
\Up^{\dag}(j)=\wt\Up(j)\,J\,,
 $$
in terms of the symplectic metric, $J=\left(\matrix{0&1\cr-1&0}\right)$, the eigenvalue
polynomial is, therefore, the matrix element,
  $$\eqalign{
  \left(\matrix{-1,&\!\! 1+\be}\right)\Up_{in}(\nu)=&\Up^{\dag}_{out}\Up_{in}(\nu)\cr
  =&\Up^{\dag}_{out}\,M(\nu)M(\nu-1)\ldots M(1)\Up_{in}\,,\cr
 }
 \eql{eigpol}
  $$
and the product of the $\la$ eigenvalues is proportional to
$\Up_{out}^{\dag}\Up_{in}(\nu)$, evaluated at $\la=0$, the constant of
proportionality being the inverse of the coefficient of the highest power of $\la$ in
(\peq{eigpol}). For very large $\la$, $M$, (\peq{em}), approximates to,
  $$
  M(j)\sim\left(\matrix{0&0\cr0 &-\la}\right)\,,
  \eql{em}
  $$
when the right--hand side of (\peq{eigpol}) becomes $(1+\be)(1+\al)(-\la)^\nu$ so the
determinant is, after scaling to the physical eigenvalues, $\la$,
  $$
  \Det_R(\ol\al,\ol\be)={1\over h^{2\nu}}\,{1\over(1+\be)(1+\al)}\,\Up^{\dag}_{out}\,
  \prod_{j=1}^\nu M_0(j)\,\Up_{in}\,,
\eql{robdet}
  $$
where $M_0$ is the matrix $M$ evaluated at $\la=0$. For future use I have also
introduced the physical constants, $\ol\al=\al/h$, $\ol\be=\be/h$.

I note that the symplectic product is just the Casoratian (discrete Wronskian),
[\pref{Elaydi}],
  $$
  \Up^{\dag}_1(j)\,\Up_2(j)\equiv\wt\Up_1(j)\,J\,\Up_2(j)\,,
  \eql{casor}
  $$
of two solutions, $\Up_1$ and $\Up_2$, of (\peq{nrec}), which is a symplectic
development because,
  $$
   \wt M\,J\,M=J\,,
  $$
and so (\peq{casor}) is uniform, \ie\ independent of $j$. This is a neater proof of this
fact than the usual ones, \eg\ Fort, [\pref{Fort}].

The equivalence of  a $2\times2$ real matrix formulation and a three--term recurrence
relation is expounded by Atkinson, [\pref{Atkinson}], \S3.5 involving a geometrical
interpretation of symplectic action.

I take up the general formalism again in \S10 and turn now to an elementary case.

\section{\bf 6. Constant potential}

As the simplest example, I again take that of a constant potential. Then $M(j)$ is,
  $$
  M(j)=M=\left(\matrix{0&1\cr-1 &2\cosh2\ga}\right)\,,
  \eql{em}
  $$
with $\ga$ as before, \ie $4\sinh^2\ga=\mu^2-\la$.

It is shown in Appendix 2 that the power $M^\nu$ is given by,
  $$
  M^\nu=\left(\matrix{-U_{\nu-2}&U_{\nu-1}\cr-U_{\nu-1}&U_\nu}\right)
  $$
in terms of Chebyshev polynomials, (\peq{cheb}), of argument $\cosh2\ga$. Then,
   $$\eqalign{
  \Up^{\dag}_{out}\, M^\nu\,\Up_{in}&=(-1,1+\be)
  \left(\matrix{-U_{\nu-2}&U_{\nu-1}\cr-U_{\nu-1}&U_\nu}\right)\,
  \left(\matrix{1\cr1+\al}\right)\cr
  \noalign{\vskip5truept}
 &=(\al+\be+\al\be)\,U_\nu-(\al+\be-\la)U_{\nu-1}\,,
  }
  \eql{me}
  $$
which is to be substituted into (\peq{robdet}), after setting $\la=0$ to give the
determinant. The formula is symmetric under interchange of $\al$ and $\be$, as it
should be by geometric symmetry.

As another check, the Dirichlet choice,
  $$
\Up^D_{in}=\left(\matrix{0\cr 1}\right)\,,
\quad \Up_{out}^D\propto\left(\matrix{1\cr 0}\right)\,,
\eql{dcond}
  $$
reproduces (\peq{detd2}).

Incidentally, if $M$ were more general, rather than diagonalisation, it would be easier, for
iteration purposes,  to set,
  $$
  M^\nu=a{\bf1}+b\,M\,,
  \eql{emnu}
  $$
and compute $a$ and $b$ in terms of the eigenvalues of $M$, \eg\ Goldberg,
[\pref{Goldberg}].

\section {\bf 7. The characteristic polynomial and Euler--Rayleigh sums}

The Euler--Rayleigh sums, analogous to (\peq{cosecs}), arise from the expansion of
(\peq{me}) in powers of $\la$, \ie of $-4\sinh^2\ga$ (for simplicity I set $\mu$ to zero)
which is easily accomplished by, say, using Bromwich, [\pref{Bromwich}] chap.IX or the
relation to Chebychev polynomials. I find,
  $$\eqalign{
  \Up^{\dag}_{out}\,& M^\nu\,\Up_{in}=\cr
  &=(\al\,\be+\al+\be)\sum_{s=0}^\nu {\nu\over2\nu-s}\comb{2\nu-s}{s}
  (-\la)^{\nu-s}\cr
  &+\bigg(\al\,\be-{1\over2}(\al\,\be+\al+\be+2)\la\bigg)
  \sum_{s=0}^{\nu-1}\comb{2\nu-s-1}{s}(-\la)^{\nu-s-1}\,.
  }
  \eql{me2}
  $$
The $\nu$ eigenvalues, $\la_n$, are the roots of this characteristic polynomial and it is
next required to expand its logarithm, which, for low powers, can be done directly by
hand. As the simplest case I give
  $$
  \sum_{n=0}^{\nu-1}\cosec^2(\th_n)=2{3\nu^2(\al\be+\al+\be)
  +\nu(\nu^2-1)\al\be+3\nu(\al\be+\al+\be+2)
  \over3\big((1+\nu)  \al\be+\al+\be\big)}\,,
 \eql{raysum1}
  $$
where I have set
  $$
  \la_n=4\sin^2\th_n\,,
  $$
which defines $\th_n$.

The continuum limit $h\to0$, $\nu\to\infty$ is not without interest and is discussed in a
section 9.

\section {\bf  8. Neumann conditions revisited}
As the free $N$ case is not given in Fort, [\pref{Fort}], I give, for pedagogic
completeness,  the conventional calculation by noting, first, that the $N$ conditions,
(\peq{bcs}), can be written $\De y(0)=\De y(\nu)=0$. So, defining $\phi(j)=\De y(j)$,
one has, from (\peq{lap}),
   $$
  -{1\over h^2}\,\nabla\De\,\phi(j)=\ol\la\,\phi(j)
  \eql{lap2}
  $$
with $\phi(0)=\phi(\nu)=0$, which is a $D$ problem on $\nu+1$ vertices but with the
original spacing, $h$. The $D$--eigenfunctions are
   $$
  \phi(j)=\sin {n\pi j\over \nu}\,,\quad n=1,\ldots,\nu-1\,,
  $$
and hence the $N$--eigenfunctions are,\footnote{ If you use Jordan, [\pref{Jordan}], be
aware that there is an error on p.117 that is carried forward. For example, on p.124 the
sum of $\cos(x+b)\phi$ is incorrect. The upper limit should be $n-1$.}
  $$\eqalign{
  y(j)&=\De^{-1}\,\phi(j)=\De^{-1}\,\sin {n\pi j\over \nu}\cr
  &\approx\cos {n\pi(2j-1)\over2\nu}\,,\quad n=0,1,\ldots,\nu-1\,,
  }
  \eql{Neigenf}
  $$
up to a numerical factor and possible additional constant.  The eigenvalues are,
  $$
  \la_n={4\over h^2}\sin^2 {\pi n\over 2\nu}\,,\quad n=0,\ldots,\nu-1\,.
  \eql{Neigen}
  $$

Again, there are $\nu$ modes, including the uniform zero one, $n=0$, which corresonds
to a constant of integration in (\peq{Neigenf}). ($n=\nu$ gives a vanishing mode.)

Before going on, it would be best to see, as a check, if the pure Neumann determinant,
for the free case with mass, agrees with the above mode structure and the Robin
expression, (\peq{me}). Effectively I start again. The initial condition that fixes the
Gel'fand--Yaglom function is $z(0)=1$ and $z(1)=1$, \ie\ $\De z(0)=0$. (This is
Forman's $z$.) The general solution is again,
$$
z(j)=Ae^{2\ga j}+Be^{-2\ga j}\,,
  $$
and the conditions imply,
  $$\eqalign{
  A+B=1\cr
  Ae^{2\ga }+Be^{-2\ga }=1\,,
  }
  $$
which solve to,
  $$
  A={e^{-\ga}\over2\cosh\ga}\,,\quad  B={e^{\ga}\over2\cosh\ga}\,,
  $$
so that,
  $$\eqalign{
  z(j)&={\cosh (2j-1)\ga\over\cosh\ga}=V_{j-1}=\nabla U_{j-1}\cr
   \De z(j)&=4\sinh^2\ga {\sinh2\ga j\over\sinh2\ga}=\De\nabla U_{j-1}\,.
   }
  \eql{gyn}
  $$
where $V_j$ is a Chebyshev polynomial of the third kind, [\pref{MandH}], and all
Chebyshev arguments are $\cosh2\ga$. Appendix 1 contains some relations for
Chebyshev polynomials couched in the two--matrix language and (\peq{gyn}) can be
obtained more rapidly using this.

Applying the eigenvalue restriction, $\De z(\nu,\la)\equiv\De z(\nu)=0$,  yields the
condition $\ga=\ga_n=n\pi i/\nu$, $n=0,1,\ldots \nu-1$, and the eigenvalues are,
  $$
  \la_n=\mu^2+{4\over h^2}\sin^2 {\pi n\over 2\nu}\,,\quad n=0,\ldots,\nu-1\,,
  \eql{Neigen2}
  $$
consistent with (\peq{Neigen}). The eigenfunctions, (\peq{Neigenf}), also follow trivially
from (\peq{gyn}).

One sees from (\peq{gyn})  that the eigenvalue condition is the same as the Dirichlet
one, except for the replacement $\nu\to\nu-1$ and  for the factor
$4\sinh^2\ga=\mu^2-\la$. This factor is responsible for the $n=0$ mode which, in the
massless case, is a zero mode.

The Neumann determinant is then,\footnote{ This appears to differ by a factor of $1/4$
from Forman's formula, [\pref{Forman}].}
  $$
  \Det_N(\mu)={1\over h^{2\nu}}\,\De z(\nu,0)\,,
  $$
where the numerical factor follows on the limit,
  $$\eqalign{
  &4\sinh^2\ga {\sinh2\ga\nu\over\sinh2\ga}
  \to4\sinh^2\ga(2\cosh2\ga)^{\nu-1}\sim(-\la)^\nu\,.
  }
  $$
Equating the two forms of the determinant yields the same product formula as in the D
case.

The determinant also agrees with the Robin formula, from (\peq{robdet}), for
$\al=\be=0$. (This is, of course, simply a check of algebra.)

\section{\bf 9. The continuum limit}

Comparisons with known results can also be obtained by considering the continuum limit,
an historical motivation for discretisation. Again as an example, I consider the Robin
determinant (\peq{robdet}) with (\peq{me}) in the limit $h\to0$. To get the leading
divergence, the lowest power of $h$ is required in the expression multiplying
$1/h^{2\nu}$. As $h\to 0$ one has the limiting behaviours,
  $$\eqalign{
  &2\sinh \ga_0\sim 2\ga_0\sim\mu=h\ol\mu\,,\quad2\ga_0\nu\sim h\ol\mu\nu\sim
  \ol\mu L\,,\cr
}
  $$
and therefore by inspection of (\peq{me}), one sees that the leading term is of order
$h$. Extracting this gives,
  $$\eqalign{
  \Up^{\dag}_{out}\,& M_0^\nu\,\Up_{in}\cr
  &\to(\ol\al+\ol\be)\cosh\ol\mu L+(\ol\al\ol\be+\ol\mu^2){\sinh\ol\mu L\over
  \ol\mu}\,,
  }
  \eql{mel}
  $$
which agrees with an expression in [\pref{Dowrob}] for the continuum case.

Related is the limit of the simplest Euler--Rayleigh eigenvalue sum, (\peq{raysum1}).
Reverting to physical quantities,
  $$
  h^2\sum_{n=0}^{\nu-1}{1\over\la_n}=\sum_{n=0}^{\nu-1}
  {1\over\ol\la_n}\to{3(\ol\al+\ol\be)+\ol\al\ol\be+6\over 6(\ol\al\ol\be+\ol\al+\ol\be)}\,,
 \eql{raysum2}
  $$
which is also given in [\pref{Dowrob}].

\section{\bf 10. Non--uniform potential. The propagator}

Difference equation Sturm--Liouville  theory is well developed and can be pursued by
analogy to the continuum version, \eg\ Fort, [\pref{Fort}], Levy and Baggott,
[\pref{LandB}]. In fact Sturm obtained many continuum results via a discrete route,
although this was never published.

In this section I wish to develope and summarize the previous matrix formulation, see
(\peq{matrix}), (\peq{nrec}), (\peq{em}). I consider (\peq{nrec}) as a Schr\"odinger
equation for a two--state system with a discrete time labelled by $j$, and rewrite it by
defining a matrix `propagator' $K(\la;j,j')$,
  $$
  K(\la;j,j')=\th(j,j')\,\prod_{k=j'+1}^j M(k)\,
  \eql{prop}
  $$
as
  $$
  \Up(j)=K(\la;j,j')\Up(j')\,,\quad j\ge j'\,,
  \eql{trans}
  $$
which propagates forwards from $j'$ to $j$ and acts as a transfer $2\times2$  matrix. In
the simplest case, the matrix $M$ is given by (\peq{em}). The form, (\peq{prop}), is an
equivalent of the time--ordered exponential solution in time--dependent perturbation
theory, but here `vertex--ordered'. The propagator, $K(\la;j,0)$ is sometimes referred to
as the state transition matrix. I denote it by $K(\la;j)$. The basic theory is given by
Elaydi, [\pref{Elaydi}] \S3.2, but my treatment is modified a little and also deals,
particularly, with a symplectic invariant propagation.

For consistency, the initial condition, (\ie the first empty product in (\peq{prop})),
  $$
  K(\la;j,j)={\bf1}\,,
  \eql{econd}
  $$
has to be taken. The step function $\th$ ensures that $K(j,j')=0$ for $j<j'$,
corresponding to causal propagation. The semi--group property,
  $$
  K(\la;j,j')K(\la;j',j'')=K(\la;j,j'')\,,
  $$
(no sum on $j'$)  and symplectic invariance,
  $$
  \wt K(\la;j,j')\,J\,K(\la;j,j')=J\,,
  \eql{sympinv}
  $$
also hold.

$K$ satisfies the equation of motion,
    $$
    K(\la;j,j')\equiv EK(\la;j-1,j')={\bf 1}\de_{j,j'}+M(j)K(\la;j-1,j')\,,
    \eql{eom}
    $$
where the first term arises from the $\th$ factor in (\peq{prop}). A matrix which
satisfies (\peq{eom}) is a {\it fundamental matrix}.

Iteration of (\peq{eom}) gives a power series expansion,
   $$
    K(\la;j,j')={\bf 1}\de_{j,j'}+M(j)\de_{j,j'+1}+M(j)M(j-1)\de_{j,j'+2}+\ldots\,,
    \eql{pwrs}
   $$
which is quite equivalent to (\peq{prop}). It also follows from the decomposition,
   $$\eqalign{
    \th(j,j')&=\de_{j,j'}+\de_{j,j'+1}+\de_{j,j'+2}\,+\ldots\cr
    &=\De^{-1}\de_{j,j'}\,,
    }
    \eql{comp1}
   $$
obvious graphically, arithmetically and in $(\nu+1)\times(\nu+1)$ matrix form. It is the
discrete version of the distributional operator statement that the $\th$--function is the
integral of the $\de$--function.

If $M(j)$ is constant, then, trivially $K(\la;j,j')=\th(j,j')M^{j-j'}$, either from
(\peq{prop}) or read off from (\peq{pwrs}).

If an `unperturbed' propagator, $K_0$, is defined by,
  $$
    K_0(j,j')={\bf 1}\de_{j,j'}+M_0(j)K_0(j-1,j')\,,
    \eql{eom2}
    $$
then,
  $$
  K(\la;j,j'')=K_0(j,j'')+K_0(j,j')\big(M(j')-M_0(j')\big)K(\la;j'-1,j'')\,,
  $$
where $j'$ is summed over from $1$ to $\nu$, can be considered as a perturbation
expansion. If $M_0$ is constant,
  $$
  K(\la;j,j'')=M_0^{j-j''}+M_0^{j-j'}\big(M(j')-M_0\big)K(\la;j'-1,j'')\,.
  \eql{kprop}
  $$

The propagator, $K(\la;j,j')$ is defined independently of any boundary conditions which
are incorporated, in my approach, by constructing the symplectic scalar products,
  $$\eqalign{
  P(\la)&=\Up^{\dag}_{out}(j)\,K(\la;j,j')\Up_{in}(j')\cr
  &=\Up^{\dag}_{out}(j)\,\Up_{in}(j)\,,\quad \forall j\,.
  }
  \eql{poly}
  $$
These are polynomials in $\la$ and, because of the uniformity of the Casoratian, are
independent of $j$. The boundary conditions are given by $\Up_{in}(0)=\Up_{in}$ and
$\Up_{out}(\nu)=\Up_{out}$, as given in (\peq{robbcs}).  $\Up_{in}(j)$ is the
solution of (\peq{trans}) for the `in' condition and $\Up_{out}(j)$ that for the `out'
one.

The vanishing of $P(\la)$ determines the $\nu$ eigenvalues, $\la_n$,
($n=0,\ldots,\nu-1$). This characteristic polynomial reads, in the extreme cases,
  $$
  P(\la)=\Up^{\dag}_{out}(\nu)\,\Up_{in}(\nu)
  =\Up^{\dag}_{out}(0)\,\Up_{in}(0)\,.
  $$
The $\la$ dependence is contained in $\Up_{in}(\nu)$ or in $\Up_{out}^{\dag}(0)$.

The full determinant is the normalised $P(0)$,\footnote{ $P(\la)$ is the analogue of an
$S$--matrix element.}
  $$
   \Det={P(0)\over \Up^{\dag}_{out}A\Up_{in}}\, \,.
   \eql{gendet}
   $$
All this we have had before in particular cases.

To expose the parameter $\la$, and to enlarge on the formalism, it is helpful to split the
driving matrix $M$ as,
  $$
  M(j)=B(j)-\la\,A(j)\,,
  \eql{msplit}
  $$
where,
  $$\eqalign{
  B(j)&=\left(\matrix{0&1\cr-1 &V(j)+2}\right)\,,\quad
  A(j)=A=\left(\matrix{0&0\cr0 &1}\right)\,,
  }
  \eql{eba}
  $$
with
  $$
  \wt B\,J\,B=J,\quad \wt A\,J\,A=0,\quad \wt A\,J\,B=-A\,.
  \eql{symps}
  $$

Then consider two fundamental matrices, $K(\la;j)$ and $K(\mu;j)$, and make the usual
construction, \cf\ [\pref{Atkinson}],
  $$\eqalign{
  \wt K(\mu;j+1)\,J\,K(\la;j&+1)-\wt K(\mu;j)\,J\,K(\la;j)\cr
  &=\wt K(\mu;j)\big((\wt B(j)-\mu\wt A)\,J\, (\wt B(j)-\la A)-J\big)K(\la;j)\cr
  &=(\la-\mu)\wt K(\mu;j)\,A\,K(\la;j)\,.
  }
  $$

Summing over $j$ from $0$ to $\nu-1$ (\ie\ performing the inverse $\De^{-1}$), one
gets
  $$
  \wt K(\mu;\nu)\,J\,K(\la;\nu)-J=(\la-\mu)\sum_{j=0}^{\nu-1}
  \wt K(\mu;j)\,A\,K(\la;j)\,.
  \eql{christdarb}
  $$

In this equation, $\la$ and $\mu$ are any two parameters. I now restrict them to being
eigenvalues, that is, solutions of the polynomial equation $  P(\la)=0$, or,
  $$
  \Up_{out}^{\dag}(j)\Up_{in}(j)=0\,,\quad \forall j\,,
  $$
which implies that the `out' eigenvector $\Up_{out}(j)$ is the same (up to a constant
factor) as the `in' one, $\Up_{in}(j)$, for each eigenvalue\footnote{ Equivalently, the
eigenvalues are simple.}and I can denote both of them by $\Up\,(j,\la)$. In particular,
$\Up_{out}(0)$ and $\Up_{in}(\nu)$ are independent of the eigenvalue.

In this case, mutiplying (\peq{christdarb}) by the boundary (eigenvalue independent)
vectors $\Up_{in}$, on the right, and $\wt\Up_{out}$ on the left, the left--hand side
vanishes,
  $$\eqalign{
  \wt\Up_{out}\,\wt K(\mu;\nu)\,J\,&  K(\la;\nu)\,\Up_{in}-
  \wt \Up_{out}\,J\,\Up_{in}\cr
  &= \wt \Up_{out}(0,\mu)\,J\,\Up_{in}(\nu,\la)
  - \wt \Up_{out}(\nu)\,J\,\Up_{in}(0)\cr
  &= \wt \Up(0)\,J\,\Up(\nu)- \wt \Up(\nu)\,J\,\Up(0)\cr
  &=0
  }
  $$
and so,
  $$
  (\la-\mu)\sum_{j=0}^{\nu-1}
  \wt \Up(j,\mu)\,A\,\Up(j,\la)=0\,,
  $$
with the usual conclusion that eigenvectors with different eigenvalues are orthogonal,
  $$
  \sum_{j=0}^{\nu-1}
  \wt \Up(j,\mu)\,A\,\Up(j,\la)=0\,,\quad \mu\ne\la.
  $$

In the circumstances of the present paper, the matrix $A$ is the projection onto the
lower components of $\Up(j)$, \ie onto $y(j+1)$, and so I have regained the usual
orthogonality,
  $$
  \sum_{j=1}^{\nu} y(j,\la_n)\,\,y(j,\la_m)=\rho_n\de_{nm}\,,\quad 0\le n,m\le \nu-1\,,
  \eql{orthog}
  $$
where $\rho_n$ is a normalisation.

A standard procedure then yields the completeness (or dual orthogonality) relation,
  $$
  \sum_{n=0}^{\nu-1} y(j,\la_n)\,\,y(j',\la_n)\rho_n^{-1}
  =\de_{jj'}\,,\quad 1\le j,j'\le \nu\,.
  $$

The appearance in (\peq{orthog}) of a sum over the vertices, $j$, leads us to the
traditional matrix approach, (\eg\ Rayleigh, [\pref{Rayleigh}], Atkinson,
[\pref{Atkinson}], Chap.6, Gantmacher and Krein, [\pref{GandK}] ), which takes the
entire set of (dynamic) values, $y(j)$ ($1\le j\le \nu$) as the components of a column
$\nu$--vector, $y$, and writes the collection of difference equations, (\peq{lin3}), as a
$\nu\times\nu$ matrix equation of the familiar eigenproblem form,
  $$
  L\,y=\la y\,.
  $$
The polynomials, $y(j,\la)$, are then related to the Jacobi determinant
$\det(L-\la{\bf1})$. I extend my present formalism a little to reflect this perspective,
which has already shown up in (\peq{pwrs}) and (\peq{comp1}).

It is sometimes convenient to employ the operator formalism as in finite dimensional
quantum mechanics, due to Schwinger and Weyl, [\pref{Weylqm}], \cf\ Floratos
[\pref{Floratos}], and set \eg,
  $$
  \me{j}{K}{j'}=K(j,j')\,.
  $$
I retain $K$ as a $2\times2$ {\it matrix} in phase space. Then the recurrence is,
  $$
  \Up=M\Up\,,
  $$
with $M$ a subdiagonal matrix,\mgn{Check size of matrices}
  $$
  \me jM{j'}=M(j)\,\de_{j,j'+1}\,,
  $$
and the series (\peq{pwrs}) translates into the simple operator equation,
  $$
  K={\bf1}+M\,K\,,
  $$
or, formally,
  $$
   K={{\bf1}\over {\bf1}-M}={{\bf1}\over {\bf1}-B-\la A}\,.
  $$
The elements (which are matrices) of the powers of $M$ are correctly vertex ordered.

I will not pursue this formulation any further at this time except to say that the stepping
matrix has just ones along the subdiagonal  and represents the translation operator,
often denoted by $E$ in finite difference calculus. The Heaviside matrix, $\Th$, having
$\th(j,j')$ as elements, is triangular with ones in the left--hand part, and on the
diagonal. It is related to $E$ by $E\Th=\Th-{\bf1}$.

\section {\bf 11. The $\de$ potential on the interval}

A very basic example of a variable potential is one that is non--zero at only one vertex,
\ie $V(j)=v\,\de_{jk'}$. Then, in the product form, (\peq{prop}), of the propagator, only
one term $k=k'$ will be different from the rest. The remaining products (powers) can be
dealt with as before, in \S6, and an explicit expression found for the transition operator
$K(\la;\nu)$, say.

As this is just meant for illustrative purposes, I choose a value of $k'$, namely $k'=2$,
that results in a simple formula.

In this case, for Dirichlet conditions with (\peq{dcond}), I find the polynomial in $\la$,
  $$\eqalign{
  \Up_{out}^{\dag}K(\la;\nu)&\Up_{in}=U_{\nu}(1-\la/2)+v(2-\la) U_{\nu-2}(1-\la/2)\cr
  &=U_{\nu}(1-\la/2)+v\big(U_{\nu-1}(1-\la/2)+U_{\nu-3}(1-\la/2)\big)\,,
  }
  \eql{pert}
  $$
in terms of Chebychev polynomials (the `unperturbed' functions), (\peq{cheb}) with
$2\cosh2\ga=2-\la$. The eigenvalues are easily determined numerically and, for a small
number of vertex points, even analytically as functions of the strength of the potential.

Equation (\peq{pert}) is proved and explored in Appendix 2 where it is extended to the
case when all the $v_j$ are populated.

As particular quantities, the sums of the inverse eigenvalue powers can again be
computed by expanding the logarithm of this polynomial, which, to lowest orders is,
  $$
  \Up_{out}^{\dag}K(\la;\nu)\Up_{in}=(\nu+1)+2v(\nu-1)-{\nu+1\over6}\bigg(
  \nu(\nu+2)+2v(\nu^2-3\nu+3)\bigg)\la\,+\ldots
  \eql{kexp}
  $$
on using,
  $$
   U_{\nu}(\cosh2\ga)=(\nu+1)\bigg(1-{1\over6}\nu(\nu+2)\,\la+\ldots\bigg)\,.
  $$
One then finds, exact in $v$,
  $$
  \sum_{n=0}^{\nu-1}{1\over\la_n}={\nu+1\over6}
  {\big(\nu(\nu+2)+2v(\nu^2-3\nu+3)\big)\over\nu+1+2v(\nu-1)}\,,
  $$
which generalises (\peq{cosecs}).

This identity is an example of a general class of identities discussed in the interesting
work by Annaby and Asharabi, [\pref{AandA}], where other references can be found.

The determinant is just the constant term in the polynomial, (\peq{kexp}),
  $$
  \Det_D=\nu+1+2v(\nu-1)\,,
  $$
and a zero  mode, $\la_0=0$, occurs when $v=-(\nu+1)/2(\nu-1)$. When $v$ takes the
same value with the opposite sign, the final eigenvalue, $\la_{\nu-1}$, equals 4.

I remark that perturbation theory on the lattice has been considered by Actor {\it et al},
[\pref{ABR}].
\section{\bf 12. The vacuum energy}

A rather different technical eigenvalue problem is the calculation of the Casimir energy
and I present a quick treatment as a simple and explicit use of the eigenvalues.

The Dirichlet vacuum energy of a free scalar field on $T\times I_\nu$ can be evaluated in
closed form as,
  $$\eqalign{
  E_D&\equiv{1\over2}\sum_{\ol\la} \ol\la^{1/2}\cr
  &={1\over h}\sum_{n=1}^\nu \sin {\pi n\over2(\nu+1)}\cr
  &={1\over h}\bigg(\cot {\pi\over4(\nu+1)}-1\bigg)\cr
  &= {2L\over\pi h^2}-{1\over2h}-{\pi\over24L}+\ldots\,,\quad h\to0\,.
  }
  \eql{vacend}
  $$

If one views the lattice calculation as a regularisation of the continuum one, the
$-\pi/24L$ term is recognised as the value given by the \zf\ technique while the first
two, ultimately divergent terms, being non--universal, dependent on the regularisation,
should be discarded in some way, if one is concerned just with the interval, $[0,L]$ on its
own.

The paper [\pref{ABR}] contains a full discussion of the expression (\peq{vacend}) and
I will not enter into any more details. This reference also contains other arrangements,
including a discrete version of the Casimir piston.

The Neumann energy is, likewise,
  $$\eqalign{
  E_N&={1\over h}\sum_{n=1}^{\nu-1} \sin {\pi n\over2\nu}\cr
  &={1\over h}\big(\cot {\pi\over4\nu}-1\big)\cr
  &= {2L\over\pi h^2}-{2\over\pi h}-{1\over2h}-{\pi\over24L}+\ldots\,,\quad h\to0\,.
  }
  \eql{vacenn}
  $$

The other boundary condition  usually considered is the periodic one. This is given in Fort,
[\pref{Fort}], Chap.XV. It is convenient, this time, to arrange the $\nu$ points, $0\le
j\le\nu-1$ on the unit circle and impose the periodicity conditions $ y(\nu)=y(0)$,
$y(-1)=y(\nu-1)$ which relate values outside the proper range of $j$ to those inside.

The analysis is slightly different depending on whether $\nu$ is even, $\nu=2k+2$, or
odd, $\nu=2k+1$. In both cases there are degenerate modes, $\cos(2n\pi j/\nu)$ and
$\sin(2n\pi j/\nu)$, for $0\le n\le k$ with eigenvalues,
  $$
 \ol \la={4\over h^2}\sin^2{\pi n\over\nu}\,,
  $$
where the gap, $h=2\pi/\nu$. The value $n=0$ gives the one uniform zero mode. If
$\nu$ is even, the single mode $\cos \pi j$ must also be added. (This alternates between
plus and minus one as the points around the circle are traversed and corresponds to a
wave of infinite frequency in the continuum limit.) The total number of modes is always
$\nu$.

In exactly the same way as above, the vacuum energies are,
  $$\eqalign{
  E_{2k+2}&={2k+2\over\pi}\bigg(\sum_{n=1}^k \sin{\pi n\over 2k+2}+{1\over2}\bigg)\cr
  &={2k+2\over2\pi}\,\,\cot {\pi\over2(2k+2)}\cr
  E_{2k+1}&={2k+1\over\pi}\sum_{n=1}^k \sin{\pi n\over 2k+1}\cr
  &={2k+1\over2\pi}\,\,\cot {\pi\over2(2k+1)}
  }
  $$
or
  $$
  E_P={1\over h}\,\cot{h\over4}\to {4\over h^2}-{1\over12}+\ldots\,,
  \eql{vacenp}
  $$
in both cases, as expected. Again, one sees the continuum zeta value of
$\ze_R(-1/2)=-1/12$ appearing as $h$ tends to zero.\footnote { There is a puzzle here.
In the continuous case the periodic modes on a circle are the union of Dirichlet and
Neumann modes on an interval of size {\it half} the circumference. One might, therefore,
expect to see evidence of this, even in the discrete case, as $h\to0$. In fact this works
for the terms of order $h^{-2}$ and $h^0$ in (\peq{vacend}), (\peq{vacenn}) and
(\peq{vacenp}) but not for those of order $h^{-1}$. In order for it to work, the relevant
term in (\peq{vacenn}) should read just $1/2h$ to cancel that in (\peq{vacend}), on
addition, to give (\peq{vacenp}) but I could not achieve this.}

Fort, [\pref{Fort}], also discusses anti--periodic (real) functions. However I will be a little
more general and analyse a system that, in the continuous limit, amounts to an
Aharonov-Bohm flux running through the circle. This is mimicked by imposing a phase
change on circulating the flux and leaving the equations of motion unchanged.

In the quantum case, the wave function is complex and exponential functions are very
convenient.\footnote{ It is, of course, possible to retain a real description by doubling up
the fibre to an SO(2) one.} I therefore consider a function, $\psi(j)$, defined on the
points, $j$, and satisfying the twisted periodicity condition,
  $$
  \psi(\nu)=e^{2\pi i\al}\psi(0)\,,\quad\psi(\nu-1)=e^{2\pi i\al}\psi(-1)\,.
  \eql{twbc}
  $$

The modes on the discrete circle are,
    $$
   \psi^{\al}_n(j)=e^{2\pi i(n+\al)j /\nu}\,,\quad n=0,\ldots,\nu-1\,,\quad 0<\al\le1\,,
    $$
with corresponding eigenvalues ($h=2\pi/\nu$),
  $$
  \ol\la={4\over h^2}\sin^2{\pi (n+\al)\over\nu}\,,
  \eql{Peigen}
  $$
and vacuum energy (with a factor of two from the complexification),
   $$\eqalign{
  E(\al)&={2\over h}\sum_{n=0}^{\nu-1} \sin {\pi (n+\al)\over\nu}\cr
  &={2\over h}\,\cosec{h\over4}\,\,\cos {h\over4}(2\al-1)\cr
  &= {8\over h^2}-\big({1\over6}-\al+
  \al^2\big)+\ldots\,,\quad h\to0\,.
  }
  \eql{vacentw}
  $$
$E(\al)$ must be extended beyond $\al=1$ using periodicity, \ie $E(1+\al)=E(\al)$.

The constant term agrees with the result (the periodic Bernoulli polynomial, $\wt B_2$)
that arises in the continuous circle limit, [\pref{DandB}]. When $\al=0$ one regains
twice the real periodic value (\peq{vacenp}). It might be of interest to give the full
formal expansion,
  $$
  E(\al)=2\sum_{m=0}^\infty {(-1)^m\over (2m)!}\wt B_{2m}
  (\al)\bigg({h\over2}\bigg)^{2m-2}\,,
  $$
which I have not seen elsewhere.

\section{\bf 13. Direct determination of determinants}

It is helpful to have specific values for comparison or limit purposes and I proceed to
 evaluate the determinants of the free systems directly from the eigenvalues which have
just been used. I also look at the continuum limit  and some zeta regularised values.

An organisational point is perhaps required. In deriving the answers, I have used known
expressions for some infinite products (\eg\ Bromwich, [\pref{Bromwich}]), which could
be obtained, according to the results of the previous sections, {\it from} the
determinants and so these evaluations might be considered superfluous, at one level.
However I present them as independent checks. I also include the twisted periodic
values.

To repeat,  Dalembert's equation is,
  $$
  -{1\over h^2}\nabla\De y(j)+\ol\mu^2y(j)-\ol \la\,y(j)=0\,.
  \eql{dal}
  $$

For the interval, I will again use the real form of the eigenfunctions and the calculation is
immediate.

The D--determinant on the $L$--interval  using the eigenvalues (\peq{Deigen}) is,
  $$\eqalign{
  \Det_D(\ol\mu)&=\bigg({2\over h}\bigg)^{2\nu}\prod_{n=1}^\nu
  \bigg(\sin^2{\pi n\over2(\nu+1)}+{1\over4}\mu^2\bigg)\,,\quad \mu=h\ol\mu\cr
  &=\bigg({2\over h}\bigg)^{2\nu}\bigg[{\prod_{n=1}^{2\nu+1}}'
  \bigg(\sin^2{\pi n\over2(\nu+1)}+{1\over4}\mu^2\bigg)\bigg]^{1/2}\cr
  &={1\over h^{2\nu}}{\sinh(\nu+1)2\ga\over\sinh2\ga}\,.\cr
  }
  \eql{Ddet}
  $$
The dash on the second product means that the $n=\nu+1$ term is to be excluded and I
have set $\mu=2\sinh\ga$.

The massless values  are,
  $$\eqalign{
  \Det_D(0)&={1\over h^{2\nu}}\,(\nu+1)\cr
  & ={1\over h^{2\nu+1}}\,{1\over2}\, 2L\,.
  }
  $$

The N--determinant is, using (\peq{Neigen}),
  $$\eqalign{
  \Det_N(\ol\mu)&=\bigg({2\over h}\bigg)^{2\nu}\prod_{n=0}^{\nu-1}
  \bigg(\sin^2{\pi n\over2\nu}+{1\over4}\mu^2\bigg)\cr
  &=\bigg({2\over h}\bigg)^{2\nu}\mu^2\bigg[\prod_{n=1}^{2\nu-1'}
  \bigg(\sin^2{\pi n\over2\nu}+{1\over4}\mu^2\bigg)\bigg]^{1/2}\cr
  &={1\over h^{2\nu}}2\tanh\ga\sinh2\ga\nu\,.\cr
  }
  \eql{Ndet}
  $$
In the massless limit the determinant vanishes and it is conventional to remove the
offending zero mode giving the modified determinant,

  $$\eqalign{
 \Det'_N(0) &={1\over h^{2\nu-2}}\,\nu={1\over h^{2\nu-2}}{\nu\over L}{1\over2}\, 2L\cr
  &\to{1\over h^{2\nu-1}}{1\over2}\, 2L\,\,,\quad h\to 0\,.
  }
  $$
Before discussing these results, I give the twisted periodic expressions.

From (\peq{Peigen}) ($h=2\pi/\nu$),
  $$\eqalign{
  \Det^{1/2}_P(\al,\ol\mu)&=\bigg({2\over h}\bigg)^{\!\!2\nu}\,\prod_{n=0}^{\nu-1}
  \bigg(\sin^2{\pi (n+\al)\over\nu}+{1\over4}\mu^2\bigg)\cr
  &={2\over h^{2\nu}}\big(\cosh2\ga\nu-\cos 2\pi\al\big)\cr
  \Det^{1/2}_P(\al,0)&={1\over h^{2\nu}}\,4\,\sin^2 \pi\al\,,\quad0\le \al\le1\,,\cr
 \Det'_P(0,0) &={1\over h^{2\nu+2}}\,4\, (2\pi)^2\,,\quad \al=0\,.
  }
  \eql{Pdet}
  $$
For the $\al=0$ case, I have removed the complexification squaring and, for extra
generality, I have included the mass term.

For comparison, some determinants, computed from the bare \zf\ regularisation, are well
known to be,
  $$\eqalign{
 & \Det_{\ze D}(\ol\mu)=2{\sinh\ol\mu L\over\ol\mu}\cr
  &\Det'_{\ze N}(0)=2L\cr
  &\Det^{1/2}_{\ze P}(\al,0)=4\sin^2\pi\al\cr
  &\Det'_{\ze P}(0,0)=(2\pi)^2\,,
   }
  $$
and one sees that the lattice determinants are proportional to the zeta values, in the
continuous limit. In particular,
  $$\eqalign{
  \Det_{\!D}(\ol\mu)&={1\over h^{2\nu}}{\sinh(\nu+1)2\ga\over\sinh2\ga}\cr
  &\to {1\over h^{2\nu+1}}{\sinh \ol\mu L\over\ol \mu}\,,\quad h\to0\cr
  &={1\over h^{2\nu+1}}\,{1\over2}\,\Det_{\ze\,D}(\ol\mu)\,.
  }
  $$

Forman introduces  a twisted periodic condition, denoted $B_\de$ in [\pref{Forman}],
which has a complex multiplying factor $\de$ instead of $e^{2\pi i\al}$ in (\peq{twbc})
and Theorem 2.6 gives its determinant. Evaluating  [\pref{Forman}] equation, (2.23), in
the free case I find,
   $$
   \Det_{B_\de}=- {1\over h^{2\nu}}\, {\de(1-\de)^2\over(1+|\de|^2)}\,,
   $$
which vanishes when $\de=1$, as a check, but does not agree, apart from the lattice
scaling factor, with (\peq{Pdet}) when $\de=e^{2\pi i \al}$.\footnote{ Curiously, if one
of the $\de$s is replaced by $\de^{-1}$, then agreement is found, apart from a factor of
two, which seems too much of a coincidence.}

\section{\bf 14. Conclusion}

A good deal of this paper is expository but there are some novelties. It has been
emphasised that the Gel'fand--Yaglom formula for the determinant in the discrete case is
a standard component of Sturm--Liouville and orthogonal polynomial theory. I have
rewritten this in a neat $2\times2$ symplectic matrix formulation, slightly different from
the usual one, and have calculated the determinant for Robin boundary conditions for a
constant potential, as a basic example. The continuum limits have been discussed in a
simple--minded way and comparisons made with the work of Forman, [\pref{Forman}],
revealing some minor discrepancies. For Dirichlet conditions, the determinant for a $\de$
potential was evaluated exactly, highlighting the significance of Chebychev polynomials
as `unperturbed' Sturm--Liouville solutions which is further explored in the Appendices
and which contain some technical advances.

The calculations could be broadened to include the general Sturm--Liouville operator, and
higher order equations.

\section{\bf Added note}

The Green functions for free propagation on the discrete interval (or `path'), for various
boundary conditions, have been obtained in terms of Chebyshev polynomials by Bass,
[\pref{Bass}], Chung and Yau, [\pref{CandY}] and Bendito, Encinas and Carmona,
[\pref{BEC}].
\section{\bf Appendix 1. Chebyshev Polynomials}

I give some basic results for Chebyshev polynomials in a way that reflects the procedures
of this paper. These polynomials occur in the Chebyshev--Gauss scheme for mechanical
quadratures. Probably the most economical way of defining them is through the
recursion,
  $$
  P_{n+1}(x)-2xP_n(x)+P_{n-1}(x)=0\,,
  \eql{rec}
  $$
or
  $$
  \nabla\De P_n(x)=-(2-2x)P_n(x)\,,
  \eql{rec2}
  $$
the different kinds being selected by the `initial' values, $P_0$ and $P_1$, \eg\
[\pref{MandH}].

As a slight novelty, I use the matrix description adopted in the main body of this paper.
So, introducing the two--vector,
  $$
  \Pi_{n+1}(x)\equiv\left(\matrix{P_n(x)\cr P_{n+1}(x)}\right)\,,
  $$
the three--term recursion (\peq{rec}) becomes the two--term matrix one,
  $$
  \Pi_{n}(x)=C(x)\,\Pi_{n-1}(x)\,,
  $$
\ie
  $$
  \Pi_{n}(x)=C^n(x)\,\Pi_0(x)\,,
  $$
where\footnote { $C$ is what I call $M$ in the main body of this paper.}
  $$
  C(x)=\left(\matrix{0&1\cr-1&2x}\right)\,.
  \eql{cee}
  $$
This allows me to specify the initial conditions neatly. For example, the `Dirichlet' vector,
  $$
  \Pi^D_0(x)\equiv\left(\matrix{0\cr 1}\right)\,,
  $$
produces Chebyshev polynomials of the second kind, $P_n(x)=U_n(x)$ while the
Neumann one,
  $$
  \Pi^N_0(x)\equiv\left(\matrix{1\cr 1}\right)\,,
  $$
gives third kind polynomials,  $P_n(x)=V_n(x)$, [\pref{MandH}], [\pref{AAR}].

Many relations between the various kinds can be obtained by combining initial conditions.

The first iteration of the initial vector $ \left(\matrix{1\cr 0}\right)$ gives
$\left(\matrix{0\cr -1}\right)$ \ie minus the Dirichlet one and therefore yields
$-U_{n-1}$ after $n$ iterations. So, writing  $\left(\matrix{1\cr 1}\right)=
\left(\matrix{0\cr 1}\right)+ \left(\matrix{1\cr 0}\right)$, one obtains the relation
$V_n=U_n-U_{n-1}=\De U_{n-1}$. From which, for example, $\De V_n=\nabla\De
U_n$, helpfully relating Neumann and Dirichlet.

The Robin choice,
  $$
  \Pi^R_0(x)\equiv\left(\matrix{1\cr 1+\al}\right)\,,
  $$
likewise generates the D--N combination $V_n+\al U_n=(1+\al)U_n-U_{n-1}$.

The different polynomial combinations are encapsulated in the form of the power, $C^n$,
  $$
  C^n=\left(\matrix{-U_{n-2}&U_{n-1}\cr-U_{n-1}&U_n}\right)\,,
  \eql{iter}
  $$
($U_{-1}=0$) which can be written,
  $$
  C^n=  U_n\,A +U_{n-1}\,J +U_{n-2}\,A'\,,
  \eql{iter2}
  $$
with $A'\equiv A-{\bf1}$. From (\peq{iter}),  by taking the determinant, one finds the
Chebyshev identity,
  $$
  U_{n-1}^2-U_n\,U_{n-2}=1>0\,,
  $$
which is a (known) statement of a Tur\'{a}n inequality.

From the $\oZ$ group composition rule, $C^m\,C^n=C^{m+n}$,  the combination
relation,
  $$
  U_{m+n}=U_m\,U_n-U_{m-1}\,U_{n-1}\,,
  $$
can be deduced and the SU(2) character Clebsch--Gordan series,
  $$
   U_m\,U_n=\sum_{k=|m-n|}^{m+n}U_k\,,
  $$
readily follows therefrom by iteration. All these relations have trigonometric derivations.

The generating function can also be transcribed as follows. The matrix equation,
  $$\eqalign{
  ({\bf 1}-C\,t)^{-1}&=\left(\matrix{1&-t\cr t&1-2tx}\right)^{-1}\cr
  &={1\over 1-2tx+t^2}\left(\matrix{1-2tx&t\cr -t&1}\right)\cr
  &=\sum_{n=0}^\infty C^n \,t^n\,,
  }
  $$
and (\peq{iter}), yield the normal generating function for $U_n$. (Incidentally, to check
the top left entry, one has to use $U_{-2}=-1,\,U_{-1}=0$.) Other identities can be
deduced in a similar fashion.

\section{\bf Appendix 2. Calculation of matrix elements}

I give some details of the computation of the matrix element polynomial,
$P(\la)=\Up_{out}^{\dag}K(\la;\nu)\Up_{in}$, whose vanishing determines the
eigenvalues.

I write, in dyadic form,
  $$
  P(\la)=\Tr \big(K\,\Up_{in}\!\otimes\!\Up^{\dag}_{out}\big)\equiv
  \Tr \big(K\,Q\big)\,,
  \eql{polyn}
  $$
where the matrix $Q$ takes the specific forms in the D and N cases,
  $$
  Q_D=\left(\matrix{0&0\cr0&1}\right)=A\,,\quad
  Q_N=\left(\matrix{-1&1\cr-1&1}\right)\,.
  $$

This is a trace formula for the Gel'fand--Yaglom function. Other expressions are
determinant ones, [\pref{Atkinson}], p.95, and also, \eg, [\pref{Forman}],
[\pref{KandM}].

The general form of $K(\la;\nu)$ is a product of $\nu$ matrices, (\peq{prop}), but as a
first example I treat the case discussed in \S11 where all matrices are identical, except
one. I use the notation of the previous Appendix and write $K$ as,
  $$
  K(\la;\nu)=C^m(x)\,C(y)\,C^n(x)\,,\quad m+n+1=\nu\,,
  $$
where $C(x)$ and $C(y)$ are given by (\peq{cee}) with $2x=2-\la$ and $2y=v+2-\la$.
Then,
  $$\eqalign{
  K&=C^m(x)\,C(y)\,C^n(x)=C^\nu(x)+ C^m(x)\,\big(C(y)-C(x)\big)\,C^n(x)\cr
  &=C^\nu(x)+ v\,C^m(x)\,A\,C^n(x)
  }
  $$
and hence for (\peq{polyn}) in the case of D conditions,
  $$\eqalign{
  P_D(\la)&=\Tr\big(C^\nu(x)\,A)+v\Tr\big(C^m(x)\,A\,C^n(x)A\big)\cr
  &=U_\nu+v U_n\,U_m\cr
  &=U_\nu+v \big(U_{n+m}+U_{n-1}\,U_{m-1}\big)\cr
  &=U_\nu+v \big(U_{\nu-1}+U_{n-1}\,U_{\nu-n-2}\big)\,,
  }
  \eql{pert2}
  $$
where I have used (\peq{iter}). Setting $n=1$ gives the result (\peq{pert}).

Equation (\peq{pert2}) gives the first terms of a perturbation expansion, to elucidate the
general nature of which, in a direct way, I consider the case when there are two
distinguished matrices in the product (\peq{prop}), and a more systematic notation is
required.

The vertices of the interval have been labelled by $j$ which runs from $0$ to $\nu+1$
with the vertices $1$ to $\nu$ being dynamic.\footnote{ In graph theory language these
would be internal vertices.} In this range, let the vertices $j_1$ and $j_2$ be singled out
to correspond to matrices $C(y_{j_1})$ and $C(y_{j_2})$. Then, taking $j_1>j_2$, and
noting that $C(y_{j_i})=C+v_{j_i}\,A$, with $C\equiv C(x)$, I find,
  $$\eqalign{
  K(\la;\nu)A&=C^{\nu-j_1}\,C(y_{j_1})\,C^{j_1-j_2-1}C(y_{j_2})\,C^{j_2-1}A\cr
  &=C^{\nu-j_1}\,(C+v_{j_1}A)\,C^{j_1-j_2-1}(C+v_{j_2})\,C^{j_2-1}A\cr
  &=C^\nu\,A+v_{j_1}\,C^{\nu-j_1}\,A\,C^{j_1-1}\,A
  +v_{j_2}\,C^{\nu-j_2}\,A\,C^{j_2-1}A\cr
  &\hspace{****}+ v_{j_1}v_{j_2}\,C^{\nu-j_1}
  \,A\,C^{j_1-j_2-1}\,A\,C^{j_2-1}A\,.
  }
  $$

I have added the post factor of $A$ to bring out the fact that it is the combination
$C^n\,A$ that enters, and this takes the form,
  $$
  C^n\,A=\left(\matrix{0&U_{n-1}\cr0& U_n}\right)\,,
  $$
which is preserved under multiplication,
  $$
  C^n\,A\,C^m A=\left(\matrix{0&U_m\,U_{n-1}\cr0&U_m\, U_n}\right)\,.
  $$

The final act of taking the trace picks out just the lower right corner term and so,
  $$\eqalign{
  \Tr\big(K(\la;\nu)A\big)&=U_\nu+U_{\nu-j_1}\,v_{j_1}\,U_{j_1-1}
  +U_{\nu-j_2}\,v_{j_2}\,U_{j_2-1}\cr
  &\hspace{**********}+\,U_{\nu-j_1}\, v_{j_1}
  \,U_{j_1-j_2-1}\,v_{j_2}\,U_{j_2-1}\,.
  }
  \eql{pert3}
  $$

The structure when more vertices are marked is clear and the general expression is,
  $$\eqalign{
  \Tr\big(K(\la;\nu)A\big)&=U_\nu+\sum_{j_1=1}^\nu U_{\nu-j_1}\,v_{j_1}\,U_{j_1-1}
  +\sum_{j_1>j_2=1}^\nu U_{\nu-j_1}\, v_{j_1}
  \,U_{j_1-j_2-1}\,v_{j_2}\,U_{j_2-1}\cr
  &+\sum_{j_1>j_2>j_3=1}^\nu U_{\nu-j_1}\, v_{j_1}\,U_{j_1-j_2-1}
  \,v_{j_2}\,U_{j_2-j_3-1}\,v_{j_3}\,U_{j_3-1}\cr
  &+\ldots\cr
  &+ v_1 v_2\ldots v_{\nu}\,.
  }
  \eql{pert4}
  $$

This equation also follows (equivalently) from the complete iteration of the
conventional--looking  (\peq{kprop}), which I rewrite here, in the notation of these
Appendices,
  $$
  K(\la;j)=C^j(x)+\sum_{j_1=1}^\nu C^{j-j_1}(x)\,A\,v_{j_1}\,K(\la;j_1-1,0)\,,
  \quad 2x=2-\la.
  \eql{sumeq}
  $$
This can be given the usual propagation interpretation of a perturbation series (although
finite and exact) with the Chebyshev polynomials acting as free propagators. An obvious
graphical representation can be set up.

Setting $\la$ to zero in (\peq{pert4}) gives the determinant,
  $$\eqalign{
  \Det_D&=\nu+1+\sum_{j_1=1}^\nu (\nu-j_1+1)\,j_1\,v_{j_1}
  +\sum_{j_1=1}^\nu \sum_{j_2=1}^{j_1}(\nu-j_1+1)\,
  \,(j_1-j_2)\,j_2\,\, v_{j_1}\,v_{j_2}\cr
  &+\sum_{j_1=1}^\nu\sum_{j_2=1}^{j_1}\sum_{j_3=1}^{j_2}
  (\nu-j_1+1)\,(j_1-j_2)
  \,(j_2-j_3)\,j_3\,\, v_{j_1}\,v_{j_2}\,v_{j_3}\cr
  &+\ldots\cr
  &+ v_1 v_2\ldots v_{\nu}\,.
  }
  \eql{pert5}
  $$
The upper limits have been extended to the diagonal values using the vanishing of the
summands there, but I make no use of this at the present time.

As a numerical illustration, the characteristic polynomial when $\nu=3$ is,
  $$
   P(\la)=-\la^3+\la^2(6+S_1)-\la\big(10+4S_1+S_2\big)+4+v_2+3S_1+2S_2+S_3\,,
  $$
where
  $$
  S_1=v_1+v_2+v_3\,,\quad S_2=v_1v_2+v_1v_3+v_2v_3\,,\quad S_3=v_1v_2v_3\,.
  $$

It is interesting to note that, for a symmetric potential ($v_1=v_3$), $P(\la)$ has the
linear factor, $(\la-v_1-2)$, which is related, presumably, to Borg's reconstruction
theorem.

Neumann boundary conditions can be handled in a like manner. The required polynomial
turns out to be
  $$\eqalign{
  \Tr\big(K(\la;\nu)\,Q_N\big)&=\De V_{\nu-1}
  + \sum_{j_1=1}^\nu V_{\nu-j_1}\,v_{j_1}\,V_{j_1-1}\cr
  &+\sum_{j_1>j_2=1}^\nu V_{\nu-j_1}\, v_{j_1}
  \,U_{j_1-j_2-1}\,v_{j_2}\,V_{j_2-1}\cr
  &+\sum_{j_1>j_2>j_3=1}^\nu V_{\nu-j_1}\, v_{j_1}\,U_{j_1-j_2-1}
  \,v_{j_2}\,U_{j_2-j_3-1}\,v_{j_3}\,V_{j_3-1}\cr
  &+\ldots\ldots+ (v_1 v_2\ldots v_{\nu})\,.
  }
  \eql{pert6}
  $$
The end point propagators are third kind polynomials, $V_n$, while internal ones are
second kind, $U_n$. The determinant is
  $$\eqalign{
  \Det_N=& \sum_{j_1=1}^\nu \,v_{j_1}
  +\sum_{j_1>j_2=1}^\nu \, v_{j_1}
  \,(j_1-j_2)\,v_{j_2}\cr
  &+\sum_{j_1>j_2>j_3=1}^\nu  v_{j_1}\,(j_1-j_2)
  \,v_{j_2}\,(j_2-j_3)\,v_{j_3}
  +\ldots\ldots+ (v_1 v_2\ldots v_{\nu})\,,
  }
  \eql{pert7}
  $$
since $V_n(x)$ is unity when $\la=0$. $\Det_N$ correctly vanishes when all the $v_j$
do due to the resulting zero mode.

These expressions  can be used to discuss a Borg--Levinson inverse theorem (\eg\ Hald,
[\pref{Hald}]) and Dikii trace identities, but such further analysis and manipulations
must be postponed.

 \vglue 20truept

\noin{\bf References.} \vskip5truept

\begin{putreferences}
   \ref{Bass}{Bass,R. \jmp{26}{1985}{3068}.}
   \ref{GandK}{Gantmacher,F.R. and Krein,M.G. {\it Oszillazionsmatrizen}
   (Akad.-Verlag.Berlin, 1960).}
   \ref{Hald}{Hald,O.H. {\it Numer.Math.} {\bf27} (1977) 249.}
   \ref{CandY}{Chung,F. and Yau S.--T. {\it J.Comb.Theory} {\bf A91} (2000) 191.}
   \ref{BEC}{Bendito,E, Encinas,A.M. and Carmona,A. {\it Appl.Anal.Disc.Math.}{\bf 3}
   (2009) 282.}
   \ref{AandA}{Annaby,M.H. and Asharabi,R.M. {\it Acta.Math.Scientia} {\bf 31B} (2011)408.}
   \ref{MandH}{Mason,J.C. and Handscomb,D.C. {\it Chebyshev Polynomials} (Chapman and Hall,
   Boca Raton, 2002).}
   \ref{LandB}{Levy,H. and Baggott,E.A. \phm{18}{1934}{177}.}
   \ref{BlandM}{Bleich,Fr. and Melan,E. {\it Die Gew\"onlichen und Partiellen
   Differenzengleichungen der Baustatik}, (Springer, Berlin, 1927).}
   \ref{Spiegel}{Spiegel,M.R. {\it Schaum's Outline of Calculus of Finite Differences},
   (McGraw-Hill, New York, 1971).}
   \ref{Porter}{Porter,M.B. \aom{3}{1901}{55}.}
    \ref{Dunne}{Dunne,G. \jpa{41}{2008}{304006}.}
    \ref{MandF}{Morse,P.M. and  Feshbach,H. {\it Methods of Theoretical Physics},
    (McGraw--Hill, New York, 1953).}
    \ref{KandM}{Kirsten, K. and McKane,A. \aop{308}{2003}{502}.}
    \ref{LandS}{Levit,S. and Smilansky,U. \pams{65}{1977}{299}.}
    \ref{GeandY}{Gel'fand, I.M. and Yaglom,A.M. \jmp {1}{1960}{48}.}
    \ref{ABR}{Actor,A., Bender,C. and Reingruber,J., {\it Fortschr.Phys.} {\bf48} (2000) 4.}
    \ref{Rayleigh}{Rayleigh, Lord, {\it Theory of Sound}, (MacMillan, London, 1894).}
    \ref{Floratos}{Floratos,E.G. \plb{228}{1989}{335}.}
    \ref{deverdiere}{de Verdi\`ere, C. {\it Ann. Inst. Fourier} {\bf 49} (1999) 861.}
    \ref{Goldberg}{Goldberg,S. {\it Introduction to Difference Equations},
    (Wiley, New York, 1958).}
    \ref{Jordan}{Jordan,C. {\it Calculus of Finite Differences}, (Budapest, 1939).}
    \ref{Fort}{Fort,T. {\it Finite Differences}, (Clarendon Press, Oxford,1948). }
    \ref{Atkinson}{Atkinson,F.V. {\it Discrete and Continuous Boundary Problems},
    (Academic Press, New York,1964).}
   \ref{Elaydi}{Elaydi,S.N. {\it An Introduction to Difference Equations}, (Springer, New York,
   1999.}
   \ref{Forman}{Forman,R. \cmp{147}{1992}{485}.}
    \ref{Boole}{Boole, G. {\it Calculus of Finite Differences}, (MacMillan, Cambridge,
1860).}
     \ref{Cayley5}{Cayley,A.  {\it Phil. Trans. Roy. Soc. Lond.} {\bf 148} (1858) 47.}
     \ref{Milne-Thomson}{Milne-Thomson, L.M. {\it The Calculus of Finite Differences},
     (MacMillan, London, 1933).}
    \ref{Herschel}{Herschel, J.F.W.  {\it Phil. Trans. Roy. Soc. Lond.} {\bf 106} (1816) 25.}
    \ref{Littlewood2}{Littlewood,D.E. {\it The Theory of Group Characters}
    (Clarendon Press, Oxford, 1950).}
    \ref{Wright}{Wright,E.M. \amm{68}{1961}{144}.}
\ref{Carlitz}{Carlitz,L. \dmj{27}{1960}{401}.}
     \ref{Netto}{Netto,E. {\it Lehrbuch der Combinatorik} 2nd Edn. (Teubner, Leipzig, 1927).}
    \ref{FdeB}{Fa\`{a} de Bruno, F. {\it Th\'eorie des Formes Binaires} (Brero, Turin,1876).}
    \ref{Ehrhart}{Ehrhart,E. \jram{227}{25}{1967}.}
    \ref{Bell}{Bell,E.T.\ajm{65}{1943}{382}.}
    \ref{BandR}{Beck,M. and Robins,S. {\it Computing the Continuous Discretely,}
    (Springer, New York, 2007).}
    \ref{BandR2}{Beck, M. and Robins,S. {\it Discrete and Comp. Geom.} {\bf 27}(2002) 443.}
    \ref{Harmer}{Harmer,M. {\it J.Australian Math.Soc.} {\bf 84}(2008)217.}
    \ref{RandF}{Rubinstein, B.Y. and Fel,L.G., {\it Ramanujan J.} {\bf11}(2006)331.}
    \ref{BGK}{Beck,M., Gessel, I.M. and Komatsu,T. {\it Electronic Journal of Combinatorics}
    {\bf8}(2001) 1.}
    \ref{Sylvester}{Sylvester,J.J. \qjpam{1}{1858}{81}.}
    \ref{Sylvester2}{Sylvester,J.J. \qjpam{1}{1858}{142}.}
    \ref{Sylvester3}{Sylvester,J.J. \ajm{5}{1882}{79}.}
    \ref{Sylvester4}{Sylvester,J.J. \plms{28}{1896}{33}.}
    \ref{Dowgta}{J.S.Dowker, {\it Group theory aspects of spectral problems on spherical factors},
   ArXiv.Math.DG: 0907.1309.}
    \ref{BDR}{Beck,M., Diaz and Robins,S. {\it J.Numb.Theory} {\bf 96} (2002) 1.}
    \ref{PandS}{P\'{o}lya, G. and Szeg\H{o},G. {\it Aufgaben und Lehrs\"atze aus der Analysis}
    (Springer--Verlag, Berlin, 1925).}
    \ref{EOS}{Elizalde,E., Odintsov, S.D. and Saharian, A.A. \prD{79}{2009}{065023}.}
    \ref{Cavalcanti}{Cavalcanti,R.M. \prD{69}{2004}{065015}.}
    \ref{MWK}{Milton, K.A., Wagner,J. and Kirsten,K. \prD{80}{2009}{125028}.}
    \ref{EBM2}{Ellingsen,S.A., Brevik,I. and Milton,K.A. \prE{81}{2010}{065031}.}
    \ref{EBM}{Ellingsen,S.A., Brevik,I. and Milton,K.A. \prE{80}{2009}{021125}.}
    \ref{BEM}{Brevik,I., Ellingsen,S.A. and Milton,K.A. \prE{79}{2009}{041120}.}
    \ref{FKW}{Fulling,S.A, Kaplan L. and Wilson,J.H. \prA{76}{2007}{012118}.}
    \ref{Lukosz}{Lukosz,W, {\it Physica} {\bf 56} (1971) 109; \zfp{258}{1973}{99}
    ;\zfp{262}{1973}{327}.}
    \ref{Gromes}{Gromes, D. \mz{94}{1966}{110}.}
    \ref{FandK1}{Kirsten,K. and Fulling,S.A. \prD{79}{2009}{065019} .}
    \ref{FandK2}{Fucci,G. and Kirsten,K, JHEP (2011), 1103:016.}
    \ref{dowgjms}{Dowker,J.S. {\it Determinants and conformal anomalies
    of GJMS operators on spheres}, ArXiv: 1007.3865.}
    \ref{Dowcascone}{dowker,J.S. \prD{36}{1987}{3095}.}
    \ref{Dowcos}{dowker,J.S. \prD{36}{1987}{3742}.}
    \ref{Dowtherm}{Dowker,J.S. \prD{18}{1978}{1856}.}
    \ref{Dowgeo}{Dowker,J.S. \cqg{11}{1994}{L55}.}
    \ref{ApandD2}{Dowker,J.S. and Apps,J.S. \cqg{12}{1995}{1363}.}
   \ref{HandW}{Hertzberg,M.P. and Wilczek,F. {\it Some calculable contributions to
   Entanglement Entropy}, ArXiv:1007.0993.}
   \ref{KandB}{Kamela,M. and Burgess,C.P. \cjp{77}{1999}{85}.}
   \ref{Dowhyp}{Dowker,J.S. \jpa{43}{2010}{445402}; ArXiv:1007.3865.}
   \ref{LNST}{Lohmayer,R., Neuberger,H, Schwimmer,A. and Theisen,S.
   \plb{685}{2010}{222}.}
   \ref{Allen2}{Allen,B. PhD Thesis, University of Cambridge, 1984.}
   \ref{MyandS}{Myers,R.C. and Sinha,A. {\it Seeing a c-theorem with
   holography}, ArXiv:1006.1263}
   \ref{MyandS2}{Myers,R.C. and Sinha,A. {\it Holographic c-theorems in
   arbitrary dimensions},\break ArXiv: 1011.5819.}
   \ref{RyandT}{Ryu,S. and Takayanagi,T. JHEP {\bf 0608}(2006)045.}
   \ref{CaandH}{Casini,H. and Huerta,M. {\it Entanglement entropy
   for the n--sphere},\break arXiv:1007.1813.}
   \ref{CaandH3}{Casini,H. and Huerta,M. \jpa {42}{2009}{504007}.}
   \ref{Solodukhin}{Solodukhin,S.N. \plb{665}{2008}{305}.}
   \ref{Solodukhin2}{Solodukhin,S.N. \plb{693}{2010}{605}.}
   \ref{CaandW}{Callan,C.G. and Wilczek,F. \plb{333}{1994}{55}.}
   \ref{FandS1}{Fursaev,D.V. and Solodukhin,S.N. \plb{365}{1996}{51}.}
   \ref{FandS2}{Fursaev,D.V. and Solodukhin,S.N. \prD{52}{1995}{2133}.}
   \ref{Fursaev}{Fursaev,D.V. \plb{334}{1994}{53}.}
   \ref{Donnelly2}{Donnelly,H. \ma{224}{1976}{161}.}
   \ref{ApandD}{Apps,J.S. and Dowker,J.S. \cqg{15}{1998}{1121}.}
   \ref{FandM}{Fursaev,D.V. and Miele,G. \prD{49}{1994}{987}.}
   \ref{Dowker2}{Dowker,J.S.\cqg{11}{1994}{L137}.}
   \ref{Dowker1}{Dowker,J.S.\prD{50}{1994}{6369}.}
   \ref{FNT}{Fujita,M.,Nishioka,T. and Takayanagi,T. JHEP {\bf 0809}
   (2008) 016.}
   \ref{Hund}{Hund,F. \zfp{51}{1928}{1}.}
   \ref{Elert}{Elert,W. \zfp {51}{1928}{8}.}
   \ref{Poole2}{Poole,E.G.C. \qjm{3}{1932}{183}.}
   \ref{Bellon}{Bellon,M.P. {\it On the icosahedron: from two to three
   dimensions}, arXiv:0705.3241.}
   \ref{Bellon2}{Bellon,M.P. \cqg{23}{2006}{7029}.}
   \ref{McLellan}{McLellan,A,G. \jpc{7}{1974}{3326}.}
   \ref{Boiteaux}{Boiteaux, M. \jmp{23}{1982}{1311}.}
   \ref{HHandK}{Hage Hassan,M. and Kibler,M. {\it On Hurwitz
   transformations} in {Le probl\`eme de factorisation de Hurwitz}, Eds.,
   A.Ronveaux and D.Lambert (Fac.Univ.N.D. de la Paix, Namur, 1991),
   pp.1-29.}
   \ref{Weeks2}{Weeks,Jeffrey \cqg{23}{2006}{6971}.}
   \ref{LandW}{Lachi\`eze-Rey,M. and Weeks,Jeffrey, {\it Orbifold construction of
   the modes on the Poincar\'e dodecahedral space}, arXiv:0801.4232.}
   \ref{Cayley4}{Cayley,A. \qjpam{58}{1879}{280}.}
   \ref{JMS}{Jari\'c,M.V., Michel,L. and Sharp,R.T. {\it J.Physique}
   {\bf 45} (1984) 1. }
   \ref{AandB}{Altmann,S.L. and Bradley,C.J.  {\it Phil. Trans. Roy. Soc. Lond.}
   {\bf 255} (1963) 199.}
   \ref{CandP}{Cummins,C.J. and Patera,J. \jmp{29}{1988}{1736}.}
   \ref{Sloane}{Sloane,N.J.A. \amm{84}{1977}{82}.}
   \ref{Gordan2}{Gordan,P. \ma{12}{1877}{147}.}
   \ref{DandSh}{Desmier,P.E. and Sharp,R.T. \jmp{20}{1979}{74}.}
   \ref{Kramer}{Kramer,P., \jpa{38}{2005}{3517}.}
   \ref{Klein2}{Klein, F.\ma{9}{1875}{183}.}
   \ref{Hodgkinson}{Hodgkinson,J. \jlms{10}{1935}{221}.}
   \ref{ZandD}{Zheng,Y. and Doerschuk, P.C. {\it Acta Cryst.} {\bf A52}
   (1996) 221.}
   \ref{EPM}{Elcoro,L., Perez--Mato,J.M. and Madariaga,G.
   {\it Acta Cryst.} {\bf A50} (1994) 182.}
    \ref{PSW2}{Prandl,W., Schiebel,P. and Wulf,K.
   {\it Acta Cryst.} {\bf A52} (1999) 171.}
    \ref{FCD}{Fan,P--D., Chen,J--Q. and Draayer,J.P.
   {\it Acta Cryst.} {\bf A55} (1999) 871.}
   \ref{FCD2}{Fan,P--D., Chen,J--Q. and Draayer,J.P.
   {\it Acta Cryst.} {\bf A55} (1999) 1049.}
   \ref{Honl}{H\"onl,H. \zfp{89}{1934}{244}.}
   \ref{PSW}{Patera,J., Sharp,R.T. and Winternitz,P. \jmp{19}{1978}{2362}.}
   \ref{LandH}{Lohe,M.A. and Hurst,C.A. \jmp{12}{1971}{1882}.}
   \ref{RandSA}{Ronveaux,A. and Saint-Aubin,Y. \jmp{24}{1983}{1037}.}
   \ref{JandDeV}{Jonker,J.E. and De Vries,E. \npa{105}{1967}{621}.}
   \ref{Rowe}{Rowe, E.G.Peter. \jmp{19}{1978}{1962}.}
   \ref{KNR}{Kibler,M., N\'egadi,T. and Ronveaux,A. {\it The Kustaanheimo-Stiefel
   transformation and certain special functions} \lnm{1171}{1985}{497}.}
   \ref{GLP}{Gilkey,P.B., Leahy,J.V. and Park,J-H, \jpa{29}{1996}{5645}.}
   \ref{Kohler}{K\"ohler,K.: Equivariant Reidemeister torsion on
   symmetric spaces. Math.Ann. {\bf 307}, 57-69 (1997)}
   \ref{Kohler2}{K\"ohler,K.: Equivariant analytic torsion on ${\bf P^nC}$.
   Math.Ann.{\bf 297}, 553-565 (1993) }
   \ref{Kohler3}{K\"ohler,K.: Holomorphic analytic torsion on Hermitian
   symmetric spaces. J.Reine Angew.Math. {\bf 460}, 93-116 (1995)}
   \ref{Zagierzf}{Zagier,D. {\it Zetafunktionen und Quadratische
   K\"orper}, (Springer--Verlag, Berlin, 1981).}
   \ref{Stek}{Stekholschkik,R. {\it Notes on Coxeter transformations and the McKay
   correspondence.} (Springer, Berlin, 2008).}
   \ref{Pesce}{Pesce,H. \cmh {71}{1996}{243}.}
   \ref{Pesce2}{Pesce,H. {\it Contemp. Math} {\bf 173} (1994) 231.}
   \ref{Sutton}{Sutton,C.J. {\it Equivariant isospectrality
   and isospectral deformations on spherical orbifolds}, ArXiv:math/0608567.}
   \ref{Sunada}{Sunada,T. \aom{121}{1985}{169}.}
   \ref{GoandM}{Gornet,R, and McGowan,J. {\it J.Comp. and Math.}
   {\bf 9} (2006) 270.}
   \ref{Suter}{Suter,R. {\it Manusc.Math.} {\bf 122} (2007) 1-21.}
   \ref{Lomont}{Lomont,J.S. {\it Applications of finite groups} (Academic
   Press, New York, 1959).}
   \ref{DandCh2}{Dowker,J.S. and Chang,Peter {\it Analytic torsion on
   spherical factors and tessellations}, arXiv:math.DG/0904.0744 .}
   \ref{Mackey}{Mackey,G. {\it Induced representations}
   (Benjamin, New York, 1968).}
   \ref{Koca}{Koca, {\it Turkish J.Physics}.}
   \ref{Brylinski}{Brylinski, J-L., {\it A correspondence dual to McKay's}
    ArXiv alg-geom/9612003.}
   \ref{Rossmann}{Rossman,W. {\it McKay's correspondence
   and characters of finite subgroups of\break SU(2)} {\it Progress in Math.}
      Birkhauser  (to appear) .}
   \ref{JandL}{James, G. and Liebeck, M. {\it Representations and
   characters of groups} (CUP, Cambridge, 2001).}
   \ref{IandR}{Ito,Y. and Reid,M. {\it The Mckay correspondence for finite
   subgroups of SL(3,C)} Higher dimensional varieties, (Trento 1994),
   221-240, (Berlin, de Gruyter 1996).}
   \ref{BandF}{Bauer,W. and Furutani, K. {\it J.Geom. and Phys.} {\bf
   58} (2008) 64.}
   \ref{Luck}{L\"uck,W. \jdg{37}{1993}{263}.}
   \ref{LandR}{Lott,J. and Rothenberg,M. \jdg{34}{1991}{431}.}
   \ref{DoandKi} {Dowker.J.S. and Kirsten, K. {\it Analysis and Appl.}
   {\bf 3} (2005) 45.}
   \ref{dowtess1}{Dowker,J.S. \cqg{23}{2006}{1}.}
   \ref{dowtess2}{Dowker,J.S. {\it J.Geom. and Phys.} {\bf 57} (2007) 1505.}
   \ref{MHS}{De Melo,T., Hartmann,L. and Spreafico,M. {\it Reidemeister
   Torsion and analytic torsion of discs}, ArXiv:0811.3196.}
   \ref{Vertman}{Vertman, B. {\it Analytic Torsion of a  bounded
   generalized cone}, ArXiv:0808.0449.}
   \ref{WandY} {Weng,L. and You,Y., {\it Int.J. of Math.}{\bf 7} (1996)
   109.}
   \ref{ScandT}{Schwartz, A.S. and Tyupkin,Yu.S. \np{242}{1984}{436}.}
   \ref{AAR}{Andrews, G.E., Askey,R. and Roy,R. {\it Special functions}
   (CUP, Cambridge, 1999).}
   \ref{Tsuchiya}{Tsuchiya, N.: R-torsion and analytic torsion for spherical
   Clifford-Klein manifolds.: J. Fac.Sci., Tokyo Univ. Sect.1 A, Math.
   {\bf 23}, 289-295 (1976).}
   \ref{Tsuchiya2}{Tsuchiya, N. J. Fac.Sci., Tokyo Univ. Sect.1 A, Math.
   {\bf 23}, 289-295 (1976).}
  \ref{Lerch}{Lerch,M. \am{11}{1887}{19}.}
  \ref{Lerch2}{Lerch,M. \am{29}{1905}{333}.}
  \ref{TandS}{Threlfall, W. and Seifert, H. \ma{104}{1930}{1}.}
  \ref{RandS}{Ray, D.B., and Singer, I. \aim{7}{1971}{145}.}
  \ref{RandS2}{Ray, D.B., and Singer, I. {\it Proc.Symp.Pure Math.}
  {\bf 23} (1973) 167.}
  \ref{Jensen}{Jensen,J.L.W.V. \aom{17}{1915-1916}{124}.}
  \ref{Rosenberg}{Rosenberg, S. {\it The Laplacian on a Riemannian Manifold}
  (CUP, Cambridge, 1997).}
  \ref{Nando2}{Nash, C. and O'Connor, D-J. {\it Int.J.Mod.Phys.}
  {\bf A10} (1995) 1779.}
  \ref{Fock}{Fock,V. \zfp{98}{1935}{145}.}
  \ref{Levy}{Levy,M. \prs {204}{1950}{145}.}
  \ref{Schwinger2}{Schwinger,J. \jmp{5}{1964}{1606}.}
  \ref{Muller}{M\"uller, \lnm{}{}{}.}
  \ref{VMK}{Varshalovich.}
  \ref{DandWo}{Dowker,J.S. and Wolski, A. \prA{46}{1992}{6417}.}
  \ref{Zeitlin1}{Zeitlin,V. {\it Physica D} {\bf 49} (1991).  }
  \ref{Zeitlin0}{Zeitlin,V. {\it Nonlinear World} Ed by
   V.Baryakhtar {\it et al},  Vol.I p.717,  (World Scientific, Singapore, 1989).}
  \ref{Zeitlin2}{Zeitlin,V. \prl{93}{2004}{264501}. }
  \ref{Zeitlin3}{Zeitlin,V. \pla{339}{2005}{316}. }
  \ref{Groenewold}{Groenewold, H.J. {\it Physica} {\bf 12} (1946) 405.}
  \ref{Cohen}{Cohen, L. \jmp{7}{1966}{781}.}
  \ref{AandW}{Argawal G.S. and Wolf, E. \prD{2}{1970}{2161,2187,2206}.}
  \ref{Jantzen}{Jantzen,R.T. \jmp{19}{1978}{1163}.}
  \ref{Moses2}{Moses,H.E. \aop{42}{1967}{343}.}
  \ref{Carmeli}{Carmeli,M. \jmp{9}{1968}{1987}.}
  \ref{SHS}{Siemans,M., Hancock,J. and Siminovitch,D. {\it Solid State
  Nuclear Magnetic Resonance} {\bf 31}(2007)35.}
 \ref{Dowk}{Dowker,J.S. \prD{28}{1983}{3013}.}
 \ref{Heine}{Heine, E. {\it Handbuch der Kugelfunctionen}
  (G.Reimer, Berlin. 1878, 1881).}
  \ref{Pockels}{Pockels, F. {\it \"Uber die Differentialgleichung $\De
  u+k^2u=0$} (Teubner, Leipzig. 1891).}
  \ref{Hamermesh}{Hamermesh, M., {\it Group Theory} (Addison--Wesley,
  Reading. 1962).}
  \ref{Racah}{Racah, G. {\it Group Theory and Spectroscopy}
  (Princeton Lecture Notes, 1951). }
  \ref{Gourdin}{Gourdin, M. {\it Basics of Lie Groups} (Editions
  Fronti\'eres, Gif sur Yvette. 1982.)}
  \ref{Clifford}{Clifford, W.K. \plms{2}{1866}{116}.}
  \ref{Story2}{Story, W.E. \plms{23}{1892}{265}.}
  \ref{Story}{Story, W.E. \ma{41}{1893}{469}.}
  \ref{Poole}{Poole, E.G.C. \plms{33}{1932}{435}.}
  \ref{Dickson}{Dickson, L.E. {\it Algebraic Invariants} (Wiley, N.Y.
  1915).}
  \ref{Dickson2}{Dickson, L.E. {\it Modern Algebraic Theories}
  (Sanborn and Co., Boston. 1926).}
  \ref{Hilbert2}{Hilbert, D. {\it Theory of algebraic invariants} (C.U.P.,
  Cambridge. 1993).}
  \ref{Olver}{Olver, P.J. {\it Classical Invariant Theory} (C.U.P., Cambridge.
  1999.)}
  \ref{AST}{A\v{s}erova, R.M., Smirnov, J.F. and Tolsto\v{i}, V.N. {\it
  Teoret. Mat. Fyz.} {\bf 8} (1971) 255.}
  \ref{AandS}{A\v{s}erova, R.M., Smirnov, J.F. \np{4}{1968}{399}.}
  \ref{Shapiro}{Shapiro, J. \jmp{6}{1965}{1680}.}
  \ref{Shapiro2}{Shapiro, J.Y. \jmp{14}{1973}{1262}.}
  \ref{NandS}{Noz, M.E. and Shapiro, J.Y. \np{51}{1973}{309}.}
  \ref{Cayley2}{Cayley, A. {\it Phil. Trans. Roy. Soc. Lond.}
  {\bf 144} (1854) 244.}
  \ref{Cayley3}{Cayley, A. {\it Phil. Trans. Roy. Soc. Lond.}
  {\bf 146} (1856) 101.}
  \ref{Wigner}{Wigner, E.P. {\it Gruppentheorie} (Vieweg, Braunschweig. 1931).}
  \ref{Sharp}{Sharp, R.T. \ajop{28}{1960}{116}.}
  \ref{Laporte}{Laporte, O. {\it Z. f. Naturf.} {\bf 3a} (1948) 447.}
  \ref{Lowdin}{L\"owdin, P-O. \rmp{36}{1964}{966}.}
  \ref{Ansari}{Ansari, S.M.R. {\it Fort. d. Phys.} {\bf 15} (1967) 707.}
  \ref{SSJR}{Samal, P.K., Saha, R., Jain, P. and Ralston, J.P. {\it
  Testing Isotropy of Cosmic Microwave Background Radiation},
  astro-ph/0708.2816.}
  \ref{Lachieze}{Lachi\'eze-Rey, M. {\it Harmonic projection and
  multipole Vectors}. astro- \break ph/0409081.}
  \ref{CHS}{Copi, C.J., Huterer, D. and Starkman, G.D.
  \prD{70}{2003}{043515}.}
  \ref{Jaric}{Jari\'c, J.P. {\it Int. J. Eng. Sci.} {\bf 41} (2003) 2123.}
  \ref{RandD}{Roche, J.A. and Dowker, J.S. \jpa{1}{1968}{527}.}
  \ref{KandW}{Katz, G. and Weeks, J.R. \prD{70}{2004}{063527}.}
  \ref{Waerden}{van der Waerden, B.L. {\it Die Gruppen-theoretische
  Methode in der Quantenmechanik} (Springer, Berlin. 1932).}
  \ref{EMOT}{Erdelyi, A., Magnus, W., Oberhettinger, F. and Tricomi, F.G. {
  \it Higher Transcendental Functions} Vol.1 (McGraw-Hill, N.Y. 1953).}
  \ref{Dowzilch}{Dowker, J.S. {\it Proc. Phys. Soc.} {\bf 91} (1967) 28.}
  \ref{DandD}{Dowker, J.S. and Dowker, Y.P. {\it Proc. Phys. Soc.}
  {\bf 87} (1966) 65.}
  \ref{DandD2}{Dowker, J.S. and Dowker, Y.P. \prs{}{}{}.}
  \ref{Dowk3}{Dowker,J.S. \cqg{7}{1990}{1241}.}
  \ref{Dowk5}{Dowker,J.S. \cqg{7}{1990}{2353}.}
  \ref{CoandH}{Courant, R. and Hilbert, D. {\it Methoden der
  Mathematischen Physik} vol.1 \break (Springer, Berlin. 1931).}
  \ref{Applequist}{Applequist, J. \jpa{22}{1989}{4303}.}
  \ref{Torruella}{Torruella, \jmp{16}{1975}{1637}.}
  \ref{Weinberg}{Weinberg, S.W. \pr{133}{1964}{B1318}.}
  \ref{Meyerw}{Meyer, W.F. {\it Apolarit\"at und rationale Curven}
  (Fues, T\"ubingen. 1883.) }
  \ref{Ostrowski}{Ostrowski, A. {\it Jahrsb. Deutsch. Math. Verein.} {\bf
  33} (1923) 245.}
  \ref{Kramers}{Kramers, H.A. {\it Grundlagen der Quantenmechanik}, (Akad.
  Verlag., Leipzig, 1938).}
  \ref{ZandZ}{Zou, W.-N. and Zheng, Q.-S. \prs{459}{2003}{527}.}
  \ref{Weeks1}{Weeks, J.R. {\it Maxwell's multipole vectors
  and the CMB}.  astro-ph/0412231.}
  \ref{Corson}{Corson, E.M. {\it Tensors, Spinors and Relativistic Wave
  Equations} (Blackie, London. 1950).}
  \ref{Rosanes}{Rosanes, J. \jram{76}{1873}{312}.}
  \ref{Salmon}{Salmon, G. {\it Lessons Introductory to the Modern Higher
  Algebra} 3rd. edn. \break (Hodges,  Dublin. 1876.)}
  \ref{Milnew}{Milne, W.P. {\it Homogeneous Coordinates} (Arnold. London. 1910).}
  \ref{Niven}{Niven, W.D. {\it Phil. Trans. Roy. Soc.} {\bf 170} (1879) 393.}
  \ref{Scott}{Scott, C.A. {\it An Introductory Account of
  Certain Modern Ideas and Methods in Plane Analytical Geometry,}
  (MacMillan, N.Y. 1896).}
  \ref{Bargmann}{Bargmann, V. \rmp{34}{1962}{300}.}
  \ref{Maxwell}{Maxwell, J.C. {\it A Treatise on Electricity and
  Magnetism} 2nd. edn. (Clarendon Press, Oxford. 1882).}
  \ref{BandL}{Biedenharn, L.C. and Louck, J.D.
  {\it Angular Momentum in Quantum Physics} (Addison-Wesley, Reading. 1981).}
  \ref{Weylqm}{Weyl, H. {\it The Theory of Groups and Quantum Mechanics}
  (Methuen, London. 1931).}
  \ref{Robson}{Robson, A. {\it An Introduction to Analytical Geometry} Vol I
  (C.U.P., Cambridge. 1940.)}
  \ref{Sommerville}{Sommerville, D.M.Y. {\it Analytical Conics} 3rd. edn.
   (Bell, London. 1933).}
  \ref{Coolidge}{Coolidge, J.L. {\it A Treatise on Algebraic Plane Curves}
  (Clarendon Press, Oxford. 1931).}
  \ref{SandK}{Semple, G. and Kneebone. G.T. {\it Algebraic Projective
  Geometry} (Clarendon Press, Oxford. 1952).}
  \ref{AandC}{Abdesselam A., and Chipalkatti, J. {\it The Higher
  Transvectants are redundant}, arXiv:0801.1533 [math.AG] 2008.}
  \ref{Elliott}{Elliott, E.B. {\it The Algebra of Quantics} 2nd edn.
  (Clarendon Press, Oxford. 1913).}
  \ref{Elliott2}{Elliott, E.B. \qjpam{48}{1917}{372}.}
  \ref{Howe}{Howe, R. \tams{313}{1989}{539}.}
  \ref{Clebsch}{Clebsch, A. \jram{60}{1862}{343}.}
  \ref{Prasad}{Prasad, G. \ma{72}{1912}{136}.}
  \ref{Dougall}{Dougall, J. \pems{32}{1913}{30}.}
  \ref{Penrose}{Penrose, R. \aop{10}{1960}{171}.}
  \ref{Penrose2}{Penrose, R. \prs{273}{1965}{171}.}
  \ref{Burnside}{Burnside, W.S. \qjm{10}{1870}{211}. }
  \ref{Lindemann}{Lindemann, F. \ma{23} {1884}{111}.}
  \ref{Backus}{Backus, G. {\it Rev. Geophys. Space Phys.} {\bf 8} (1970) 633.}
  \ref{Baerheim}{Baerheim, R. {\it Q.J. Mech. appl. Math.} {\bf 51} (1998) 73.}
  \ref{Lense}{Lense, J. {\it Kugelfunktionen} (Akad.Verlag, Leipzig. 1950).}
  \ref{Littlewood}{Littlewood, D.E. \plms{50}{1948}{349}.}
  \ref{Fierz}{Fierz, M. {\it Helv. Phys. Acta} {\bf 12} (1938) 3.}
  \ref{Williams}{Williams, D.N. {\it Lectures in Theoretical Physics} Vol. VII,
  (Univ.Colorado Press, Boulder. 1965).}
  \ref{Dennis}{Dennis, M. \jpa{37}{2004}{9487}.}
  \ref{Pirani}{Pirani, F. {\it Brandeis Lecture Notes on
  General Relativity,} edited by S. Deser and K. Ford. (Brandeis, Mass. 1964).}
  \ref{Sturm}{Sturm, R. \jram{86}{1878}{116}.}
  \ref{Schlesinger}{Schlesinger, O. \ma{22}{1883}{521}.}
  \ref{Askwith}{Askwith, E.H. {\it Analytical Geometry of the Conic
  Sections} (A.\&C. Black, London. 1908).}
  \ref{Todd}{Todd, J.A. {\it Projective and Analytical Geometry}.
  (Pitman, London. 1946).}
  \ref{Glenn}{Glenn. O.E. {\it Theory of Invariants} (Ginn \& Co, N.Y. 1915).}
  \ref{DowkandG}{Dowker, J.S. and Goldstone, M. \prs{303}{1968}{381}.}
  \ref{Turnbull}{Turnbull, H.A. {\it The Theory of Determinants,
  Matrices and Invariants} 3rd. edn. (Dover, N.Y. 1960).}
  \ref{MacMillan}{MacMillan, W.D. {\it The Theory of the Potential}
  (McGraw-Hill, N.Y. 1930).}
   \ref{Hobson}{Hobson, E.W. {\it The Theory of Spherical
   and Ellipsoidal Harmonics} (C.U.P., Cambridge. 1931).}
  \ref{Hobson1}{Hobson, E.W. \plms {24}{1892}{55}.}
  \ref{GandY}{Grace, J.H. and Young, A. {\it The Algebra of Invariants}
  (C.U.P., Cambridge, 1903).}
  \ref{FandR}{Fano, U. and Racah, G. {\it Irreducible Tensorial Sets}
  (Academic Press, N.Y. 1959).}
  \ref{TandT}{Thomson, W. and Tait, P.G. {\it Treatise on Natural Philosophy}
   (Clarendon Press, Oxford. 1867).}
  \ref{Brinkman}{Brinkman, H.C. {\it Applications of spinor invariants in
atomic physics}, North Holland, Amsterdam 1956.}
  \ref{Kramers1}{Kramers, H.A. {\it Proc. Roy. Soc. Amst.} {\bf 33} (1930) 953.}
  \ref{DandP2}{Dowker,J.S. and Pettengill,D.F. \jpa{7}{1974}{1527}}
  \ref{Dowk1}{Dowker,J.S. \jpa{}{}{45}.}
  \ref{Dowk2}{Dowker,J.S. \aop{71}{1972}{577}}
  \ref{DandA}{Dowker,J.S. and Apps, J.S. \cqg{15}{1998}{1121}.}
  \ref{Weil}{Weil,A., {\it Elliptic functions according to Eisenstein
  and Kronecker}, Springer, Berlin, 1976.}
  \ref{Ling}{Ling,C-H. {\it SIAM J.Math.Anal.} {\bf5} (1974) 551.}
  \ref{Ling2}{Ling,C-H. {\it J.Math.Anal.Appl.}(1988).}
 \ref{BMO}{Brevik,I., Milton,K.A. and Odintsov, S.D. \aop{302}{2002}{120}.}
 \ref{KandL}{Kutasov,D. and Larsen,F. {\it JHEP} 0101 (2001) 1.}
 \ref{KPS}{Klemm,D., Petkou,A.C. and Siopsis {\it Entropy
 bounds, monoticity properties and scaling in CFT's}. hep-th/0101076.}
 \ref{DandC}{Dowker,J.S. and Critchley,R. \prD{15}{1976}{1484}.}
 \ref{AandD}{Al'taie, M.B. and Dowker, J.S. \prD{18}{1978}{3557}.}
 \ref{Dow1}{Dowker,J.S. \prD{37}{1988}{558}.}
 \ref{Dowrob}{Dowker,J.S. \cqg{13}{1996}{585}.}
 \ref{Dow30}{Dowker,J.S. \prD{28}{1983}{3013}.}
 \ref{DandK}{Dowker,J.S. and Kennedy,G. \jpa{}{1978}{895}.}
 \ref{Dow2}{Dowker,J.S. \cqg{1}{1984}{359}.}
 \ref{DandKi}{Dowker,J.S. and Kirsten, K. {\it Comm. in Anal. and Geom.
 }{\bf7} (1999) 641.}
 \ref{DandKe}{Dowker,J.S. and Kennedy,G.\jpa{11}{1978}{895}.}
 \ref{Gibbons}{Gibbons,G.W. \pl{60A}{1977}{385}.}
 \ref{Cardy}{Cardy,J.L. \np{366}{1991}{403}.}
 \ref{ChandD}{Chang,P. and Dowker,J.S. \np{395}{1993}{407}.}
 \ref{DandC2}{Dowker,J.S. and Critchley,R. \prD{13}{1976}{224}.}
 \ref{Camporesi}{Camporesi,R. \prp{196}{1990}{1}.}
 \ref{BandM}{Brown,L.S. and Maclay,G.J. \pr{184}{1969}{1272}.}
 \ref{CandD}{Candelas,P. and Dowker,J.S. \prD{19}{1979}{2902}.}
 \ref{Unwin1}{Unwin,S.D. Thesis. University of Manchester. 1979.}
 \ref{Unwin2}{Unwin,S.D. \jpa{13}{1980}{313}.}
 \ref{DandB}{Dowker,J.S. and Banach,R. \jpa{11}{1978}{2255}.}
 \ref{Obhukov}{Obhukov,Yu.N. \pl{109B}{1982}{195}.}
 \ref{Kennedy}{Kennedy,G. \prD{23}{1981}{2884}.}
 \ref{CandT}{Copeland,E. and Toms,D.J. \np {255}{1985}{201}.}
  \ref{CandT2}{Copeland,E. and Toms,D.J. \cqg {3}{1986}{431}.}
 \ref{ELV}{Elizalde,E., Lygren, M. and Vassilevich,
 D.V. \jmp {37}{1996}{3105}.}
 \ref{Malurkar}{Malurkar,S.L. {\it J.Ind.Math.Soc} {\bf16} (1925/26) 130.}
 \ref{Glaisher}{Glaisher,J.W.L. {\it Messenger of Math.} {\bf18}
(1889) 1.} \ref{Anderson}{Anderson,A. \prD{37}{1988}{536}.}
 \ref{CandA}{Cappelli,A. and D'Appollonio,\pl{487B}{2000}{87}.}
 \ref{Wot}{Wotzasek,C. \jpa{23}{1990}{1627}.}
 \ref{RandT}{Ravndal,F. and Tollesen,D. \prD{40}{1989}{4191}.}
 \ref{SandT}{Santos,F.C. and Tort,A.C. \pl{482B}{2000}{323}.}
 \ref{FandO}{Fukushima,K. and Ohta,K. {\it Physica} {\bf A299} (2001) 455.}
 \ref{GandP}{Gibbons,G.W. and Perry,M. \prs{358}{1978}{467}.}
 \ref{Dow4}{Dowker,J.S..}
  \ref{Rad}{Rademacher,H. {\it Topics in analytic number theory,}
Springer-Verlag,  Berlin,1973.}
  \ref{Halphen}{Halphen,G.-H. {\it Trait\'e des Fonctions Elliptiques},
  Vol 1, Gauthier-Villars, Paris, 1886.}
  \ref{CandW}{Cahn,R.S. and Wolf,J.A. {\it Comm.Mat.Helv.} {\bf 51}
  (1976) 1.}
  \ref{Berndt}{Berndt,B.C. \rmjm{7}{1977}{147}.}
  \ref{Hurwitz}{Hurwitz,A. \ma{18}{1881}{528}.}
  \ref{Hurwitz2}{Hurwitz,A. {\it Mathematische Werke} Vol.I. Basel,
  Birkhauser, 1932.}
  \ref{Berndt2}{Berndt,B.C. \jram{303/304}{1978}{332}.}
  \ref{RandA}{Rao,M.B. and Ayyar,M.V. \jims{15}{1923/24}{150}.}
  \ref{Hardy}{Hardy,G.H. \jlms{3}{1928}{238}.}
  \ref{TandM}{Tannery,J. and Molk,J. {\it Fonctions Elliptiques},
   Gauthier-Villars, Paris, 1893--1902.}
  \ref{schwarz}{Schwarz,H.-A. {\it Formeln und
  Lehrs\"atzen zum Gebrauche..},Springer 1893.(The first edition was 1885.)
  The French translation by Henri Pad\'e is {\it Formules et Propositions
  pour L'Emploi...},Gauthier-Villars, Paris, 1894}
  \ref{Hancock}{Hancock,H. {\it Theory of elliptic functions}, Vol I.
   Wiley, New York 1910.}
  \ref{watson}{Watson,G.N. \jlms{3}{1928}{216}.}
  \ref{MandO}{Magnus,W. and Oberhettinger,F. {\it Formeln und S\"atze},
  Springer-Verlag, Berlin 1948.}
  \ref{Klein}{Klein,F. {\it Lectures on the Icosohedron}
  (Methuen, London. 1913).}
  \ref{AandL}{Appell,P. and Lacour,E. {\it Fonctions Elliptiques},
  Gauthier-Villars,
  Paris. 1897.}
  \ref{HandC}{Hurwitz,A. and Courant,C. {\it Allgemeine Funktionentheorie},
  Springer,
  Berlin. 1922.}
  \ref{WandW}{Whittaker,E.T. and Watson,G.N. {\it Modern analysis},
  Cambridge. 1927.}
  \ref{SandC}{Selberg,A. and Chowla,S. \jram{227}{1967}{86}. }
  \ref{zucker}{Zucker,I.J. {\it Math.Proc.Camb.Phil.Soc} {\bf 82 }(1977)
  111.}
  \ref{glasser}{Glasser,M.L. {\it Maths.of Comp.} {\bf 25} (1971) 533.}
  \ref{GandW}{Glasser, M.L. and Wood,V.E. {\it Maths of Comp.} {\bf 25}
  (1971)
  535.}
  \ref{greenhill}{Greenhill,A,G. {\it The Applications of Elliptic
  Functions}, MacMillan. London, 1892.}
  \ref{Weierstrass}{Weierstrass,K. {\it J.f.Mathematik (Crelle)}
{\bf 52} (1856) 346.}
  \ref{Weierstrass2}{Weierstrass,K. {\it Mathematische Werke} Vol.I,p.1,
  Mayer u. M\"uller, Berlin, 1894.}
  \ref{Fricke}{Fricke,R. {\it Die Elliptische Funktionen und Ihre Anwendungen},
    Teubner, Leipzig. 1915, 1922.}
  \ref{Konig}{K\"onigsberger,L. {\it Vorlesungen \"uber die Theorie der
 Elliptischen Funktionen},  \break Teubner, Leipzig, 1874.}
  \ref{Milne}{Milne,S.C. {\it The Ramanujan Journal} {\bf 6} (2002) 7-149.}
  \ref{Schlomilch}{Schl\"omilch,O. {\it Ber. Verh. K. Sachs. Gesell. Wiss.
  Leipzig}  {\bf 29} (1877) 101-105; {\it Compendium der h\"oheren
  Analysis}, Bd.II, 3rd Edn, Vieweg, Brunswick, 1878.}
  \ref{BandB}{Briot,C. and Bouquet,C. {\it Th\`eorie des Fonctions
  Elliptiques}, Gauthier-Villars, Paris, 1875.}
  \ref{Dumont}{Dumont,D. \aim {41}{1981}{1}.}
  \ref{Andre}{Andr\'e,D. {\it Ann.\'Ecole Normale Superior} {\bf 6} (1877)
  265;
  {\it J.Math.Pures et Appl.} {\bf 5} (1878) 31.}
  \ref{Raman}{Ramanujan,S. {\it Trans.Camb.Phil.Soc.} {\bf 22} (1916) 159;
 {\it Collected Papers}, Cambridge, 1927}
  \ref{Weber}{Weber,H.M. {\it Lehrbuch der Algebra} Bd.III, Vieweg,
  Brunswick 190  3.}
  \ref{Weber2}{Weber,H.M. {\it Elliptische Funktionen und algebraische
  Zahlen},
  Vieweg, Brunswick 1891.}
  \ref{ZandR}{Zucker,I.J. and Robertson,M.M.
  {\it Math.Proc.Camb.Phil.Soc} {\bf 95 }(1984) 5.}
  \ref{JandZ1}{Joyce,G.S. and Zucker,I.J.
  {\it Math.Proc.Camb.Phil.Soc} {\bf 109 }(1991) 257.}
  \ref{JandZ2}{Zucker,I.J. and Joyce.G.S.
  {\it Math.Proc.Camb.Phil.Soc} {\bf 131 }(2001) 309.}
  \ref{zucker2}{Zucker,I.J. {\it SIAM J.Math.Anal.} {\bf 10} (1979) 192,}
  \ref{BandZ}{Borwein,J.M. and Zucker,I.J. {\it IMA J.Math.Anal.} {\bf 12}
  (1992) 519.}
  \ref{Cox}{Cox,D.A. {\it Primes of the form $x^2+n\,y^2$}, Wiley,
  New York, 1989.}
  \ref{BandCh}{Berndt,B.C. and Chan,H.H. {\it Mathematika} {\bf42} (1995)
  278.}
  \ref{EandT}{Elizalde,R. and Tort.hep-th/}
  \ref{KandS}{Kiyek,K. and Schmidt,H. {\it Arch.Math.} {\bf 18} (1967) 438.}
  \ref{Oshima}{Oshima,K. \prD{46}{1992}{4765}.}
  \ref{greenhill2}{Greenhill,A.G. \plms{19} {1888} {301}.}
  \ref{Russell}{Russell,R. \plms{19} {1888} {91}.}
  \ref{BandB}{Borwein,J.M. and Borwein,P.B. {\it Pi and the AGM}, Wiley,
  New York, 1998.}
  \ref{Resnikoff}{Resnikoff,H.L. \tams{124}{1966}{334}.}
  \ref{vandp}{Van der Pol, B. {\it Indag.Math.} {\bf18} (1951) 261,272.}
  \ref{Rankin}{Rankin,R.A. {\it Modular forms} C.U.P. Cambridge}
  \ref{Rankin2}{Rankin,R.A. {\it Proc. Roy.Soc. Edin.} {\bf76 A} (1976) 107.}
  \ref{Skoruppa}{Skoruppa,N-P. {\it J.of Number Th.} {\bf43} (1993) 68 .}
  \ref{Down}{Dowker.J.S. {\it Nucl.Phys.}B (Proc.Suppl) ({\bf 104})(2002)153;
  also Dowker,J.S. hep-th/ 0007129.}
  \ref{Eichler}{Eichler,M. \mz {67}{1957}{267}.}
  \ref{Zagier}{Zagier,D. \invm{104}{1991}{449}.}
  \ref{Lang}{Lang,S. {\it Modular Forms}, Springer, Berlin, 1976.}
  \ref{Kosh}{Koshliakov,N.S. {\it Mess.of Math.} {\bf 58} (1928) 1.}
  \ref{BandH}{Bodendiek, R. and Halbritter,U. \amsh{38}{1972}{147}.}
  \ref{Smart}{Smart,L.R., \pgma{14}{1973}{1}.}
  \ref{Grosswald}{Grosswald,E. {\it Acta. Arith.} {\bf 21} (1972) 25.}
  \ref{Kata}{Katayama,K. {\it Acta Arith.} {\bf 22} (1973) 149.}
  \ref{Ogg}{Ogg,A. {\it Modular forms and Dirichlet series} (Benjamin,
  New York,
   1969).}
  \ref{Bol}{Bol,G. \amsh{16}{1949}{1}.}
  \ref{Epstein}{Epstein,P. \ma{56}{1903}{615}.}
  \ref{Petersson}{Petersson.}
  \ref{Serre}{Serre,J-P. {\it A Course in Arithmetic}, Springer,
  New York, 1973.}
  \ref{Schoenberg}{Schoenberg,B., {\it Elliptic Modular Functions},
  Springer, Berlin, 1974.}
  \ref{Apostol}{Apostol,T.M. \dmj {17}{1950}{147}.}
  \ref{Ogg2}{Ogg,A. {\it Lecture Notes in Math.} {\bf 320} (1973) 1.}
  \ref{Knopp}{Knopp,M.I. \dmj {45}{1978}{47}.}
  \ref{Knopp2}{Knopp,M.I. \invm {}{1994}{361}.}
  \ref{LandZ}{Lewis,J. and Zagier,D. \aom{153}{2001}{191}.}
  \ref{DandK1}{Dowker,J.S. and Kirsten,K. {\it Elliptic functions and
  temperature inversion symmetry on spheres} hep-th/.}
  \ref{HandK}{Husseini and Knopp.}
  \ref{Kober}{Kober,H. \mz{39}{1934-5}{609}.}
  \ref{HandL}{Hardy,G.H. and Littlewood, \am{41}{1917}{119}.}
  \ref{Watson}{Watson,G.N. \qjm{2}{1931}{300}.}
  \ref{SandC2}{Chowla,S. and Selberg,A. {\it Proc.Nat.Acad.} {\bf 35}
  (1949) 371.}
  \ref{Landau}{Landau, E. {\it Lehre von der Verteilung der Primzahlen},
  (Teubner, Leipzig, 1909).}
  \ref{Berndt4}{Berndt,B.C. \tams {146}{1969}{323}.}
  \ref{Berndt3}{Berndt,B.C. \tams {}{}{}.}
  \ref{Bochner}{Bochner,S. \aom{53}{1951}{332}.}
  \ref{Weil2}{Weil,A.\ma{168}{1967}{}.}
  \ref{CandN}{Chandrasekharan,K. and Narasimhan,R. \aom{74}{1961}{1}.}
  \ref{Rankin3}{Rankin,R.A. {} {} ().}
  \ref{Berndt6}{Berndt,B.C. {\it Trans.Edin.Math.Soc}.}
  \ref{Elizalde}{Elizalde,E. {\it Ten Physical Applications of Spectral
  Zeta Function Theory}, \break (Springer, Berlin, 1995).}
  \ref{Allen}{Allen,B., Folacci,A. and Gibbons,G.W. \pl{189}{1987}{304}.}
  \ref{Krazer}{Krazer}
  \ref{Elizalde3}{Elizalde,E. {\it J.Comp.and Appl. Math.} {\bf 118}
  (2000) 125.}
  \ref{Elizalde2}{Elizalde,E., Odintsov.S.D, Romeo, A. and Bytsenko,
  A.A and
  Zerbini,S.
  {\it Zeta function regularisation}, (World Scientific, Singapore,
  1994).}
  \ref{Eisenstein}{Eisenstein}
  \ref{Hecke}{Hecke,E. \ma{112}{1936}{664}.}
  \ref{Hecke2}{Hecke,E. \ma{112}{1918}{398}.}
  \ref{Terras}{Terras,A. {\it Harmonic analysis on Symmetric Spaces} (Springer,
  New York, 1985).}
  \ref{BandG}{Bateman,P.T. and Grosswald,E. {\it Acta Arith.} {\bf 9}
  (1964) 365.}
  \ref{Deuring}{Deuring,M. \aom{38}{1937}{585}.}
  \ref{Mordell}{Mordell,J. \prs{}{}{}.}
  \ref{GandZ}{Glasser,M.L. and Zucker, {}.}
  \ref{Landau2}{Landau,E. \jram{}{1903}{64}.}
  \ref{Kirsten1}{Kirsten,K. \jmp{35}{1994}{459}.}
  \ref{Sommer}{Sommer,J. {\it Vorlesungen \"uber Zahlentheorie}
  (1907,Teubner,Leipzig).
  French edition 1913 .}
  \ref{Reid}{Reid,L.W. {\it Theory of Algebraic Numbers},
  (1910,MacMillan,New York).}
  \ref{Milnor}{Milnor, J. {\it Is the Universe simply--connected?},
  IAS, Princeton, 1978.}
  \ref{Milnor2}{Milnor, J. \ajm{79}{1957}{623}.}
  \ref{Opechowski}{Opechowski,W. {\it Physica} {\bf 7} (1940) 552.}
  \ref{Bethe}{Bethe, H.A. \zfp{3}{1929}{133}.}
  \ref{LandL}{Landau, L.D. and Lishitz, E.M. {\it Quantum
  Mechanics} (Pergamon Press, London, 1958).}
  \ref{GPR}{Gibbons, G.W., Pope, C. and R\"omer, H., \np{157}{1979}{377}.}
  \ref{Jadhav}{Jadhav,S.P. PhD Thesis, University of Manchester 1990.}
  \ref{DandJ}{Dowker,J.S. and Jadhav, S. \prD{39}{1989}{1196}.}
  \ref{CandM}{Coxeter, H.S.M. and Moser, W.O.J. {\it Generators and
  relations of finite groups} (Springer. Berlin. 1957).}
  \ref{Coxeter2}{Coxeter, H.S.M. {\it Regular Complex Polytopes},
   (Cambridge University Press, \break Cambridge, 1975).}
  \ref{Coxeter}{Coxeter, H.S.M. {\it Regular Polytopes}.}
  \ref{Stiefel}{Stiefel, E., J.Research NBS {\bf 48} (1952) 424.}
  \ref{BandS}{Brink, D.M. and Satchler, G.R. {\it Angular momentum theory}.
  (Clarendon Press, Oxford. 1962.).}
  \ref{Rose}{Rose}
  \ref{Schwinger}{Schwinger, J. {\it On Angular Momentum}
  in {\it Quantum Theory of Angular Momentum} edited by
  Biedenharn,L.C. and van Dam, H. (Academic Press, N.Y. 1965).}
  \ref{Bromwich}{Bromwich, T.J.I'A. {\it Infinite Series},
  (Macmillan, London, 1926).}
  \ref{Ray}{Ray,D.B. \aim{4}{1970}{109}.}
  \ref{Ikeda}{Ikeda,A. {\it Kodai Math.J.} {\bf 18} (1995) 57.}
  \ref{Kennedy}{Kennedy,G. \prD{23}{1981}{2884}.}
  \ref{Ellis}{Ellis,G.F.R. {\it General Relativity} {\bf2} (1971) 7.}
  \ref{Dow8}{Dowker,J.S. \cqg{20}{2003}{L105}.}
  \ref{IandY}{Ikeda, A and Yamamoto, Y. \ojm {16}{1979}{447}.}
  \ref{BandI}{Bander,M. and Itzykson,C. \rmp{18}{1966}{2}.}
  \ref{Schulman}{Schulman, L.S. \pr{176}{1968}{1558}.}
  \ref{Bar1}{B\"ar,C. {\it Arch.d.Math.}{\bf 59} (1992) 65.}
  \ref{Bar2}{B\"ar,C. {\it Geom. and Func. Anal.} {\bf 6} (1996) 899.}
  \ref{Vilenkin}{Vilenkin, N.J. {\it Special functions},
  (Am.Math.Soc., Providence, 1968).}
  \ref{Talman}{Talman, J.D. {\it Special functions} (Benjamin,N.Y.,1968).}
  \ref{Miller}{Miller, W. {\it Symmetry groups and their applications}
  (Wiley, N.Y., 1972).}
  \ref{Dow3}{Dowker,J.S. \cmp{162}{1994}{633}.}
  \ref{Cheeger}{Cheeger, J. \jdg {18}{1983}{575}.}
  \ref{Cheeger2}{Cheeger, J. \aom {109}{1979}{259}.}
  \ref{Dow6}{Dowker,J.S. \jmp{30}{1989}{770}.}
  \ref{Dow20}{Dowker,J.S. \jmp{35}{1994}{6076}.}
  \ref{Dowjmp}{Dowker,J.S. \jmp{35}{1994}{4989}.}
  \ref{Dow21}{Dowker,J.S. {\it Heat kernels and polytopes} in {\it
   Heat Kernel Techniques and Quantum Gravity}, ed. by S.A.Fulling,
   Discourses in Mathematics and its Applications, No.4, Dept.
   Maths., Texas A\&M University, College Station, Texas, 1995.}
  \ref{Dow9}{Dowker,J.S. \jmp{42}{2001}{1501}.}
  \ref{Dow7}{Dowker,J.S. \jpa{25}{1992}{2641}.}
  \ref{Warner}{Warner.N.P. \prs{383}{1982}{379}.}
  \ref{Wolf}{Wolf, J.A. {\it Spaces of constant curvature},
  (McGraw--Hill,N.Y., 1967).}
  \ref{Meyer}{Meyer,B. \cjm{6}{1954}{135}.}
  \ref{BandB}{B\'erard,P. and Besson,G. {\it Ann. Inst. Four.} {\bf 30}
  (1980) 237.}
  \ref{PandM}{P\'{o}lya,G. and Meyer,B. \cras{228}{1948}{28}.}
  \ref{Springer}{Springer, T.A. Lecture Notes in Math. vol 585 (Springer,
  Berlin,1977).}
  \ref{SeandT}{Threlfall, H. and Seifert, W. \ma{104}{1930}{1}.}
  \ref{Hopf}{Hopf,H. \ma{95}{1925}{313}. }
  \ref{Dow}{Dowker,J.S. \jpa{5}{1972}{936}.}
  \ref{LLL}{Lehoucq,R., Lachi\'eze-Rey,M. and Luminet, J.--P. {\it
  Astron.Astrophys.} {\bf 313} (1996) 339.}
  \ref{LaandL}{Lachi\'eze-Rey,M. and Luminet, J.--P.
  \prp{254}{1995}{135}.}
  \ref{Schwarzschild}{Schwarzschild, K., {\it Vierteljahrschrift der
  Ast.Ges.} {\bf 35} (1900) 337.}
  \ref{Starkman}{Starkman,G.D. \cqg{15}{1998}{2529}.}
  \ref{LWUGL}{Lehoucq,R., Weeks,J.R., Uzan,J.P., Gausman, E. and
  Luminet, J.--P. \cqg{19}{2002}{4683}.}
  \ref{Dow10}{Dowker,J.S. \prD{28}{1983}{3013}.}
  \ref{BandD}{Banach, R. and Dowker, J.S. \jpa{12}{1979}{2527}.}
  \ref{Jadhav2}{Jadhav,S. \prD{43}{1991}{2656}.}
  \ref{Gilkey}{Gilkey,P.B. {\it Invariance theory,the heat equation and
  the Atiyah--Singer Index theorem} (CRC Press, Boca Raton, 1994).}
  \ref{BandY}{Berndt,B.C. and Yeap,B.P. {\it Adv. Appl. Math.}
  {\bf29} (2002) 358.}
  \ref{HandR}{Hanson,A.J. and R\"omer,H. \pl{80B}{1978}{58}.}
  \ref{Hill}{Hill,M.J.M. {\it Trans.Camb.Phil.Soc.} {\bf 13} (1883) 36.}
  \ref{Cayley}{Cayley,A. {\it Quart.Math.J.} {\bf 7} (1866) 304.}
  \ref{Seade}{Seade,J.A. {\it Anal.Inst.Mat.Univ.Nac.Aut\'on
  M\'exico} {\bf 21} (1981) 129.}
  \ref{CM}{Cisneros--Molina,J.L. {\it Geom.Dedicata} {\bf84} (2001)
  \ref{Goette1}{Goette,S. \jram {526} {2000} 181.}
  207.}
  \ref{NandO}{Nash,C. and O'Connor,D--J, \jmp {36}{1995}{1462}.}
  \ref{Dows}{Dowker,J.S. \aop{71}{1972}{577}; Dowker,J.S. and Pettengill,D.F.
  \jpa{7}{1974}{1527}; J.S.Dowker in {\it Quantum Gravity}, edited by
  S. C. Christensen (Hilger,Bristol,1984)}
  \ref{Jadhav2}{Jadhav,S.P. \prD{43}{1991}{2656}.}
  \ref{Dow11}{Dowker,J.S. \cqg{21}{2004}4247.}
  \ref{Dow12}{Dowker,J.S. \cqg{21}{2004}4977.}
  \ref{Dow13}{Dowker,J.S. \jpa{38}{2005}1049.}
  \ref{Zagier}{Zagier,D. \ma{202}{1973}{149}}
  \ref{RandG}{Rademacher, H. and Grosswald,E. {\it Dedekind Sums},
  (Carus, MAA, 1972).}
  \ref{Berndt7}{Berndt,B, \aim{23}{1977}{285}.}
  \ref{HKMM}{Harvey,J.A., Kutasov,D., Martinec,E.J. and Moore,G.
  {\it Localised Tachyons and RG Flows}, hep-th/0111154.}
  \ref{Beck}{Beck,M., {\it Dedekind Cotangent Sums}, {\it Acta Arithmetica}
  {\bf 109} (2003) 109-139 ; math.NT/0112077.}
  \ref{McInnes}{McInnes,B. {\it APS instability and the topology of the brane
  world}, hep-th/0401035.}
  \ref{BHS}{Brevik,I, Herikstad,R. and Skriudalen,S. {\it Entropy Bound for the
  TM Electromagnetic Field in the Half Einstein Universe}; hep-th/0508123.}
  \ref{BandO}{Brevik,I. and Owe,C.  \prD{55}{4689}{1997}.}
  \ref{Kenn}{Kennedy,G. Thesis. University of Manchester 1978.}
  \ref{KandU}{Kennedy,G. and Unwin S. \jpa{12}{L253}{1980}.}
  \ref{BandO1}{Bayin,S.S.and Ozcan,M.
  \prD{48}{2806}{1993}; \prD{49}{5313}{1994}.}
  \ref{Chang}{Chang, P., {\it Quantum Field Theory on Regular Polytopes}.
   Thesis. University of Manchester, 1993.}
  \ref{Barnesa}{Barnes,E.W. {\it Trans. Camb. Phil. Soc.} {\bf 19} (1903) 374.}
  \ref{Barnesb}{Barnes,E.W. {\it Trans. Camb. Phil. Soc.}
  {\bf 19} (1903) 426.}
  \ref{Stanley1}{Stanley,R.P. \joa {49Hilf}{1977}{134}.}
  \ref{Stanley}{Stanley,R.P. \bams {1}{1979}{475}.}
  \ref{Hurley}{Hurley,A.C. \pcps {47}{1951}{51}.}
  \ref{IandK}{Iwasaki,I. and Katase,K. {\it Proc.Japan Acad. Ser} {\bf A55}
  (1979) 141.}
  \ref{IandT}{Ikeda,A. and Taniguchi,Y. {\it Osaka J. Math.} {\bf 15} (1978)
  515.}
  \ref{GandM}{Gallot,S. and Meyer,D. \jmpa{54}{1975}{259}.}
  \ref{Flatto}{Flatto,L. {\it Enseign. Math.} {\bf 24} (1978) 237.}
  \ref{OandT}{Orlik,P and Terao,H. {\it Arrangements of Hyperplanes},
  Grundlehren der Math. Wiss. {\bf 300}, (Springer--Verlag, 1992).}
  \ref{Shepler}{Shepler,A.V. \joa{220}{1999}{314}.}
  \ref{SandT}{Solomon,L. and Terao,H. \cmh {73}{1998}{237}.}
  \ref{Vass}{Vassilevich, D.V. \plb {348}{1995}39.}
  \ref{Vass2}{Vassilevich, D.V. \jmp {36}{1995}3174.}
  \ref{CandH}{Camporesi,R. and Higuchi,A. {\it J.Geom. and Physics}
  {\bf 15} (1994) 57.}
  \ref{Solomon2}{Solomon,L. \tams{113}{1964}{274}.}
  \ref{Solomon}{Solomon,L. {\it Nagoya Math. J.} {\bf 22} (1963) 57.}
  \ref{Obukhov}{Obukhov,Yu.N. \pl{109B}{1982}{195}.}
  \ref{BGH}{Bernasconi,F., Graf,G.M. and Hasler,D. {\it The heat kernel
  expansion for the electromagnetic field in a cavity}; math-ph/0302035.}
  \ref{Baltes}{Baltes,H.P. \prA {6}{1972}{2252}.}
  \ref{BaandH}{Baltes.H.P and Hilf,E.R. {\it Spectra of Finite Systems}
  (Bibliographisches Institut, Mannheim, 1976).}
  \ref{Ray}{Ray,D.B. \aim{4}{1970}{109}.}
  \ref{Hirzebruch}{Hirzebruch,F. {\it Topological methods in algebraic
  geometry} (Springer-- Verlag,\break  Berlin, 1978). }
  \ref{BBG}{Bla\v{z}i\'c,N., Bokan,N. and Gilkey, P.B. {\it Ind.J.Pure and
  Appl.Math.} {\bf 23} (1992) 103.}
  \ref{WandWi}{Weck,N. and Witsch,K.J. {\it Math.Meth.Appl.Sci.} {\bf 17}
  (1994) 1017.}
  \ref{Norlund}{N\"orlund,N.E. \am{43}{1922}{121}.}
   \ref{Norlund1}{N\"orlund,N.E. {\it Differenzenrechnung} (Springer--Verlag, 1924, Berlin.)}
  \ref{Duff}{Duff,G.F.D. \aom{56}{1952}{115}.}
  \ref{DandS}{Duff,G.F.D. and Spencer,D.C. \aom{45}{1951}{128}.}
  \ref{BGM}{Berger, M., Gauduchon, P. and Mazet, E. {\it Lect.Notes.Math.}
  {\bf 194} (1971) 1. }
  \ref{Patodi}{Patodi,V.K. \jdg{5}{1971}{233}.}
  \ref{GandS}{G\"unther,P. and Schimming,R. \jdg{12}{1977}{599}.}
  \ref{MandS}{McKean,H.P. and Singer,I.M. \jdg{1}{1967}{43}.}
  \ref{Conner}{Conner,P.E. {\it Mem.Am.Math.Soc.} {\bf 20} (1956).}
  \ref{Gilkey2}{Gilkey,P.B. \aim {15}{1975}{334}.}
  \ref{MandP}{Moss,I.G. and Poletti,S.J. \plb{333}{1994}{326}.}
  \ref{BKD}{Bordag,M., Kirsten,K. and Dowker,J.S. \cmp{182}{1996}{371}.}
  \ref{RandO}{Rubin,M.A. and Ordonez,C. \jmp{25}{1984}{2888}.}
  \ref{BaandD}{Balian,R. and Duplantier,B. \aop {112}{1978}{165}.}
  \ref{Kennedy2}{Kennedy,G. \aop{138}{1982}{353}.}
  \ref{DandKi2}{Dowker,J.S. and Kirsten, K. {\it Analysis and Appl.}
 {\bf 3} (2005) 45.}
  \ref{Dow40}{Dowker,J.S. \cqg{23}{2006}{1}.}
  \ref{BandHe}{Br\"uning,J. and Heintze,E. {\it Duke Math.J.} {\bf 51} (1984)
   959.}
  \ref{Dowl}{Dowker,J.S. {\it Functional determinants on M\"obius corners};
    Proceedings, `Quantum field theory under
    the influence of external conditions', 111-121,Leipzig 1995.}
  \ref{Dowqg}{Dowker,J.S. in {\it Quantum Gravity}, edited by
  S. C. Christensen (Hilger, Bristol, 1984).}
  \ref{Dowit}{Dowker,J.S. \jpa{11}{1978}{347}.}
  \ref{Kane}{Kane,R. {\it Reflection Groups and Invariant Theory} (Springer,
  New York, 2001).}
  \ref{Sturmfels}{Sturmfels,B. {\it Algorithms in Invariant Theory}
  (Springer, Vienna, 1993).}
  \ref{Bourbaki}{Bourbaki,N. {\it Groupes et Alg\`ebres de Lie}  Chap.III, IV
  (Hermann, Paris, 1968).}
  \ref{SandTy}{Schwarz,A.S. and Tyupkin, Yu.S. \np{242}{1984}{436}.}
  \ref{Reuter}{Reuter,M. \prD{37}{1988}{1456}.}
  \ref{EGH}{Eguchi,T. Gilkey,P.B. and Hanson,A.J. \prp{66}{1980}{213}.}
  \ref{DandCh}{Dowker,J.S. and Chang,Peter, \prD{46}{1992}{3458}.}
  \ref{APS}{Atiyah M., Patodi and Singer,I.\mpcps{77}{1975}{43}.}
  \ref{Donnelly}{Donnelly.H. {\it Indiana U. Math.J.} {\bf 27} (1978) 889.}
  \ref{Katase}{Katase,K. {\it Proc.Jap.Acad.} {\bf 57} (1981) 233.}
  \ref{Gilkey3}{Gilkey,P.B.\invm{76}{1984}{309}.}
  \ref{Degeratu}{Degeratu.A. {\it Eta--Invariants and Molien Series for
  Unimodular Groups}, Thesis MIT, 2001.}
  \ref{Seeley}{Seeley,R. \ijmp {A\bf18}{2003}{2197}.}
  \ref{Seeley2}{Seeley,R. .}
  \ref{melrose}{Melrose}
  \ref{DandW}{Douglas,R.G. and Wojciekowski,K.P. \cmp{142}{1991}{139}.}
  \ref{Dai}{Dai,X. \tams{354}{2001}{107}.}
\end{putreferences}

\bye